\documentclass[11pt]{article}

\usepackage{latexsym,color,graphicx,amsmath,amssymb}

\def\reals{{\mathbb R}}
\def\eps{{\varepsilon}}
\def\bd{{\partial}}

\def\C{\mathcal{C}}

\newcommand{\A}{\mathcal{A}}

\newcommand{\oone}{Q_1}
\newcommand{\othree}{Q_3}

\def\reals{{\mathbb R}}
\def\eps{{\varepsilon}}
\def\bd{{\partial}}
\def\reals{{\mathbb R}}
\def\sph{{\mathbb S}}
\def\ZZ{{\mathbb Z}}
\def\A{{\cal A}}
\def\B{{\cal B}}
\def\C{{\cal C}}

\def\F{{\cal F}}

\graphicspath{{./images/}}

\newtheorem{theorem}{Theorem}[section]
\newtheorem{lemma}[theorem]{Lemma}
\newtheorem{corollary}[theorem]{Corollary}
\newtheorem{proposition}[theorem]{Proposition}
\newtheorem{claim}[theorem]{Claim}
 
\usepackage{tikz}
\usetikzlibrary{decorations.markings}
\tikzset{%
 	point/.style={circle,inner sep=1.5pt,minimum size=1.5pt,draw,fill=#1},
 }
\begin{document}

\title{Throwing a Sofa Through the Window\thanks{%
Work on this paper by DH and IY has been supported in part by the Israel Science
Foundation (grant no.~1736/19), by NSF/US-Israel-BSF (grant no. 2019754),
by the Israel Ministry of Science and Technology (grant no.~103129), 
by the Blavatnik Computer Science Research Fund, and by a grant from Yandex. 
Work by MS and IY has been supported in part by
Grant 260/18 from the Israel Science Foundation. Work by MS has also been supported
by Grant G-1367-407.6/2016 from the German-Israeli Foundation for Scientific Research and Development,
and by the Blavatnik Research Fund in Computer Science at Tel Aviv University.}}

\author{Dan Halperin\thanks{%
Tel Aviv University School of Computer Science, Israel; {\sf danha@tauex.tau.ac.il; http://acg.cs.tau.ac.il/danhalperin; https://orcid.org/0000-0002-3345-3765}}
\and
Micha Sharir\thanks{%
Tel Aviv University School of Computer Science, Israel; {\sf michas@post.tau.ac.il; https://orcid.org/0000-0002-2541-3763}}
\and
Itay Yehuda\thanks{%
Tel Aviv University School of Computer Science, Israel; {\sf itayyehuda1@mail.tau.ac.il; https://orcid.org/0000-0002-1825-0097}}
}

\date{}

\maketitle

\begin{abstract}
We study several variants of the problem of moving a convex polytope $K$, with $n$ edges, 
in three dimensions through a flat rectangular (and sometimes more general) window. Specifically: 

\noindent{\bf (i)}
We study variants where the motion is restricted to translations only, 
discuss situations where such a motion can be reduced to sliding (translation in a 
fixed direction), and present efficient algorithms for those variants, which run 
in time close to $O(n^{8/3})$. 

\noindent{\bf (ii)}
We consider the case of a \emph{gate} (or a slab, an unbounded window with two parallel 
infinite edges), and show that $K$ can pass through such a window, by any collision-free
rigid motion, if and only if it can slide 
through it, an observation that leads to an efficient algorithm for this variant too.

\noindent{\bf (iii)}
We consider arbitrary compact convex windows, and show that if $K$ can pass
through such a window $W$ (by any motion) then $K$ can slide through a gate of 
width equal to the diameter of $W$.

\noindent{\bf (iv)}
We show that if a purely translational motion for $K$ through a rectangular window $W$ exists, then $K$ can also slide through $W$ keeping the same orientation as in the translational motion. For a given fixed orientation of $K$ we can determine in linear time whether   
$K$ can translate (and hence slide) through $W$ keeping the given orientation, and if so plan the motion, also in linear time.

\noindent{\bf (v)}
We give an example of a polytope that cannot pass through a certain window by 
translations only, but can do so when rotations are allowed.

\noindent{\bf (vi)}
We study the case of a circular window $W$, and show that, for the regular tetrahedron 
$K$ of edge length $1$, there are two thresholds $1 > \delta_1\approx 0.901388 > \delta_2\approx 0.895611$,
such that (a) $K$ can slide through $W$ if the diameter $d$ of $W$ is $\ge 1$, 
(b) $K$ cannot slide through $W$ but can pass through it by a purely translational 
motion when $\delta_1\le d < 1$, 
(c) $K$ cannot pass through $W$ by a purely translational motion but can do it 
when rotations are allowed when $\delta_2 \le d < \delta_1$, and 
(d) $K$ cannot pass through $W$ at all when $d < \delta_2$.

\noindent{\bf (vii)}
Finally, we explore the general setup, where we want to plan a general motion 
(with all six degrees of freedom) for $K$ through a rectangular window $W$, and 
present an efficient algorithm for this problem, with running time close to $O(n^4)$.
\end{abstract}

\section{Introduction}

\begin{figure}[htbp]
\begin{center}
\includegraphics[scale = 0.2]{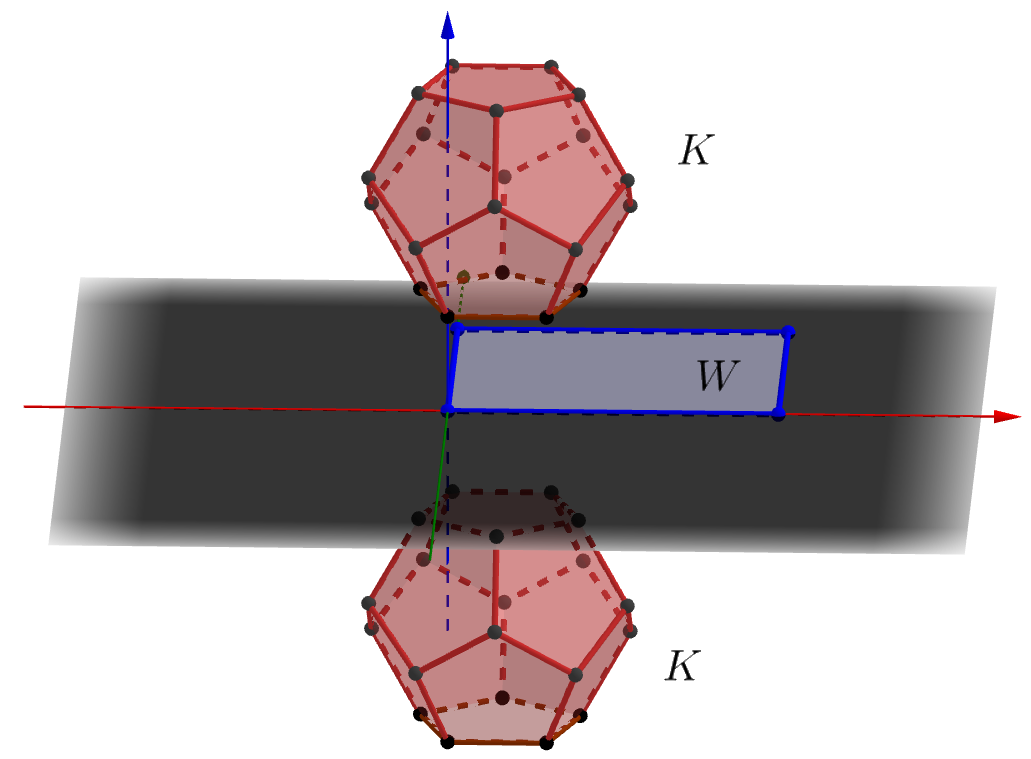}
\caption{\sf{Moving a convex polytope $K$ through a window $W$.}}
\label{throwing-sofa-problem}
\end{center}
\end{figure}

Let $K$ be a convex polytope (a `sofa') in $\reals^3$ with $n$ edges, and let $W$ be a rectangular
window, placed in the $xy$-plane in the axis-parallel position $[0,a]\times [0,b]$,
where $a$ and $b$ are the respective width and height of $W$. We assume
that the complement of $W$ in the $xy$-plane is a solid wall that $K$ must avoid. The problem is 
to determine whether $K$ can be moved, in a collision-free manner, from any position that is
fully contained in the upper halfspace $z>0$, through $W$, to any position that is fully 
contained in the lower halfspace $z<0$, and, if so, to plan such a motion (see Figure~\ref{throwing-sofa-problem}).

A continuous motion of a rigid body in three dimensions has six degrees of freedom, 
three of translation and three of rotation, and in the
general form of the problem, studied in Section~\ref{sec:6dofs},
we allow all six degrees. On the way, we will study simpler versions
where only restricted types of motion are allowed, such as purely translational motion 
(that has only three degrees of freedom), a translational motion in a fixed direction, 
which we refer to as \emph{sliding} (one degree of freedom), or a translational motion 
combined with rotations around the vertical axis only (four degrees of freedom), etc.
Some of our main results show that, in certain favorable situations, the existence of 
a general collision-free motion of $K$ through $W$ implies the existence of a restricted 
motion of one of these types.
This allows us to solve the problem in a significantly more efficient manner.

In terms of the \emph{free configuration space} $\F$ of $K$, 
all the placements of $K$ that are fully contained in the upper (resp., lower) 
halfspace are free, and form a connected subset $\F^+$ (resp., $\F^-$) of $\F$.
Our problem, in general, 
is to determine whether both $\F^+$ and $\F^-$ are contained in the same connected 
component of $\F$. This interpretation applies to the general setup, with six degrees 
of freedom, as well as to any other subclass of motions, with fewer degrees of freedom.

Motion planning is an intensively studied problem in computational geometry and robotics. 
A systematic and general way to describe the free space $\F$ is by using 
\emph{constraint surfaces}, namely surfaces describing all the configurations where 
some feature on the boundary of the moving object ($K$ in our case) touches a 
feature on the boundary of the free workspace ($\bd W$ in our case);
see, e.g.,~\cite{hss-amp-18,DBLP:books/daglib/0068760,DBLP:books/cu/L2006}.
These surfaces partition the configuration space into cells, each of which is 
either fully contained in $\F$ or fully contained in the forbidden portion of 
the configuration space. This representation is based on the arrangement 
$\A$ of the constraint surfaces, whose number in our case is $O(n)$, one surface for 
each feature (edge or vertex) of $W$ and each feature (edge, vertex or face) of $K$. 
Each cell of $\A$ is fully contained either in
$\F$ or in its complement. Hence the complexity of $\F$ is $O(n^d)$, 
where $d$ is the number of degrees of freedom (namely, the dimension of the 
configuration space)~\cite{hss-amp-18}. To exploit this representation, we 
construct and transform it into a so-called discrete \emph{connectivity graph}, 
which can be searched for the existence of a motion of the desired kind. 

One common way of doing this is to further decompose the arrangement into subcells of 
constant complexity, using \emph{vertical decomposition}~\cite{DBLP:journals/tcs/ChazelleEGS91}. 
Such constructions are easily implementable for motion planning with two degrees of 
freedom~\cite{DBLP:books/daglib/0028679}, but become significantly more involved for 
problems with three or more degrees of freedom. This has lead to the development of
alternative methods, such as sampling-based techniques (see~\cite[Chapter~7]{CBHKKLT05}
and \cite{hks-r-18}), the best known of which are PRM~\cite{KSLO96} and RRT~\cite{KL00}, 
which have dozens of variants. While extremely successful in solving practical problems, 
they trade-off the completeness of the arrangement approach with efficiency, and may fail when 
the setting contains tight passages~\cite{DBLP:journals/tase/SalzmanHH15,DBLP:journals/algorithmica/SalzmanHRH13}, 
a situation that can arise in the problems that we study in this paper.

Toussaint~\cite{T85} studied movable separability of sets, where he
collected a variety of tight-setting motion planning problems, similar in nature
to the problems studied here. These problems are interesting theoretically (see, e.g., \cite{SS94} 
for such a problem and its intriguing solution), but also from an applied perspective, 
since motion in tight settings often arises in manufacturing processes such as 
\emph{assembly planning}~\cite{DBLP:journals/algorithmica/HalperinLW00} or 
\emph{casting and molding}~\cite{DBLP:conf/case/BoseHS17}. 

It was in Toussaint's review~\cite{T85} that we encountered the problem of `throwing'
a polytope through a window. Although Toussaint's paper was published 35 years ago, 
we are not aware of any previous progress on this specific problem. We remark that 
the word \emph{sofa} in the title of the paper is borrowed from the classical 
two-dimensional \emph{moving sofa} problem (see, e.g.,~\cite{CFG91, GP14}), 
which is to find the shape of largest area that can be moved through a corner in 
an L-shaped corridor whose legs have width 1. 

%
%

\paragraph{Our results.}
We first consider, in Section~\ref{sec:1d}, sliding motions (translations in a fixed direction) of $K$. 
We characterize situations in which such a sliding motion exists, and present efficient 
algorithms, with runtime close to $O(n^{8/3})$, for finding such a motion when one exists.

We next consider, in Section~\ref{sec:unb}, the case where $W$ is an unbounded slab, enclosed between
two parallel unbounded lines (we call it a \emph{gate}). We show that if $K$ can pass through
such a gate $W$, by any collision-free rigid motion, it can also slide throgh $W$, making 
the general motion planning problem through a gate particularly easy to solve. 

In Section~\ref{sec:projection}, we consider arbitrary
compact convex windows, and show that if $K$ can move through such a winodw $W$, by an
arbitrary collision-free motion, then $K$ can slide through a gate of width equal to the
diameter of $W$, and this holds in any sliding direction. 
This requires nontrivial topological arguments, presented
in Section~\ref{sec:projection}.

We then consider, in Section~\ref{app:purely-translational}, 
purely translational motions of $K$ through 
a rectangular window $W$, and prove that the existence of such a purely-translational 
collision-free motion implies the existence of a collision-free sliding motion keeping the same orientation as in the translational motion.
For a given fixed orientation of $K$ we can determine in linear time whether   
$K$ can translate (and hence slide) through $W$ keeping the given orientation, and if so plan the motion, also in linear time.

In Section~\ref{app:rotations-needed}, we show that rotations are sometimes needed,
by giving an example of a convex polytope $K$ (actually a tetrahedron) that can move 
through a square window $W$ by a collision-free motion that includes rotation (only 
around an axis orthogonal to $W$), but there is no purely translational motion of $K$ through $W$.

In Section~\ref{sec:circular-window}, we consider the problem of passing through 
a circular window $W$, and show that, for the regular tetrahedron $K$ of edge length $1$, there are 
two thresholds
$1 > \delta_1\approx 0.901388 > \delta_2\approx 0.895611$, such that 
(i) $K$ can slide through the window $W$ if the diameter $d$ of $W$ is $\ge 1$, 
(ii) $K$ cannot slide through $W$ but can pass through it by a purely translational 
motion when $\delta_1\le d < 1$, (iii) $K$ cannot pass through $W$ by a purely 
translational motion but can do it with rotations when 
$\delta_2 \le d < \delta_1$, and (iv) $K$ cannot pass through $W$ at all 
when $d < \delta_2$.

We finally consider, in Section~\ref{sec:6dofs}, the general problem, with 
all six degrees of freedom. We present an efficient algorithm, which runs in 
time close to $O(n^4)$, for constructing the free configuration space, from 
which one can construct, within a comparable time bound, a valid motion 
through $W$ if one exists. We remark that the exponent $4$ is a significant
improvement over the `na\"ive' exponent $6$ that would arise by a general
treatment of the problem as a motion planning problem with six degrees of freedom.

\section{Translation in a fixed direction} \label{sec:1d}

In this section we address the case in which the movement is
purely translational in a single fixed direction. Such a motion, to which we
refer as a \emph{sliding motion}, has only one degree of freedom. 
In the most restricted version (which is very easy to solve), 
we are given a fixed orientation of $K$ at a fixed initial placement, 
and also the direction of motion. In this section we study a more general 
setting, in which we seek values for these parameters---orientation,
initial placement, and direction of motion, for which such a sliding motion
of $K$ through $W$ is possible (or determine that no such motion is possible).

In Section~\ref{sec:orthogonal} we observe that if a sliding motion for $K$ exists, 
then $K$ can also slide in a direction orthogonal to the plane of the window. 
Using this and other structural properties of the problem, we transform the problem at hand
into a certain range searching problem. We present an efficient novel solution
to the latter problem, which yields an algorithm for solving our original problem, 
whose running time is close to $O(n^{8/3})$.

\subsection{The existence of an orthogonal sliding motion} \label{sec:orthogonal}

For the most general version of the sliding motion, in which none of the parameters 
(orientation, initial placement, and direction of motion) is prespecified, we have:
\begin{lemma} \label{lem:vertical}
If $K$ can slide through $W$ from some starting placement in some direction, 
then $K$ can slide through $W$, possibly from some other starting placement 
(and at another orientation), by translating it in the negative $z$-direction. 
\end{lemma}

\noindent{\bf Proof.}
Let $K_0$ be the starting placement of $K$ and let $\vec{v}$ be the direction 
of motion through $W$, for which the resulting sliding motion is collision-free.
Form the infinite prism $\Pi_0 := \bigcup_{\lambda\in\reals} (K_0 + \lambda\vec{v})$ 
that $K_0$ spans in direction $\vec{v}$. The premise of the lemma implies that 
the intersection of $\Pi_0$ with the $xy$-plane is contained in $W$.

\begin{figure}[htbp]
\begin{center}
\includegraphics[scale = 0.2]{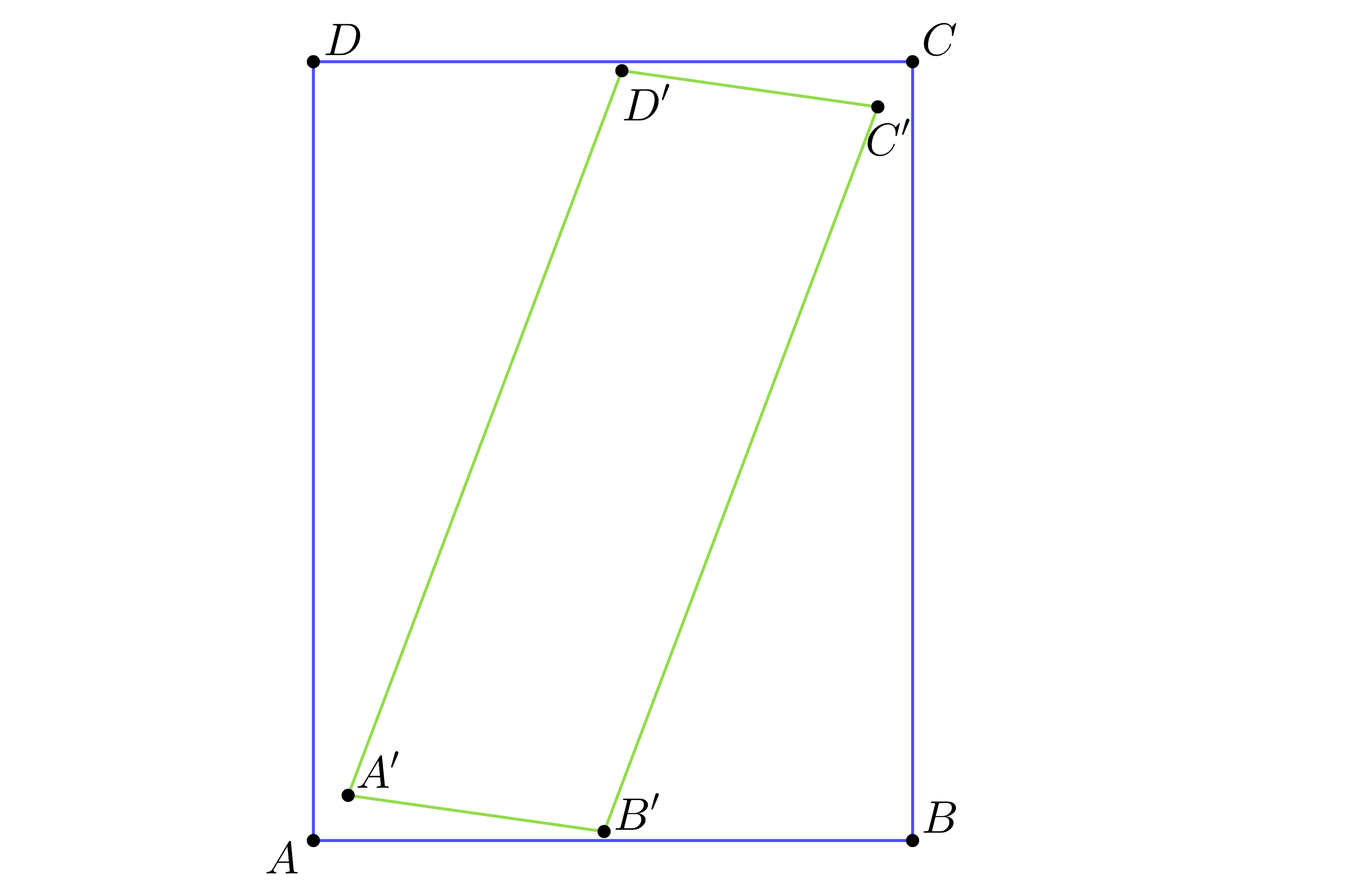}
\caption{\sf{The projection of $W$ (green) can be located in a congruent copy of $W$ (blue).}}
\label{rectangle-projection}
\end{center}
\end{figure}

Let $W_0$ be the orthogonal projection of $W$ onto some plane orthogonal to $\vec{v}$.
Note that $W_0$ is a parallelogram, and that, by construction, $K_0$ can pass through $W_0$
when translated in direction $\vec{v}$. By an old result, reviewed and proved by
Debrunner and Mani-Levitska~\cite{DML86}, it follows that, when mapped rigidly into 
the $xy$-plane, $W_0$ (the `shadow' of $W$ in direction $\vec{v}$) can be placed fully within $W$
(see Figure~\ref{rectangle-projection}).\footnote{%
  Curiously, as shown in \cite{DML86}, this property, of containing your shadows, fails in higher dimensions.}

Now rotate and translate $\reals^3$ so that $\vec{v}$ becomes the (negative) 
$z$-direction, and the image of $W_0$ is fully contained in (the former, 
untransformed copy of) $W$. Then the image of $K$ under this transformation 
can be moved vertically down through $W$, in a collision-free manner, as asserted.
$\Box$ \medskip

Debrunner and Mani-Levitska's proof is involved, and 
applies to an arbitrary planar convex shape (showing that it contains its projection 
in any direction). For the sake of completeness, we give a simple alternative proof for the case of a rectangle.

\subsection{Every rectangle can cover its shadows} \label{vertical-another-appendix}

\begin{lemma} \label{lem:vertical-another}
Let $W$ be a rectangle on some plane $h$. Let $W_0$ be the projection of $W$ on 
the $xy$-plane. Then the $xy$-plane contains a congruent copy of $W$ that contains $W_0$.
\end{lemma}

\begin{figure}[htbp]
\begin{center}
\includegraphics[scale = 0.5]{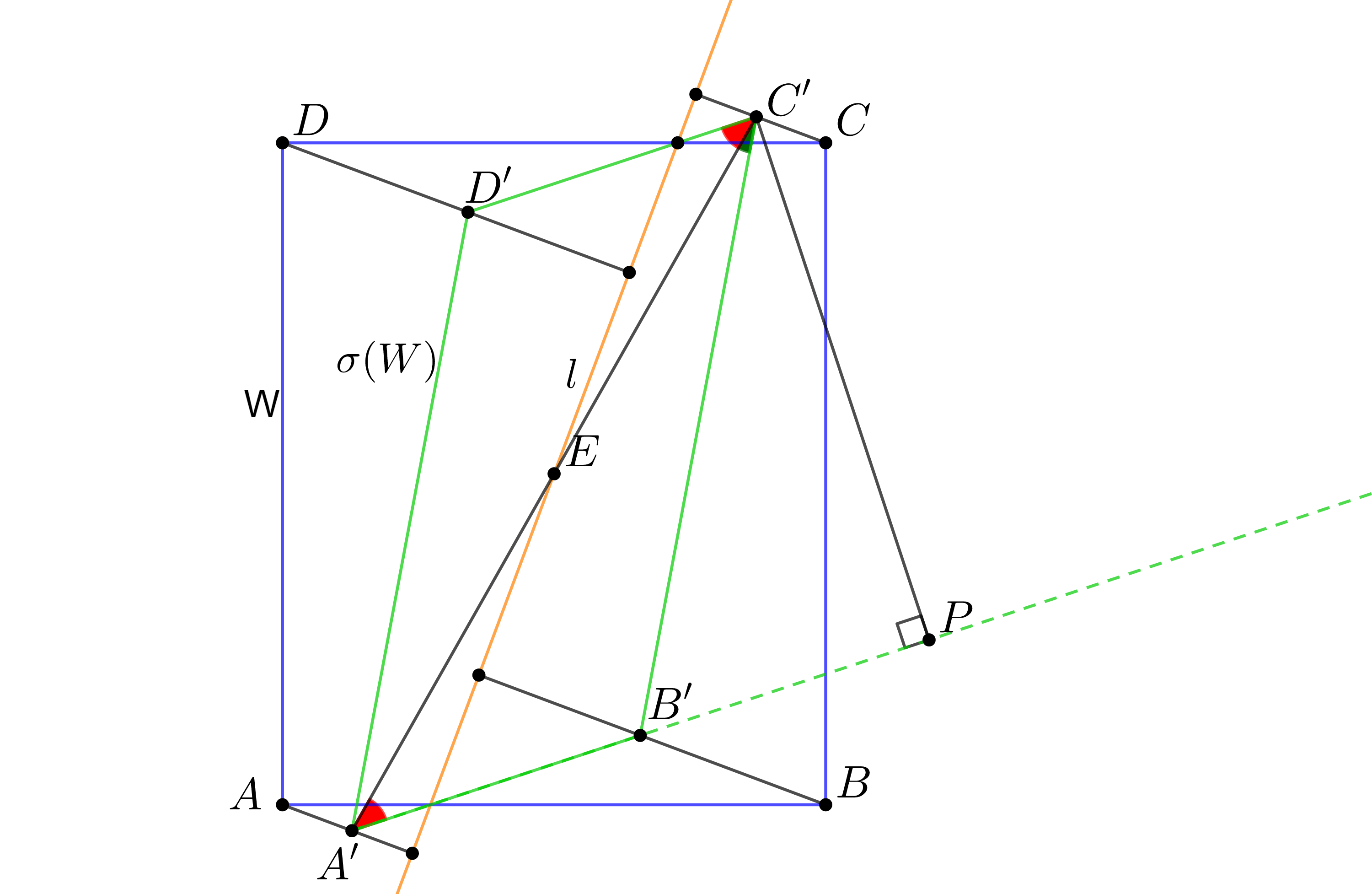}
\caption{\sf{The window $W$ (blue), the line $l$ (orange) and the image $\sigma(W)$ (green).}}
\label{vertical-i}
\end{center}
\end{figure}

\noindent{\bf Proof.} 
Denote the $xy$-projection by $\pi$. Let $l$ be the intersection line of $h$ and the $xy$-plane, 
and let $\alpha$ be the dihedral angle between these planes. Let $p$ be an arbitrary point on $h$, 
and let $d$ be the distance from $p$ to $l$. Then $\pi(p)$ lies at distance $d\cos\alpha$ from 
$l$ (with the same nearest point on $l$). Informally, $\pi$ moves every point in $h$ closer to 
$l$ by a factor of $\cos{\alpha}$. Then, instead of projecting $h$ to the $xy$ plane, we apply 
on $h$ this linear transformation that moves every point closer to $l$ by a factor of $\cos{\alpha}$. 
Denote this transformation by $\sigma$. This implies that every line segment in $h$ is transformed 
to a shorter segment or of the same length---no line segment increases its length.

Let $W=ABCD$, and let $A'=\sigma(A),B'=\sigma(B),C'=\sigma(C),D'=\sigma(D)$. Let $E$ denote the 
center of $W$ (see Figure~\ref{vertical-i}). Note that translating $W$ on $h$ keeps $\sigma(W)$ 
the same up to translation, so we may assume that $l$ passes through $E$ without loss of generality.

We use the following lemma:
\begin{lemma} \label{lem:vertical-angle}
Assume without loss of generality that $B$ and $C$ lie on one side of $l$, and that $A$ and $D$ 
lie on the other side (otherwise rename the vertices as $BCDA$), and that $l$ intersects the 
ray $\overrightarrow{BC}$, namely the ray starting at $B$ and passing through $C$ (otherwise 
rename the vertices as $DCBA$). Then $\sphericalangle A'C'B'\leq\sphericalangle ACB$.
\end{lemma}

\begin{figure}[htbp]
\begin{center}
\includegraphics[scale = 0.5]{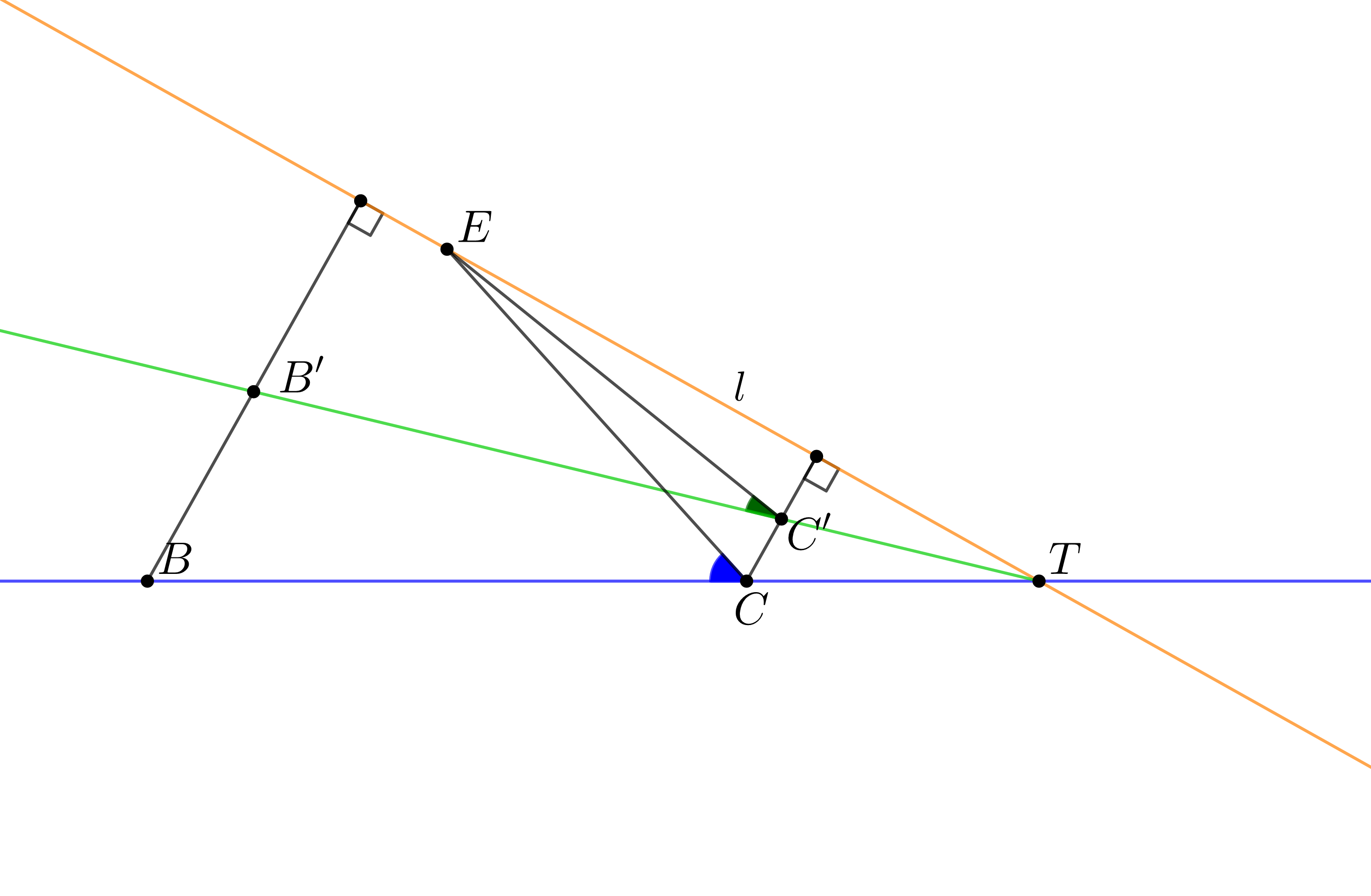}
\caption{\sf{The side $BC$ of $W$ (blue), the side $B'C'$ (green), and the line $l$ (orange).}}
\label{vertical-ii}
\end{center}
\end{figure}

\noindent{\bf Proof.} 
Denote by $T$ the intersection point of the lines $BC$ and $B'C'$ (see Figure~\ref{vertical-ii}). 
As $\sigma(BC)=B'C'$ the line $l$ must pass through $T$ since it is the only point of $BC$ that 
stays at the same location when applying $\sigma$. We then have:
$$
\sphericalangle A'C'B' = \sphericalangle EC'B' = \sphericalangle TEC'+\sphericalangle C'TE
\leq \sphericalangle TEC+\sphericalangle CTE = \sphericalangle ECB=\sphericalangle ACB.
$$
$\Box$ \medskip

Continuing with the proof of Lemma~\ref{lem:vertical-another}, there are two cases to consider:

\begin{figure}[htbp]
\begin{center}
\includegraphics[scale = 0.2]{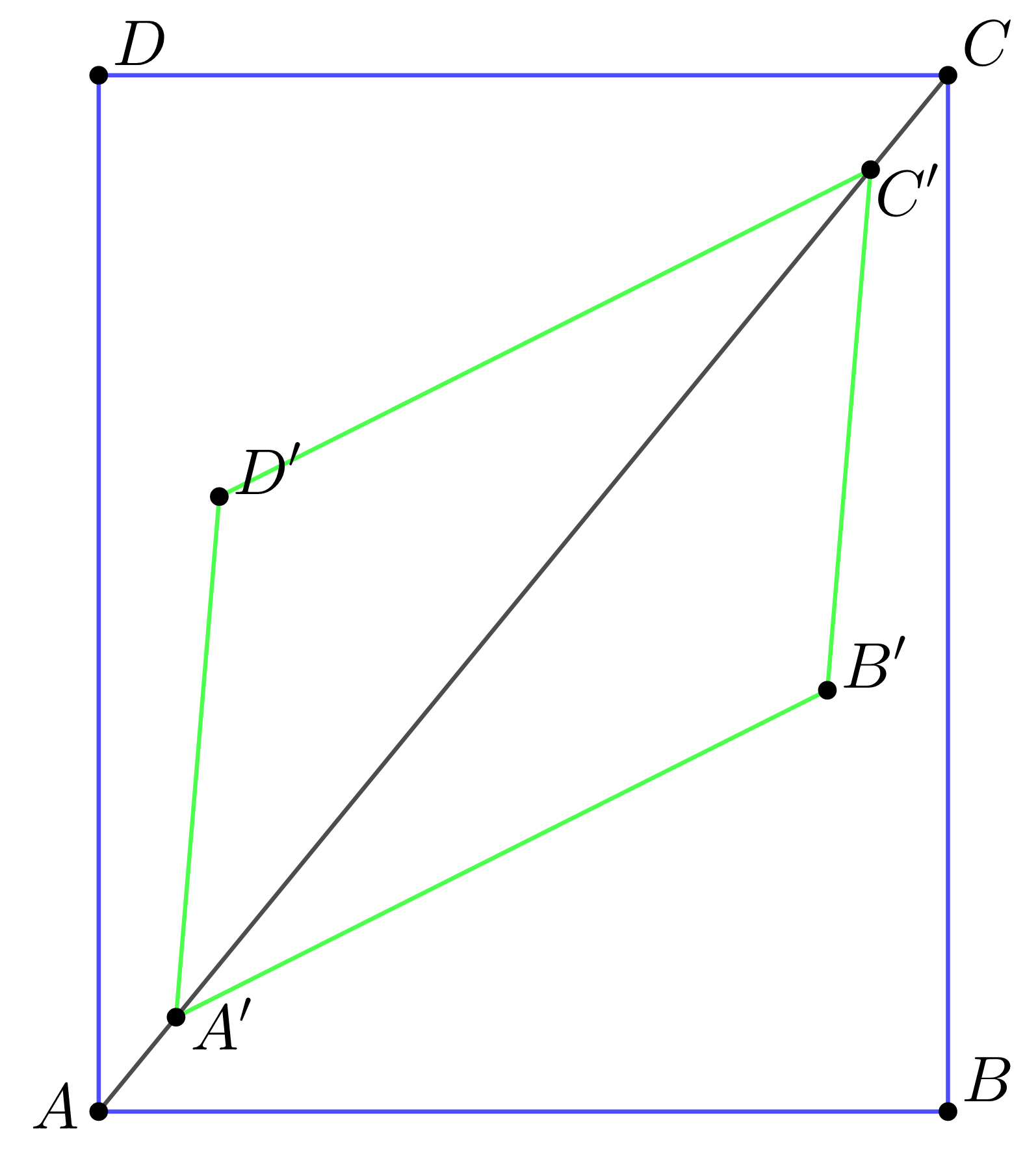}
\includegraphics[scale = 0.2]{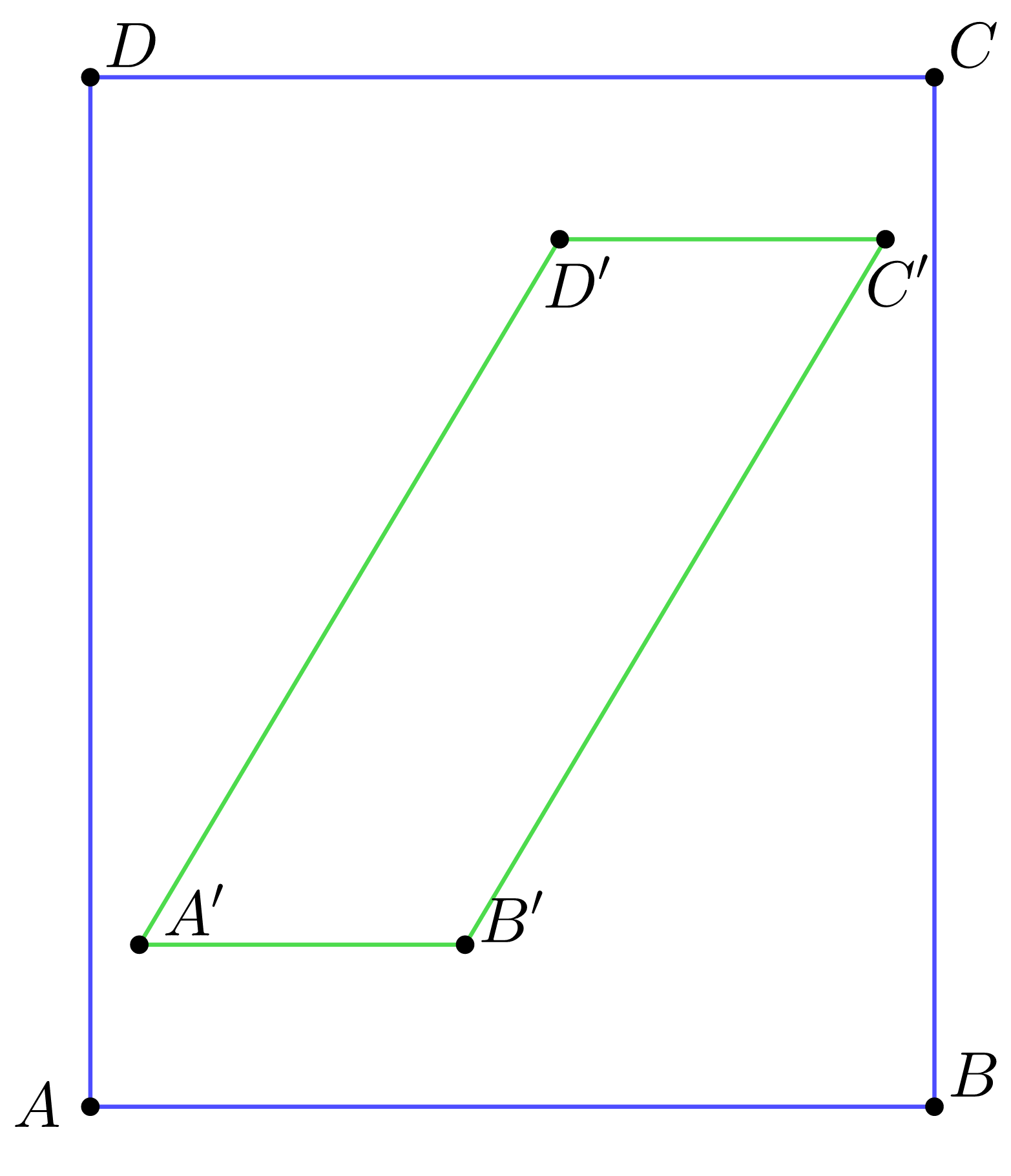}
\caption{\sf{Placing $\sigma(W)$ in a congruent copy of $W$. 
Left: Placing the diagonal $A'C'$ on the diagonal $AC$. 
Right: Placing the side $A'B'$ parallel to the side $AB$.}}
\label{parallelogram-in-rectangle}
\end{center}
\end{figure}

For any pair of points $P$ and $Q$ we denote by $\overline{PQ}$ the line through $P$ and $Q$.
If $\sphericalangle C'A'B'\leq\sphericalangle CAB$, then, since no line segment increases its 
length by applying $\sigma$, we have $A'C'\leq AC$. Denote by $m$ the line $\overline{AC}$. 
Place $A'C'$ on $m$, such that the points $A,A',C',C$ appear on $m$ in this order and $A'C'$ 
is centered at $E$ (see Figure~\ref{parallelogram-in-rectangle}, left). Note that the angle 
that $\overline{AB}$ forms with $m$ is greater than the angle that $\overline{A'B'}$ forms 
with $l$ (by assumption), and that the angle that $\overline{BC}$ forms with $m$ is greater 
than the angle that $\overline{B'C'}$ forms with $m$ (by Lemma~\ref{lem:vertical-angle}). 
Hence $B'$ is inside the triangle $ABC$. By symmetry, $D'$ is inside the triangle $CDA$, 
and therefore we successfully placed $\sigma(W)$ inside $W$.

If $\sphericalangle C'A'B'\geq\sphericalangle CAB$, draw from $C'$ a line perpendicular 
to $\overline{A'B'}$ and denote the intersection by $P$ (see Figure~\ref{vertical-i}). 
We place $\sigma(W)$ inside $W$ so that $A'B'$ is parallel to $AB$ (see 
Figure~\ref{parallelogram-in-rectangle}, right). To do so, we need to prove that 
$C'P\leq CB$ and that $A'P\leq AB$. Indeed, we have:
$$
CB\geq C'B'\geq C'P,
$$
$$
AB=CB\cdot\cot{\sphericalangle CAB}\geq C'P\cdot\cot{\sphericalangle C'A'B'}=A'P.
$$
Therefore we successfully placed $\sigma(W)$ inside $W$.
$\Box$ \medskip


\subsection{Finding a sliding motion} \label{sec:sliding-algo1}

The follwoing discussion is with respect to a fixed initial placement $K_0$ of $K$.
For a given direction $\vec{v}$, the \emph{projected silhouette} of $K_0$ in 
direction $\vec{v}$ is the boundary of the convex polygon obtained by the 
projection of $K_0$ in direction $\vec{v}$, within the image plane $h_{\vec{v}}$
(which is orthogonal to $\vec{v}$). 
The silhouette itself is the cyclic sequence of vertices and edges of $K$, 
whose projections form the projected silhouette.\footnote{%
  The silhouette is indeed such a cycle of vertices and edges of $\bd K$ for generic directions 
  $\vec{v}$. When $\vec{v}$ is parallel to a face $f$ of $K_0$, the entire $f$ is part of the silhouette.}
The silhouette and its projection do not change combinatorially, that is, when 
represented as a cyclic sequence of vertices and edges of $K$ (or of their projections), 
as long as $\vec{v}$ is not parallel to any face of $K_0$. We thus form the set
of the $O(n)$ great circles on $\sph^2$ that are parallel to the faces of $K_0$,
and construct their arrangement $\A_0$ on $\sph^2$, which is also known as the
\emph{aspect graph} of $K_0$~\cite{PD90}. In each face $\varphi$ of $\A_0$ the 
combinatorial structure of the silhouette is fixed, but the projected silhouette 
varies continuously as $\vec{v}$ moves in $\varphi$. 

A \emph{view} of $K_0$ is a pair $(\vec{v},\theta)$, where $\vec{v}$ is a direction, 
and $\theta$ is the angle of rotation of the projected silhouette within $h_{\vec{v}}$ 
(translations within that plane are ignored). The space of views 
is thus three-dimensional. A fixed view
$(\vec{v},\theta)$ fixes the uppermost, leftmost, bottommost and rightmost vertices 
$w_t$, $w_l$, $w_b$ and $w_r$ of the projected silhouette. The view is valid if
\begin{equation} \label{goodview}
x_{v_r} - x_{v_l} \le a \qquad \text{and} \qquad y_{v_t} - y_{v_b} \le b 
\end{equation}
(in the coordinate frame of $h_{\vec{v}}$ when rotated by $\theta$).

When $\vec{v}$ is fixed and $\theta$ varies, we get $O(n)$ quadruples 
$(w_t, w_l, w_b, w_r)$ of the projected silhouette. The view is valid if
the inequalities in (\ref{goodview}) (which depend on $(\vec{v},\theta)$) have a solution 
for one such quadruple, which lies in the appropriate portion of the view space
(in which $w_t, w_l, w_b, w_r$ are indeed the four extreme vertices). 
The existence of a valid view is equivalent to the
existence of a sliding motion of $K_0$ through $W$, after suitably shifting
$K_0$ and rotating it around $\vec{v}$ by $\theta$. We next show that the total number of quadruples of vertices is $O(n^3)$, 
from which we obtain an algorithm for finding a valid view, with near-cubic running time.

Returning to the aspect-graph arrangement $\A_0$, we observe that, since it
is composed of $n$ great circles, its complexity is $O(n^2)$. For all directions 
$\vec{v}$ within the same face of $\A_0$, the silhouette and its projection are fixed 
combinatorially, but the actual spatial positions of the projected vertices depend on the 
direction $\vec{v}$, and the projected silhouette can also rotate arbitrarily
within the image plane $h_{\vec{v}}$. (Note that in this discussion we completely ignore
translations of $K$, as they are irrelevant for the analysis and its conclusions.)

\begin{figure}[htbp]
\begin{center}
\includegraphics[scale = 0.3]{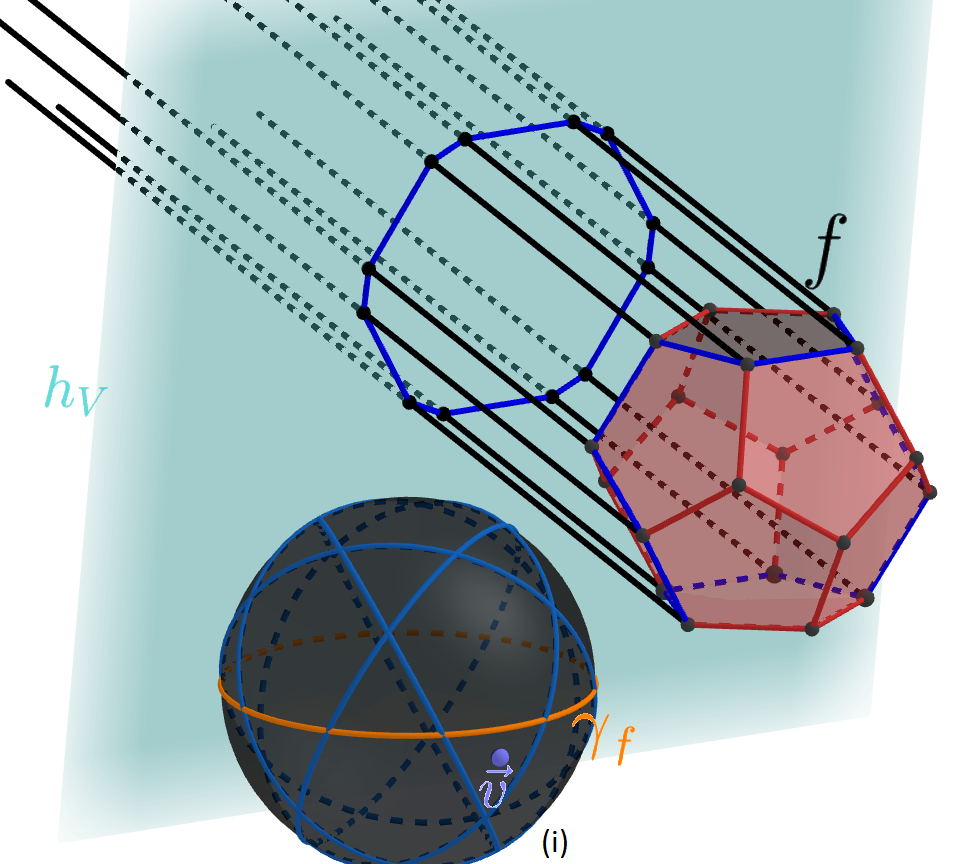}
\includegraphics[scale = 0.3]{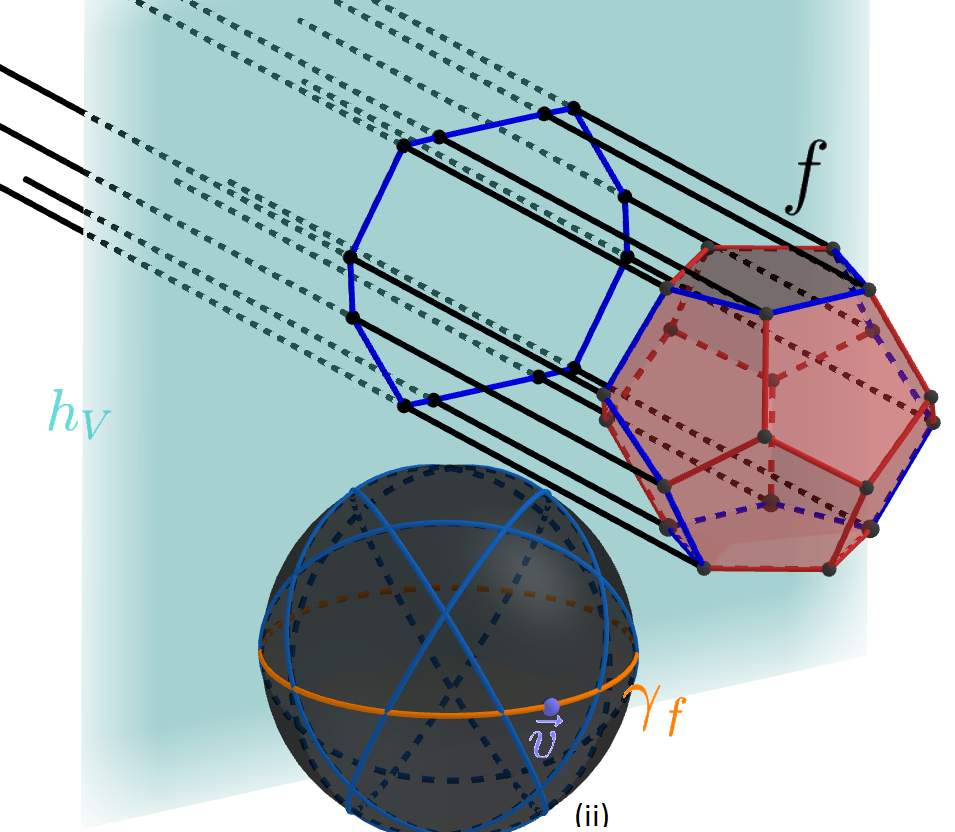}
\caption{\sf{The silhouette and its projection: (i) The case of a generic $\vec{v}$.
(ii) The case where $\vec{v}$ is parallel to a face $f$ of $K$ ($\vec{v}$ is on the great circle $\gamma_f$).}}
\label{silhouette}
\end{center}
\end{figure}

We assign some canonical coordinate frame to $h_{\vec{v}}$, and refer, for
simplicity, to its axes as the $x$- and $y$-axes (they depend on $\vec{v}$).
For example, excluding $O(1)$ problematic directions, which can be handled
separately, and easily, we can take the $x$-axis within $h_{\vec{v}}$
to be the intersection of $h_{\vec{v}}$ with the $xz$-plane, and take the
$y$-axis to be in the orthogonal direction within $h_{\vec{v}}$, oriented in
the direction that has a positive $y$-component.
The actual spatial location of the projected silhouette (up to translation,
which we ignore) of $K$ can be 
parameterized by $(\vec{v},\theta)$, where $\theta$ is the rotation of 
the projected silhouette within the image plane $h_{\vec{v}}$.
We refer to $(\vec{v},\theta)$ as the \emph{view} of $K$. See Figure~\ref{silhouette}.

\begin{figure}[htbp]
\begin{center}
\includegraphics[scale = 0.6]{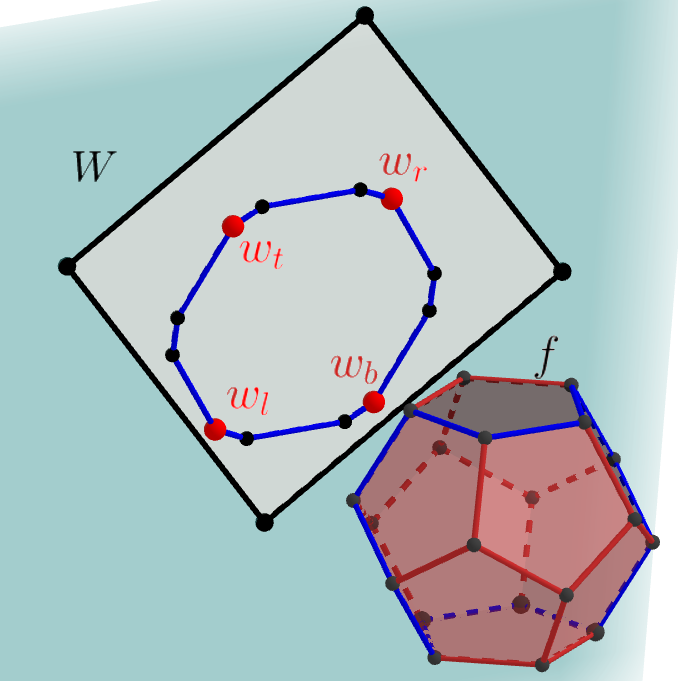}
\caption{\sf{A view of $K$. To simplify the visualization, we rotate the containing 
window $W$ rather than the projected silhouette. The leftmost, rightmost, topmost 
and bottommost vertices are highlighted.}}
\label{silhouette-in-window}
\end{center}
\end{figure}

As we vary $\vec{v}$ and $\theta$, we want to keep track of the leftmost
and rightmost vertices of the projected silhouette (in the $x$-direction), 
and of the topmost and bottommost vertices (in the $y$-direction, all with 
respect to the coordinate frame within $h_{\vec{v}}$).
We succeed when we find a projection (in direction $\vec{v}$), 
followed by a rotation (by $\theta$), for which the $x$-difference 
between the rightmost and leftmost vertices is at most $a$ and the 
$y$-difference between the topmost and bottommost vertices is at most $b$.
We reiterate that this is indeed the property that we need: It takes place
in a slanted plane $h_{\vec{v}}$ with respect to an artificial coordinate 
frame within that plane, but using a suitable rotation of $h_{\vec{v}}$ we can make 
it horizontal and its coordinate frame parallel to the standard $xy$-frame. 
A subsequent suitable translation then brings the projected silhouette 
to within $W$, as desired. 

\begin{figure}[htbp]
\begin{center}
\includegraphics[scale = 0.4]{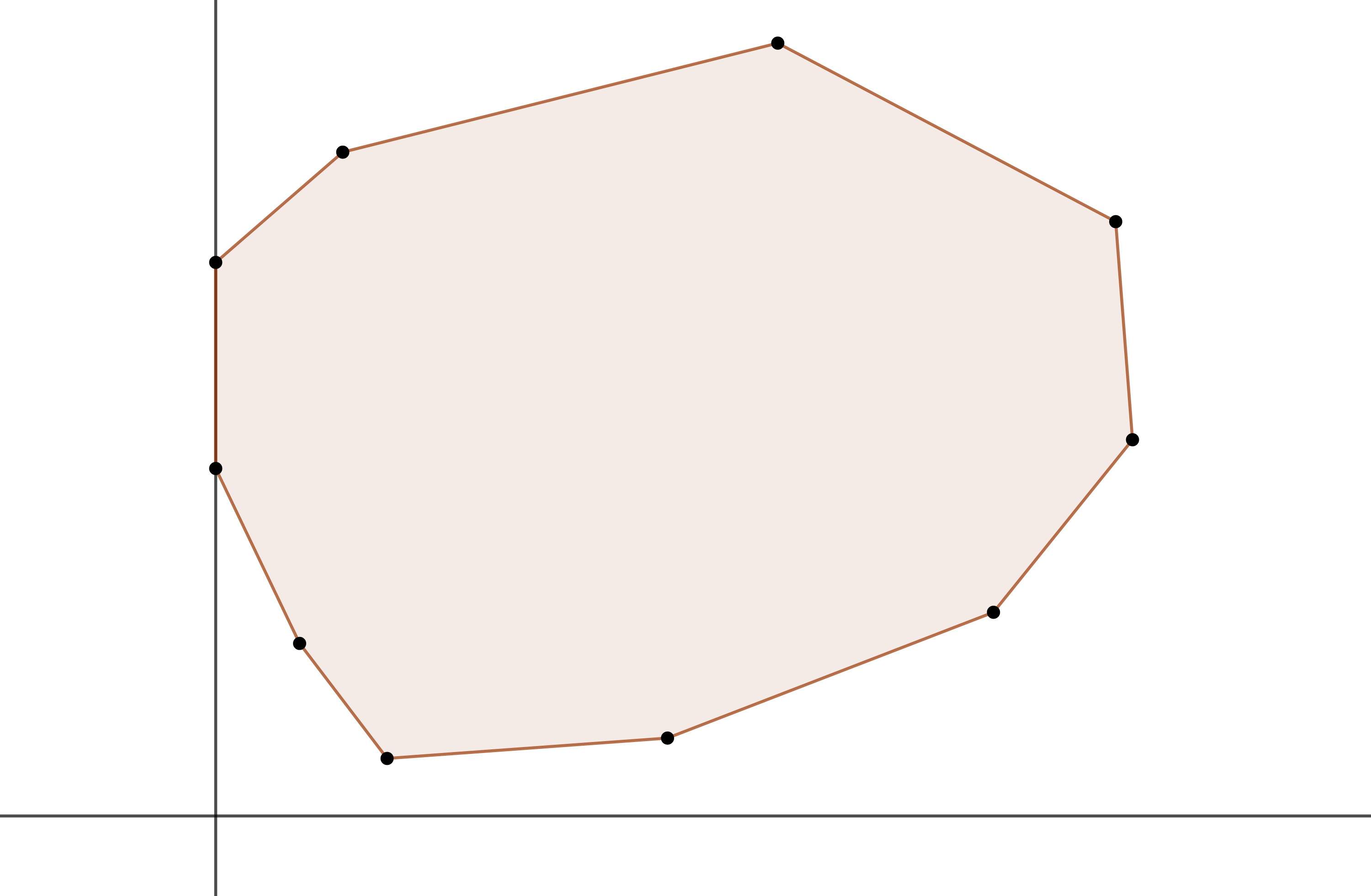}
\caption{\sf{A discrete change of the leftmost vertex of the projected silhouette.}}
\label{y-vertical}
\end{center}
\end{figure}

Fix a face $\varphi$ of $\A_0$, and let $w_1,w_2,\ldots,w_m$ denote the cyclic 
sequence of the vertices of the projected silhouette, say in counterclockwise
order, for views in $\varphi$. If the current leftmost vertex is some $w_j$, then it remains leftmost
as long as neither of the two adjacent edges $w_{j-1}w_j$ and $w_jw_{j+1}$
becomes $y$-vertical. (Recall that `leftmost' and `$y$-vertical' are with 
respect to the artificial frame within $h_{\vec{v}}$.) The views 
$(\vec{v},\theta)$ at which an edge $e$ of $K$, say, $w_{j-1}w_j$ is $y$-vertical comprise 
a two-dimensional surface $\rho_e$ in the three-dimensional 
space $V = \sph^2\times \sph^1$ of views $(\vec{v},\theta)$.
See Figure~\ref{y-vertical}.

The discussion so far has been for views that have a combinatorially fixed
silhouette. However, to make the algorithm for finding a sliding motion more efficient, we consider
all possible silhouettes `at once', using the following approach.
After forming the aspect-graph arrangement $\A_0$, as defined above, 
we replace each great circle $\gamma_f$ on $\sph^2$ by the cylindrical 
surface $\gamma^*_f = \gamma_f\times \sph^1$, and collect these surfaces 
into a set $\Gamma$, of cardinality $O(n)$. Then, for each edge $e$ of $K$ 
(regardless of whether it is a silhouette edge or not), we form the surface 
$\rho_e$, as just defined, and collect these surfaces into a set $\Sigma$, 
of cardinality $n$. We now form the three-dimensional arrangement 
$\A = \A(\Gamma\cup\Sigma)$ (note that all the surfaces of $\Gamma\cup\Sigma$ 
are two-dimensional). As is easily verified, for each three-dimensional 
cell $\tau$ of $\A$, the projected silhouette of $K$, and its 
leftmost, 
rightmost, topmost and bottommost vertices (we refer to them collectively
as the \emph{extreme} vertices of the projected sihouette) are fixed for 
all views in $\tau$. Since $|\Gamma\cup\Sigma| = O(n)$, the complexity 
of $\A$ is $O(n^3)$.

To obtain a representation that is easy to process further, we construct the 
\emph{vertical decomposition} of $\A$, which we denote as ${\rm VD}(\A)$. 
It is a decomposition of the three-dimensional cells of $\A$ into a total 
of nearly cubic number of prism-like subcells (that we simply call prisms). 
See Sharir and Agarwal~\cite[Section 8.3]{SA} for more details. A sharp bound 
on its complexity (i.e., the number of prisms) is $O(n^2\lambda_s(n))$, for 
some constant $s$ (a sharp estimation of the value of $s$ is not given in this paper), 
where $\lambda_s(n)$ is the maximum length of a Davenport--Schinzel sequence of order $s$ 
on $n$ symbols; see \cite{SA}. The vertical decomposition can be constructed in time 
$O(n^2\lambda_s(n)\log n)$~\cite{DBLP:journals/dcg/BergGH96}.

We now iterate over all prisms of ${\rm VD}(\A)$. For each prism
$\tau$, we retrieve the four extreme vertices of the projected silhouette, 
which are fixed for all views in $\tau$, and check whether there is 
a view in $\tau$ for which these vertices, and thus all of the projected silhouette, 
fit into $W$ (after suitable rotation and translation of $W$, as discussed above). 
To do so, denote these leftmost, rightmost, topmost and bottommost vertices as $w_l$, $w_r$, 
$w_t$ and $w_b$, respectively. The $x$-coordinates $x_{w_l}$, $x_{w_r}$ 
of $w_l$ and $w_r$, and the $y$-coordinates $y_{w_t}$, $y_{w_b}$ of $w_t$ 
and $w_b$ (within $h_{\vec{v}}$) are functions of $(\vec{v},\theta)$. 
We need to determine whether the region 
$$
S = S(w_l,w_r,w_t,w_b) := \{ (\vec{v},\theta) \in \sph^2\times \sph^1 \mid 
x_{w_r}(\vec{v},\theta) - x_{w_l}(\vec{v},\theta) \le a,\;\\
y_{w_t}(\vec{v},\theta) - y_{w_b}(\vec{v},\theta) \le b \} ,
$$
which is exactly the region of views $(\vec{v},\theta)$
at which $W$ contains a (rotated and translated) copy of the 
projected silhouette with these four specific vertices as the extreme vertices 
of the projection, has a nonempty intersection with $\tau$. 
Since $S$ and $\tau$ are semialgebraic regions of constant complexity, this 
test can be performed, in a suitable (and standard) model of real 
algebraic computation, in constant time~\cite{DBLP:books/daglib/0028679}.
Summing over all prisms $\tau$, the overall cost of these tests is 
proportional to the complexity of ${\rm VD}(\A)$, namely it is $O(n^2\lambda_s(n))$.

To complete the description of the algorithm, we now consider the task 
of computing the four extreme vertices $w_l$, $w_r$, $w_b$ and $w_t$
of the silhouette, or, more precisely, the four (fixed) vertices of $K$ 
that project to them, for each cell $c$ of $\A$. As an easy by-product 
of the construction of ${\rm VD}(\A)$, each of its prisms can be 
associated with the cell of $\A$ containing it, so the four extreme vertices 
will also be available for each prism of ${\rm VD}(\A)$.

By the nature of the surfaces forming $\A$, the projection of each 
cell $c$ of $\A$ onto $\sph^2$ is fully contained in a single cell
$\rho = \rho(c)$ of the two-dimensional aspect-graph arrangement 
$\A_0$. For each such cell $\rho$, the discrete nature of the 
silhouette, as a cyclic sequence of vertices (and edges) of $K$,
is fixed for every $\vec{v}\in\rho$ and for any $\theta\in\sph^1$.
Although we can do it faster, we simply iterate over the $O(n^2)$
cells of $\A_0$, and for each cell $\rho$, compute the silhouette 
in $O(n)$ time, in brute force (by picking an arbitrary point $\vec{v}$
in $\rho$, and by examining each edge of $K$ for being part of the
silhouette in direction $\vec{v}$). 
The overall cost of this step is thus $O(n^3)$.

Consider now a cell $c$ of $\A$, and let $\rho = \rho(c)$
be the cell of $\A_0$ that contains the $\sph^2$-projection of
$c$. Let $(u_1,u_2,\ldots,u_m)$ denote the cyclic counterclockwise
sequence of vertices of $K$ that forms the silhouette for directions in 
$\rho$, and let $w_i$ denote the $\sph^2$-projection of $u_i$,
for $i=1,\ldots,m$. Since the vertices of $K$ inducing  
$w_l$, $w_r$, $w_b$ and $w_t$ are fixed over $c$, it suffices
to compute them for a fixed arbitrary view in $c$. We thus fix
such a view $(\vec{v},\theta)$, and proceed as follows.

\begin{figure}[htbp]
\begin{center}
\includegraphics[scale = 0.5]{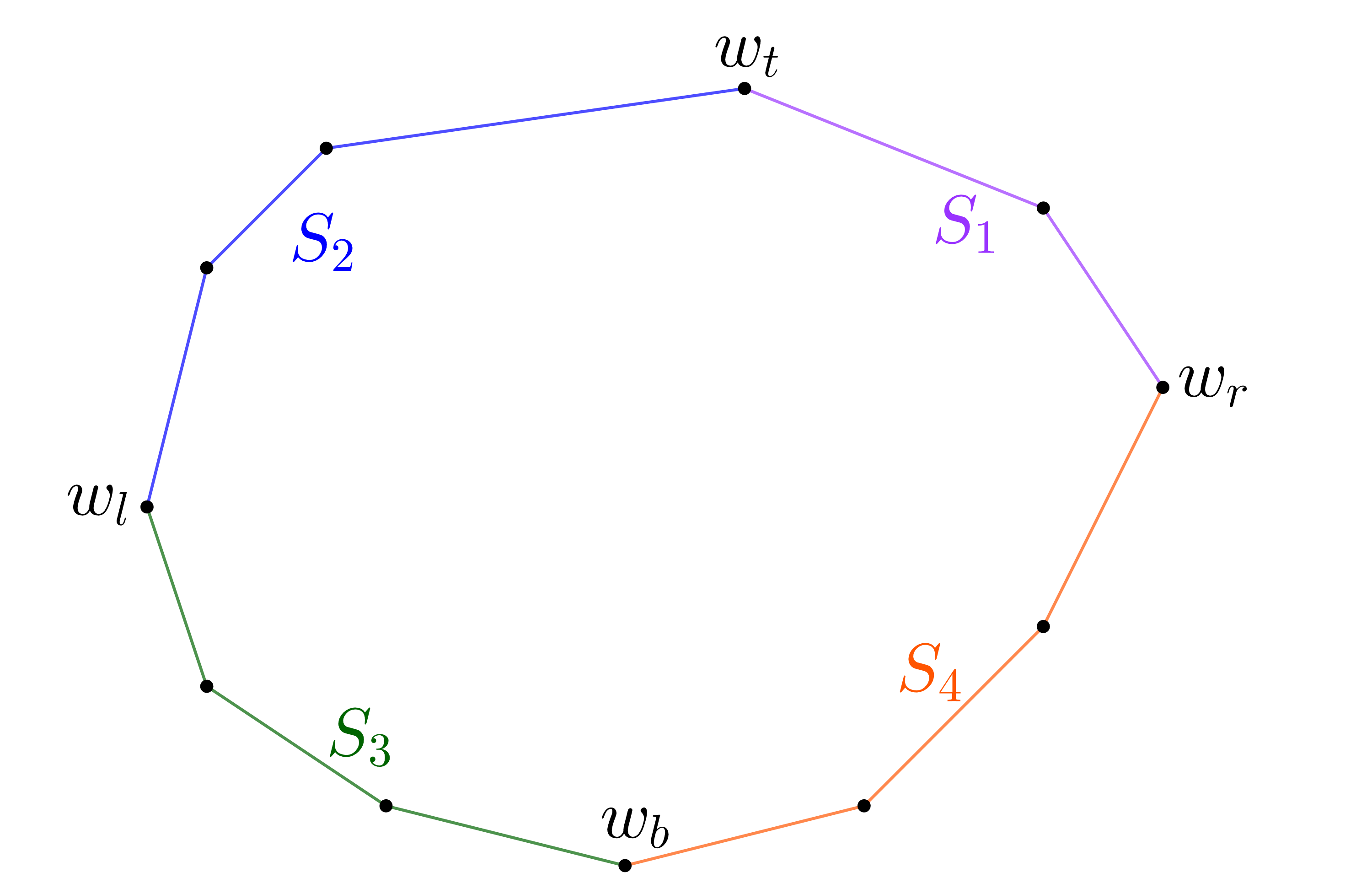}
\caption{\sf{The sub-silhouettes $S_1$ (purple), $S_2$ (blue), $S_3$ (green), $S_4$ (orange). 
The extreme vertices $w_r$, $w_t$, $w_l$, $w_b$ delimit these sub-silhouettes and are highlighted.}}
\label{extreme-vertices}
\end{center}
\end{figure}

For each $i$, define the ``derivative'' of the silhouette at $w_i$ to
be the pair of vectors 
$$
({\bf w}_i^-, {\bf w}_i^+) =
(\overrightarrow{w_{i-1}w_i},\; \overrightarrow{w_iw_{i+1}} ) ,
$$
where the vectors are represented in the coordinate
frame induced by $(\vec{v},\theta)$ in a plane orthogonal
to $\vec{v}$, and where addition and subtraction of indices
is modulo $m$. The extreme vertices $w_l$, $w_r$, $w_b$, $w_t$ 
partition the silhouette into (at most) four subsequences:
$S_1$, between $w_r$ and $w_t$, $S_2$, between $w_t$ and $w_l$,
$S_3$, between $w_l$ and $w_b$, and $S_4$, between $w_b$ and $w_r$ 
(see Figure~\ref{extreme-vertices}), so that, for $w_i\in S_1$ (resp., 
$S_2$, $S_3$, $S_4$) both vectors ${\bf w}_i^-$, ${\bf w}_i^+$ 
lie in the second (resp., third, fourth, first) quadrant.
For $w_r$ (resp., $w_t$, $w_l$, $w_b$), the vectors lie,
respectively, in the first and second (resp., second and third,
third and fourth, fourth and first) quadrants.\footnote{%
  We gloss here over the easy special cases of degeneracy,
  in which the extreme vertices are not all distinct. In such
  cases some of the sub-silhouettes $S_1,\ldots,S_4$ might
  be empty, and the rules for identifying the extreme vertices
  need to be adjusted.}
  
Using these observations, we find the four extreme vertices 
using binary search, as follows. We break the silhouette sequence
into two linear subsequences at $w_1$ and $w_{m/2}$, and find
the extreme vertices in each subsequence. Consider the subsequence
$(w_1,w_2,\ldots,w_{m/2})$. We compute the derivatives at 
$w_1$ and at $w_{m/2}$, and thereby identify the two respective
sub-silhouettes that contain these vertices. Suppose for
specificity that $w_1$ lies in $S_1$ and $w_{m/2}$ lies in $S_3$.
Then we know that our subsequence contains (only) $w_t$ and $w_l$,
and we can find each of them by a straightforward binary search,
using the derivatives to guide the search. We apply similar procedures 
in each of the other cases, and for the second subsequence
$(w_{m/2},w_{m/2+1},\ldots,w_1)$.

In conclusion, it takes $O(\log n)$ time to find the extreme vertices 
for each cell of $\A$, and thus also for each prism of ${\rm VD}(\A)$, 
for a total running time of $O(n^2\lambda_s(n)\log n)$.

\subsection{An improved algorithm for sliding motions} \label{sec:sliding-algo2}

We next present an improved, albeit more involved algorithm that solves the problem 
of finding a sliding motion of $K$, if one exists, in time $O(n^{8/3}{\rm polylog}(n))$.

The problem of finding a direction $\vec{v}$ in which we can slide $K$ through $W$ 
is equivalent to the problem of finding a placement of $W$ on some plane $h$ such 
that the projected silhouette of $K$ on $h$ is contained in $W$, which in turn is 
equivalent to verifying that all the vertices of $K$ are projected into that placement of $W$.

An equivalent way of checking for the latter characterization is to look for two unit 
vectors $x$ and $y$ (which will be the directions of the axes of $W$ in the desired 
placement; note that $h$ is spanned by $x$ and $y$) that satisfy:
\begin{description}
\item [(i)] $x$ and $y$ are perpendicular to each other.
\item [(ii)] For every segment $e$ connecting two vertices of $K$ we have $\langle x,e\rangle \le a$.
\item [(iii)] For every segment $e$ connecting two vertices of $K$ we have $\langle y,e\rangle \le b$.
\end{description}
(Note that since we go over all unordered pairs of vertices of $K$ in (ii), (iii), 
we actually require that $|\langle x,e\rangle|\le a$ and $|\langle y,e\rangle|\le b$ 
for each such segment $e$.) Every inequality in (ii) defines a halfspace that has to 
contain $x$. We intersect those $O(n^2)$ halfspaces, to obtain a convex polytope $Q$ 
of complexity $O(n^2)$, and intersect $Q$ with the unit sphere $\sph^2$ to obtain the 
admissible region $A$ of the vectors $x$ that satisfy (ii), in $O(n^2\log n)$. We apply 
the same procedure for $y$ using the suitable collection of halfspaces in (iii), and 
obtain the admissible region $B$ for the vectors $y$ that satisfy (iii), also in $O(n^2\log n)$.
To satisfy also (i), we need to check whether there exist an orthogonal pair of vectors 
$x\in A,y\in B$. We use the following lemma. 
\begin{lemma}  \label{lem:good_pair:app}
Let $S_A$ denote the set of all vertices of $A$, and let $T_A$ denote the set of the 
points that are closest locally to the north pole of $\sph^2$ along each circular arc 
of the boundary of $A$. (By choosing a generic direction for the north pole of $\sph^2$ 
we may assume that $T_A$ is finite and $\left|S_A\cup T_A\right|= O(n^2)$.) Define 
similarly the sets $S_B,T_B$. If there exist an orthogonal pair $\left(x,y\right)\in A\times B$ 
then there exist such an orthogonal pair so that either $x\in S_A\cup T_A$ or $y\in S_B\cup T_B$.
\end{lemma}

\noindent{\bf Proof.}
We refer to an orthogonal pair in $A\times B$ as a \emph{good pair}.
Let $\left(x,y\right)$ be a good pair such that $x$ is as close to the boundary of 
$A$ as possible. If there are multiple pairs with this property, pick the one in 
which $x$ is the closest to the north pole. If there are still multiple pairs, pick 
an arbitrary pair among them. By continuity and the compactness of $A$ and $B$, 
it is easy to show that such a ``minimal pair" exist.

Several cases can arise:
\begin{enumerate}
\item 
$x$ or $y$ is one of the desired vertices. In this case we are done.
\item 
Both $x$ and $y$ lie in the interiors of $A$ and $B$, respectively.
In this case they can be moved slightly together in any direction, while maintaining 
their mutual orthogonality. In particular, $x$ can get closer to the boundary of $A$ 
so $\left(x,y\right)$ is not the minimal pair.
\item 
$x$ is on the boundary of $A$, and $y$ is in the interior of $B$. 
Since we are not in Case 1, $x$ lies in the relative interior of an edge of 
$\bd A$ and is not the point on that edge that is closest to the north pole.
Then we have two available directions to move $\left(x,y\right)$ slightly such 
that $x$ remains on the same edge. One of these directions brings $x$ to a point 
closer to the north pole, so $\left(x,y\right)$ is not the minimal pair.
\item 
$y$ is on the boundary of $B$ (as in Case 3 we may assume that $y$ lies in the 
relative interior of an edge of $\bd B$). In this case we fix $x$ and move $y$ 
along the great circle $C_x$ of points perpendicular to $x$. Recall that $y$ is 
on an edge of $B$, which is a circular arc $\gamma$. Every halfspace of the 
intersection contains the origin, so $B$ is contained in the bigger portion 
$C^+$ (bigger than a hemisphere) of $\sph^2$ that is bounded by the circle $C$ 
containing $\gamma$. Since $C_x$ is a great circle, it is bigger than $C$, so 
when moving $y$ along $C_x$ in at least one of the two possible directions, 
$y$ enters $C^+$ (this is always true, regardless of the size of $C_x$, when 
the circles cross one another at $y$; the fact that $C_x$ is larger is needed 
when they are tangent at $y$), so it enters the interior of $B$. Now we are 
in one of the cases $2, 3$ that we have already settled.
\end{enumerate}
Having covered all possible cases, this completes the proof of the lemma.
$\Box$ \medskip

We iterate over the points of $S_A\cup T_A$. For each such point $v$ let $C_v$ 
be the great circle of vectors perpendicular to $v$, and let $\C$ denote the 
collection of these $O(n^2)$ great circles. We face the problem of determining 
whether any great circle in $\C$ crosses $B$. This is the same as determining 
whether any great circle in $\C$ crosses an arc of $\bd B$. This is a variant 
of the batched range searching paradigm, and we present next a detailed solution 
for this case. We apply a fully symmetric procedure to the collection of great 
circles orthogonal to the points of $S_B\cup T_B$ and to $A$. If we find a valid 
intersection it gives us a valid orthogonal pair. Otherwise, such a pair does 
not exist.

\paragraph{Detecting an intersection between the great circles of $\C$ and the boundary arcs of $B$.}
We apply a central projection (from the center of $\sph^2$) onto some plane,
say a horizontal plane $h$ lying below $\sph^2$ (with a generic choice of the coordinate 
frame, we may assume that none of the points in $S_A\cup T_A\cup S_B\cup T_B$ 
are on the great circle that is parallel to $h$). This is a bijection of the open 
lower hemisphere onto $h$, in which (the lower portions of) great circles are mapped to 
lines, and (the lower portions of) circular arcs are mapped to arcs of conic sections (ellipses,
parabolas, hyperbolas, or straight lines). This transforms the problem into a batched 
range searching problem, in which we have a set $L$ of $M = O(n^2)$
lines (which arise from the great circles orthogonal to the points of
$S_A\cup T_A$) and a set $E$ of $N = O(n^2)$ pairwise disjoint arcs of conic 
sections (which are the projections of the arcs forming the boundary
of $B$), and the goal is to determine whether any line in $L$ crosses 
any arc in $E$. We note that the halfspaces from which we obtain $B$
come in pairs that are symmetric to each other about the origin, so restricting 
the problem to the lower hemisphere incurs no loss of generality. We also 
note that there might be situations in which one of the great circles
is fully contained in $B$, but these cases are easy to detect, e.g.,
by picking an arbitrary point on each great circle and checking whether 
it belongs to $B$, using a suitable point-location data structure on $B$.

To simplify the presentation, we assume that the arcs of $E$ are elliptic 
arcs; handling the cases of parabolic or hyperbolic arcs is done in
essentially the same manner.

Orient all the lines of $L$ from left to right. We may assume that
all the arcs in $E$ are $x$-monotone (otherwise we break each arc
that is not $x$-monotone at its leftmost and rightmost points, into at 
most three $x$-monotone subarcs). We orient all these (sub)arcs also from left to 
right. We also treat separately \emph{convex} arcs,
namely arcs for which the tangent directions turn counterclockwise
as we traverse them from left to right, and \emph{concave} arcs,
for which the tangent directions turn clockwise. The treatments of
these two subfamilies are fully symmetric, so we only consider the 
case of convex arcs.

\begin{figure}[htb]
\begin{center}
\includegraphics[scale = 0.1]{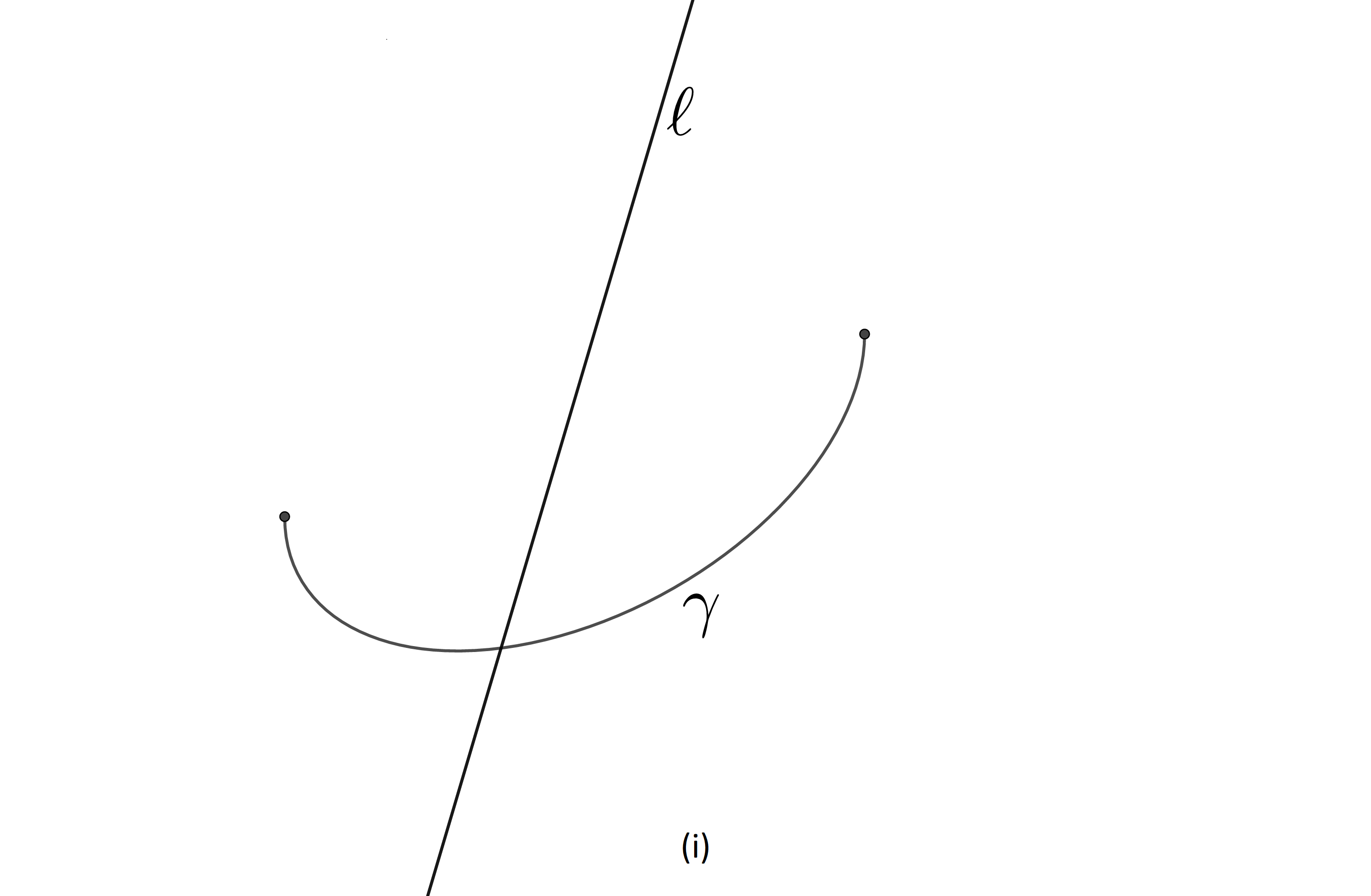}
\includegraphics[scale = 0.1]{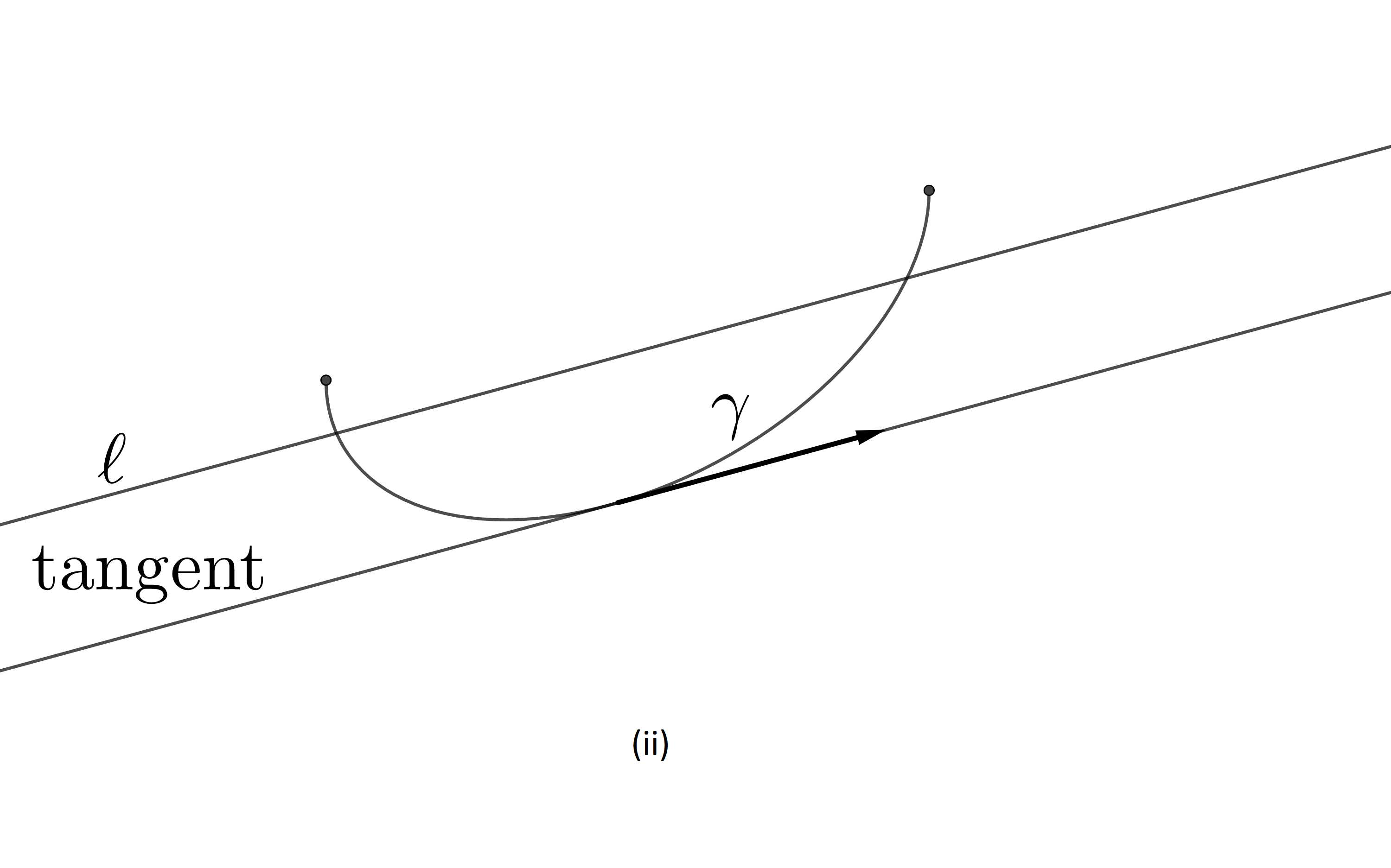}
\includegraphics[scale = 0.1]{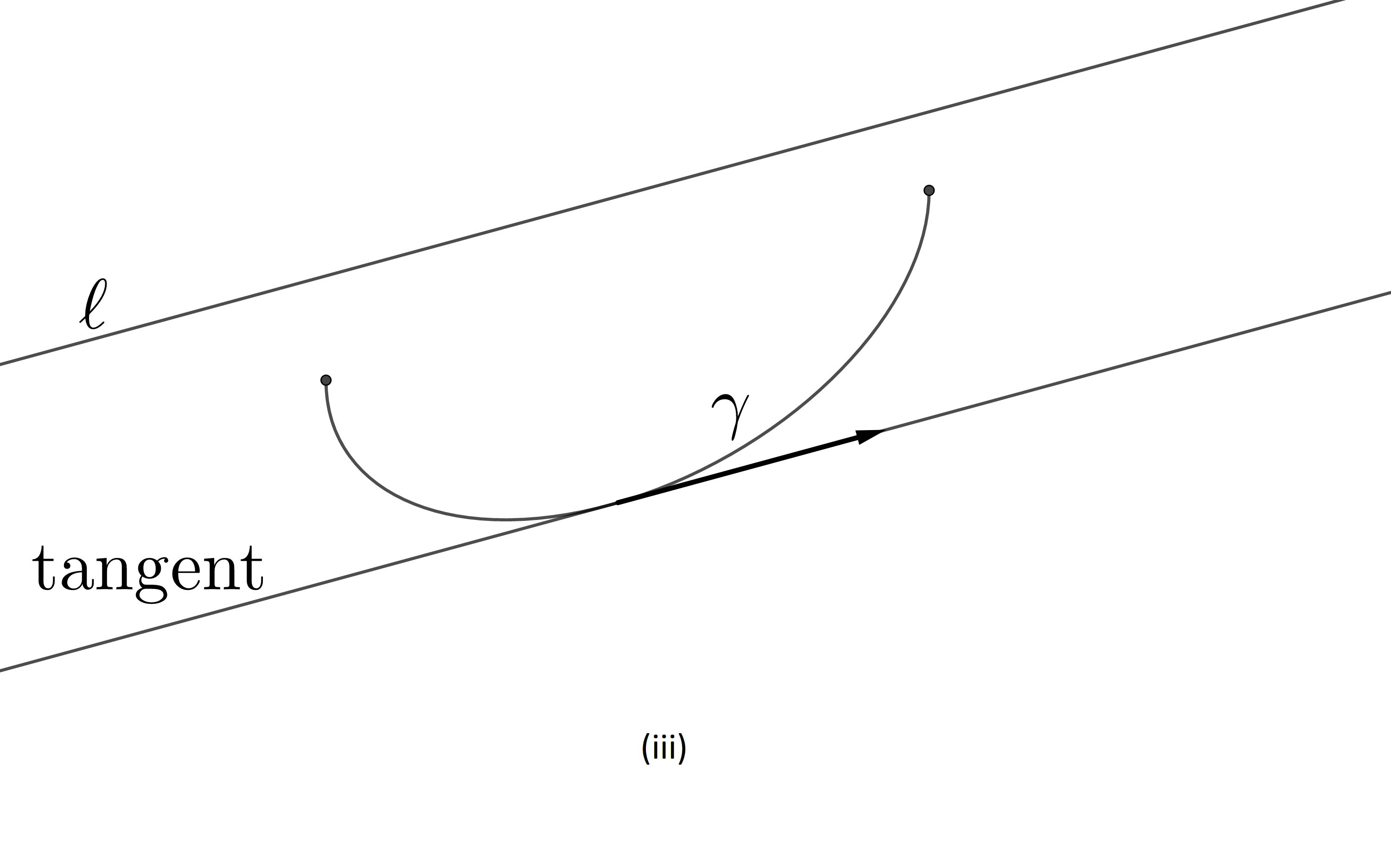}
\includegraphics[scale = 0.1]{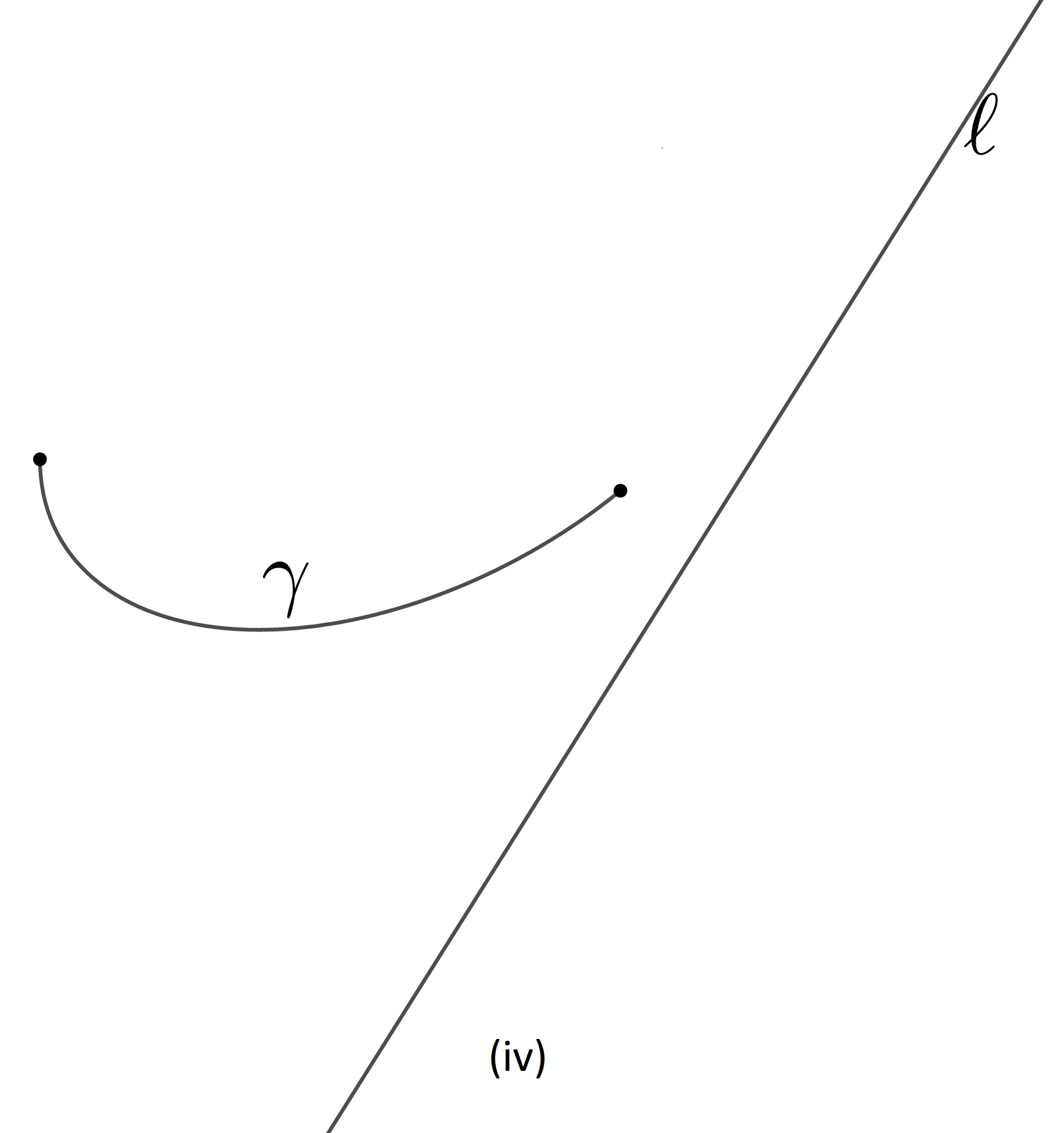}
\caption{\sf{A line $\ell$ intersecting a convex $x$-monotone elliptic arc $\gamma$: 
(i) The two endpoints of $\gamma$ lie on different sides of $\ell$.
(ii) The two endpoints lie to the left of $\ell$ and $\ell$ lies to the left of 
the parallel tangent to the arc. (iii) The two endpoints lie to the right of $\ell$ 
(and then there is no intersection). (iv) The two endpoints lie to the left of $\ell$ 
but $\gamma$ has no tangent parallel to $\ell$ (and then there is no intersection).}}
\label{le-cross-app}
\end{center}
\end{figure}

A line $\ell$ intersects a convex $x$-monotone arc $\gamma$ of some ellipse $e$, both 
oriented as above, if and only if one of the following conditions holds.
\begin{description}
\item[(i)]
The two endpoints of $\gamma$ lie on different sides of $\ell$.
See Figure~\ref{le-cross-app}(i).
\item[(ii)]
The two endpoints of $\gamma$ lie to the left of $\ell$ and $\ell$
intersects $e$. For this to happen, $\gamma$ must have a tangent
that is parallel to $\ell$. That is, the slope of $\ell$ must lie 
between the slopes of the tangents to $\gamma$ at its endpoints.
When all these conditions hold, it suffices to require that $\ell$
lies to the left of the right tangent to $e$ with the same slope
of $\ell$. See Figure~\ref{le-cross-app}(ii,iii).
\end{description}

To test for intersections of type (i), we use a two-level data structure,
where each level is a standard tree-like range searching structure for
points and halfplanes (see \cite{Ag17,AE99}). The first level collects 
the arcs that have one endpoint to the 
right of $\ell$, and the second level tests whether any of these arcs
has its other endpoint to the left of $\ell$. Using the standard machinery
for point-halfplane range searching (see, e.g.,~\cite[Theorem 6.1]{Ag17}, and also~\cite{AE99}), 
this takes time $O(M^{2/3}N^{2/3}{\rm polylog}(M+N)) = O(n^{8/3}{\rm 
polylog}(n))$.

To test for intersections of type (ii), we use a four-level data structure,
where, as before, the first two levels are standard range searching structures
for points and halfplanes, so that the first level collects the arcs that 
have their left endpoint to the left of $\ell$, and the second level collects, 
from among the arcs in the output of the first level, those arcs that 
have their right endpoint also to the left of $\ell$. The third level 
is a one-dimensional segment tree on the interval ranges of the slopes 
of the tangents to the arcs, and it collects those arcs whose tangent-slope 
range contains the slope of $\ell$. Finally, the fourth level tests whether 
any of the arcs is such that its tangent that is parallel to $\ell$ passes 
to the right of $\ell$.

To implement the fourth level, we note that the lines that are tangent 
to the ellipse $e$ and have slope $a$ can be written as $y = ax + \varphi_e^-(a)$ 
and $y = ax + \varphi_e^+(a)$, with $\varphi_e^-(a) < \varphi_e^+(a)$, 
where $\varphi_e^-(a)$ and $\varphi_e^+(a)$ are algebraic functions
of constant degree that depend on $e$. If $\ell$ has the equation
$y = ax+b$ then we need to test whether there exists an ellipse $e$ 
such that $b > \varphi_e^-(a)$. We thus compute the lower envelope of the
functions $\varphi_e^-$ in time nearly linear in the number of arcs, and then, given a line $y = ax+b$, we test
whether the point $(a,b)$ lies above the envelope, in logarithmic time.

It is easy to see that in this case too, the overall cost is $O(n^{8/3}{\rm polylog}(n))$.
In conclusion, we have shown:
\begin{theorem} \label{thm:slide:app}
Given $K$ and $W$ as above, we can determine whether $K$ can slide
through $W$ in a collision-free manner, and, if so, find such a sliding motion,
in time $O(n^{8/3}{\rm polylog}(n))$.
\end{theorem}
We are not aware of any published result that solves the specific problem at hand, 
of determining whether any great circle in $\cal C$ crosses $B$, with comparable running time.
A different solution, with a similar performance bound, was suggested to us by Pankaj Agarwal, 
and we thank him deeply for the useful interaction concerning this problem.

We end this section with the interesting challenge of improving the algorithm.
A concrete way of doing this would be to argue that not all the features of $S_A\cup T_A$
and $B$ need to be taken into account in the batched range searching step. We again thank
Pankaj Agarwal for raising this issue.

\section{Unbounded windows} \label{sec:unb}

In this section we consider the variant in which $W$ is an infinite slab 
in the $xy$-plane, bounded by, say, two vertical lines $x=0$ and $x=a$. 
We refer to such a window as a \emph{gate}. We show:
\begin{theorem} \label{thm:unbounded}
Let $K$ be a convex polytope that can be moved by some 
collision-free rigid motion through a gate $W$. Then there exists a sliding
collision-free motion of $K$ through $W$.
\end{theorem}
We can therefore apply the machinery of Theorem~\ref{thm:slide:app}, and conclude 
that we can determine whether $K$ can be moved through $W$ by a collision-free 
motion, in time $O(n^{8/3}{\rm polylog}(n))$.

\medskip
\noindent{\bf Proof.}
We start by giving a brief sketch of the proof, and then go in to the full details.
By projecting the moving polytope $K$ and $W$
onto the $xz$-plane, $W$ projects to the interval $g := [0,a]\times\{0\}$,
and $K$ projects to a time-varying convex polygon that starts from a placement
that lies in the upper halfplane $z>0$ and reaches a placement that lies in the 
lower halfplane $z<0$. For technical reasons, we approximate $K$ by a smooth convex
body, and reduce the problem to the case where $K$ is smooth and convex.

\begin{figure}[htbp]
\begin{center}
\includegraphics[scale = 0.2]{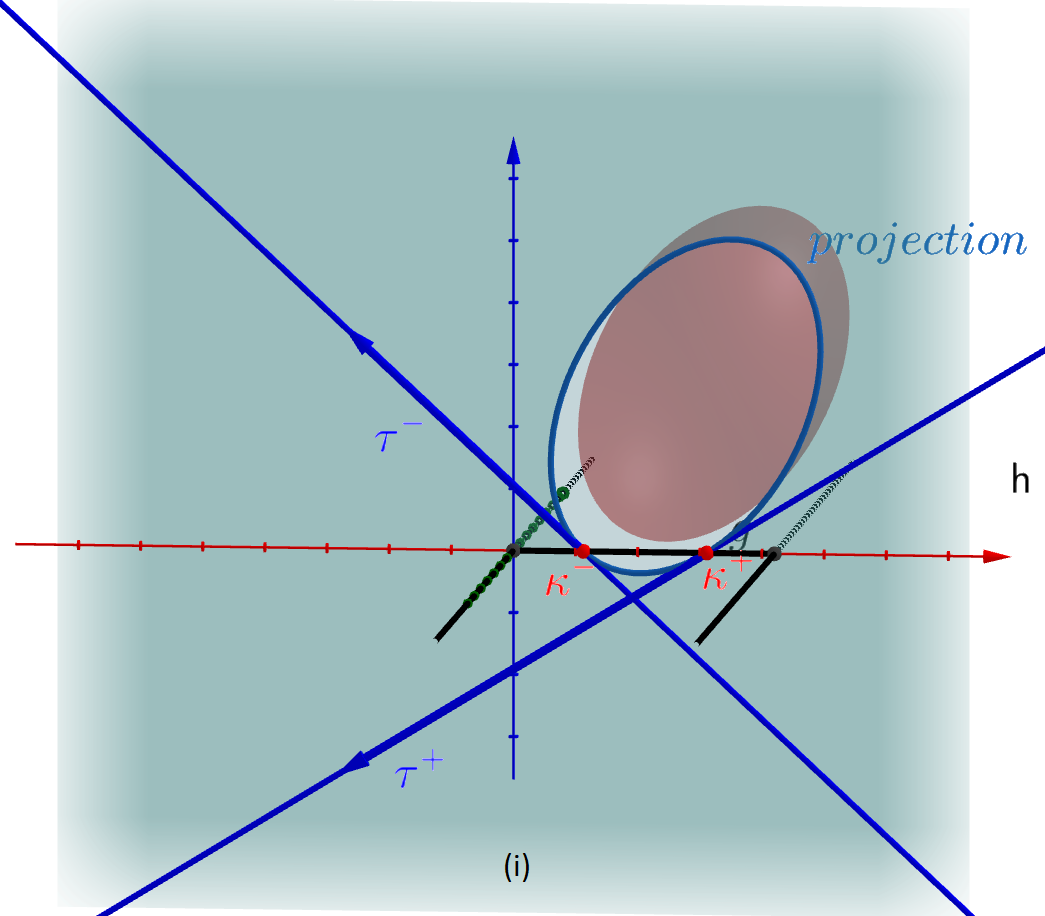}
\includegraphics[scale = 0.2]{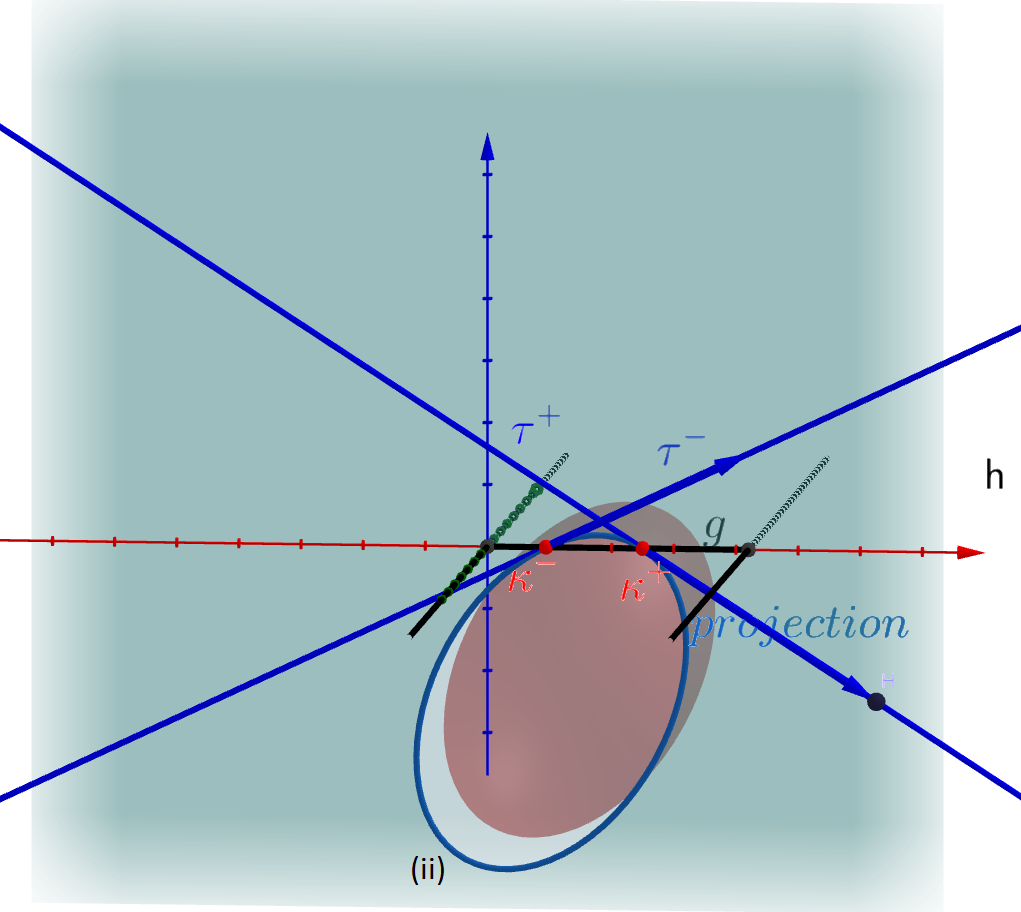}
\caption{\sf{Moving the projection of $K$ through $g$. Left: At the beginning of 
the crossing of $g$, the tangents $\tau^-(t)$ and $\tau^+(t)$ `open up' (with respect to 
their sides that contain $K(t)$). Right: At the end of the crossing, they `open down'.}}
\label{projection-moving}
\end{center}
\end{figure}

At any time $t$ during the motion, the projected planar region $\pi(K(t))$
(where $K(t)$ is the placement of $K$ at time $t$) meets $g$ at some interval
$I(t)$ (we ignore the prefix and suffix of the motion where $K(t)$ does not 
yet meet, or no longer meets $W$). We consider the two tangents to $\pi(K(t))$
at the endpoints $\kappa^-(t)$, $\kappa^+(t)$ of $I(t)$, and note that, at the 
beginning of the motion, the wedge that these tangents form and that contains 
$K(t)$ points upwards, and at the end of the motion it points downwards; 
see Figure~\ref{projection-moving}.

\begin{figure}[htbp]
\begin{center}
\includegraphics[scale = 0.2]{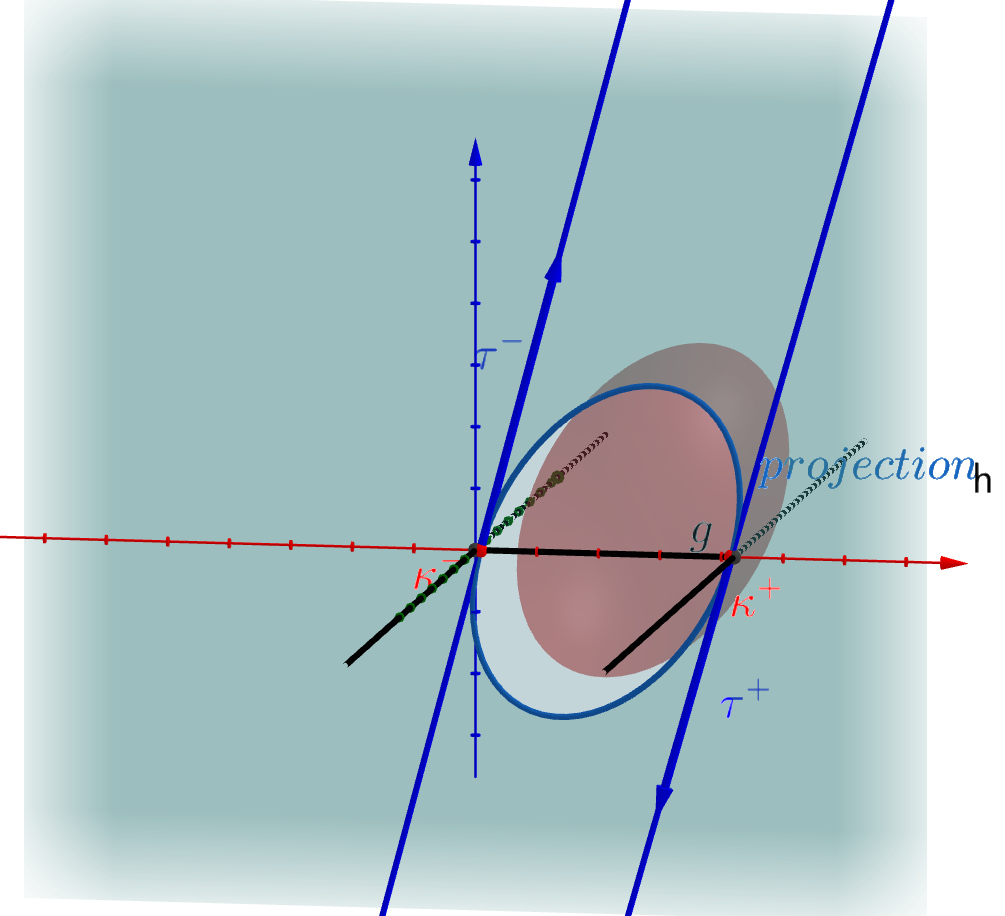}
\caption{\sf{The critical instance $t_0$ where the tangents at $\kappa^-(t_0)$ and at $\kappa^+(t_0)$ become \mbox{\sf(anti-)}parallel.}}
\label{partang}
\end{center}
\end{figure}

Since the tangents vary continuously (because $K(t)$ is always smooth),
there must be a time $t_0$ at which the two tangents are parallel to each other,
and thus span a slab $\sigma$ (in the $xz$-plane) whose width is clearly $\le a$.
See Figure~\ref{partang}.
The Cartesian product of $\sigma$ and the $y$-axis yields a slab $\sigma^*$
in $\reals^3$, whose cross-section with the $xy$-plane is contained in $W$.
This in turn implies that $K$ can slide through $W$, and completes the proof.

In more detail, we proceed as follows. Continue to assume that $K$ is smooth; 
we will later use a compactness argument to extend 
the result to convex polytopes. 

As in the short version of the proof, let then $K$ be an arbitrary 
compact convex body in $\reals^3$, let $h$ denote the $xz$-plane, 
and let $g := h\cap W$, which is the segment $0\le x\le a$, $z=0$ 
within $h$. The two complementary rays
to $g$ within the $x$-axis form the only obstacles within $h$.
Let $\pi$ denote the orthogonal projection of 3-space onto $h$.

Assume that $K$ can be moved through $W$ by an arbitrary collision-free
rigid motion, which we represent as a continuous map on $[0,1]$ (a
`time interval'), where, for each $t\in [0,1]$, $K(t)$ denotes the 
placement of $K$ at time $t$ during the motion. 
For each $t\in [0,1]$, $\bd\pi(K(t))$ is the projection of the silhouette 
of $K(t)$ on $h$. It is a time-varying convex 
region within $h$, whose shape is not rigidly fixed. For a convex polytope 
$K$, the projected silhouette $\bd\pi(K(t))$ is a time-varying convex polygon.

We have the following property, whose easy proof is omitted.

\begin{lemma} \label{lem:hproj}
The motion $t\mapsto K(t)$ is collision-free, and moves $K$ through
$W$ from a placement $K(0)$ in the upper halfspace to a placement 
$K(1)$ in the lower halfspace, if and only if the map
$t\mapsto \pi(K(t))$ is collision-free within $h$, and moves 
the (time-varying) projection $\pi(K(t))$ through $g$
from the placement $\pi(K(0))$ in the upper halfplane $z>0$
to the placement $\pi(K(1))$ in the lower halfplane $z<0$. 
\end{lemma} 

We note that in Lemma~\ref{lem:hproj} the body $K$ is not required to be smooth, 
but this requirement is needed for the proof of the following theorem.

\begin{theorem} \label{thm:smooth:app}
Let $K$ be a smooth compact convex body that can be moved, by a
collision-free rigid motion, through $W$ from a placement 
in the upper halfspace $z>0$ to a placement in the lower halfspace $z<0$.
Then there exists a sliding collision-free motion of $K$ through $W$.
\end{theorem}

\noindent{\bf Proof.}
Let $K$ be as in the theorem, and let $t\mapsto K(t)$ be a 
collision-free rigid motion that takes $K$ through $W$, as in the
theorem statement. For each $t$, $\pi(K(t))$ is also smooth
(as a planar convex region).
Put $\kappa(t) := \pi(K(t))\cap g$, which is a subsegment of
$g$ (by assumption, and by Lemma~\ref{lem:hproj}, the intersection 
of $\pi(K(t))$ with the $x$-axis is always fully contained in $g$). 
$\kappa(t)$ is empty at the begining and at the end of the
motion, namely during some prefix interval and some suffix
interval of $[0,1]$ (if the motion is `crazy' enough, $\kappa(t)$
might also be empty during some other inner intervals of $[0,1]$).
Nevertheless, since $\pi(K(t))$ crosses $g$ from side to side,
there must exist at least one closed maximal connected interval
$I = [t_1,t_2]$ within $[0,1]$ such that $\kappa(t)\ne\emptyset$
for all $t\in I$, and such that $\kappa(t_1)$ and $\kappa(t_2)$
are singletons, so that $\kappa(t_1)$ (resp., $\kappa(t_2)$)
is the $z$-lowest (resp., $z$-highest) point of $\pi(K(t_1))$
(resp., of $\pi(K(t_2))$). See Figure~\ref{projection-moving} for an illustration.


Denote, for $t\in I$, the left and right endpoints of 
$\kappa(t)$ by $\kappa^-(t)$ and $\kappa^+(t)$, respectively,
and let $\tau^-(t)$ (resp., $\tau^+(t)$) denote the tangent to
$\pi(K(t))$ at $\kappa^-(t)$ (resp., at $\kappa^+(t)$), where
we orient both tangents so that $\pi(K(t))$ lies to their right.

Since $\pi(K(t))$ is smooth, the two tangents are well defined
and unique. Moreover, since the motion of $K(t)$ is continuous,
so is the `motion' of $\pi(K(t))$, and this is easily seen
to imply that the directions $\mu^-(t)$ of $\tau^-(t)$, and 
$\mu^+(t)$ of $\tau^+(t)$ are also continuous functions of $t$. 

Consider the map $\varphi(t)$ that maps $t\in I$ to the counterclockwise
angle between $\mu^-(t)$ and $\mu^+(t)$. The map is undefined at
$t_1$ and at $t_2$, but we assume that it is defined everywhere 
in the interior of $I$ (as would be the typical situation---see the comment made earlier).
$\varphi(t)$ is clearly a continuous function. For $t$ slightly 
larger than $t_1$, $\varphi(t)$ has a small positive value, and 
for $t$ slightly smaller than $t_2$, $\varphi(t)$ is close to $2\pi$.
It follows, by continuity, that there exists $t_0\in I$ for which
$\varphi(t_0) = \pi$, that is, the two tangents at $\kappa^-(t_0)$ and at $\kappa^+(t_0)$ are parallel to each
other. This means that $\pi(K(t_0))$ is contained in
the slab $\sigma$, within $h$, bounded by the two tangent lines.
This in turn implies that $K(t_0)$ is contained in the three-dimensional
slab $S$ which is the Cartesian product of $\sigma$ and the $y$-axis.
Moreover, the intersection of $S$ with the $xy$-plane is a $y$-vertical
slab that is contained in $W$ (see Figure~\ref{partang} for an illustration). This in turn means that, if we fix
the orientation of $K$ to be that of $K(t_0)$, we can slide $K$
within $S$ through $W$ (note that there are infinitely many ways to do so, each with its own $y$-component of the sliding direction).
This completes the proof.
$\Box$ \medskip

\begin{figure}[htbp]
\begin{center}
\includegraphics[scale = 0.5]{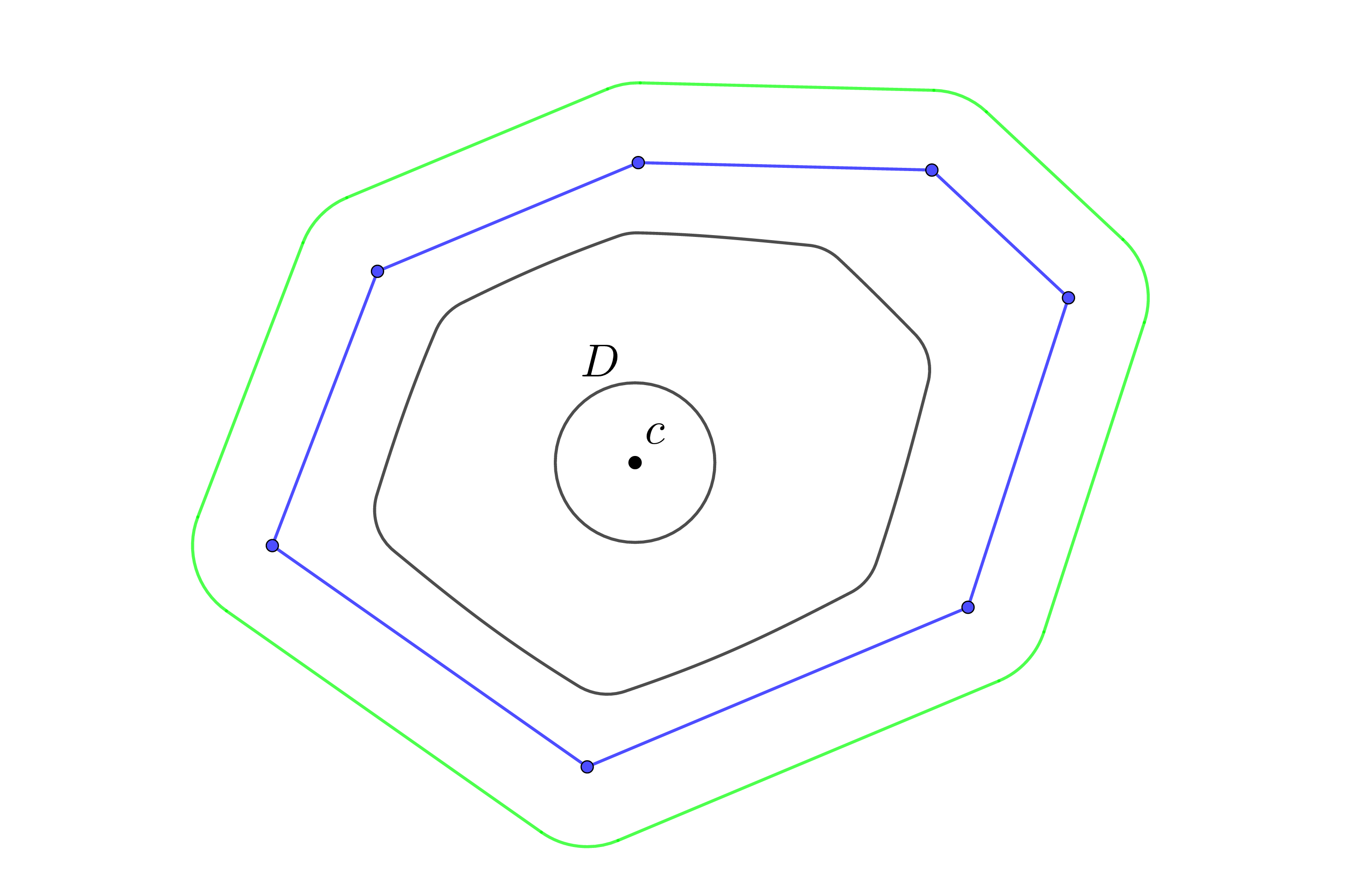}
\caption{\sf{A two-dimensional illustration of the scheme for approximating a convex 
polytope by smooth strongly convex bodies: the convex polygon (blue), Minkowski sum 
with a small circle (green) and $K_\delta$ (black).}}
\label{app:apxconv}
\end{center}
\end{figure}

We now continue with the proof of Theorem \ref{thm:unbounded}.
To extend Theorem~\ref{thm:smooth:app} to the case where $K$ is a polytope,
we use the following approximation scheme. Let $D$ be some ball fully contained 
in $K$, with center $c$ and radius $\rho$. For each $\delta > 0$, let $L_\delta$ 
be the Minkowski sum of $K$ and a ball centered at the origin with radius $\delta$, 
and define a map $f_\delta$ on $\sph^2$, so that, for each 
$\vec{v}\in\sph^2$, $f_\delta(\vec{v}) = (1-\delta)g(\vec{v}) + \delta\rho$, 
where $g(\vec{v})$ is the distance from $c$ to $\bd L_\delta$ in direction 
$\vec{v}$. Define $K_\delta$ to be
$$
 \{c+tf_\delta(\vec{v})\vec{v} \mid \vec{v} \in \sph^2, t\in [0,1]\} ,
$$
scaled down by a factor of $1+\delta$.
See Figure~\ref{app:apxconv} for an illustration.

It is easily seen that $K_\delta$ is a smooth compact
strongly convex object that is contained in $K$, and that $K_\delta\to K$ as $\delta\to 0$,
in the sense that the Hausdorff distance between $K$ and $K_\delta$ tends
to zero. Clearly, if $K$ can be moved through $W$ (by an arbitrary
collision-free rigid motion), then so can $K_\delta$.

For each $\delta>0$, apply Theorem~\ref{thm:smooth:app} to $K_\delta$,
to obtain a direction $\vec{v}_\delta$ and a rotation $\theta_\delta$ orthogonal to $\vec{v}_\delta$ so that there is a sliding
collision-free motion of $K_\delta$ in direction $\vec{v}_\delta$ from its view $(\vec{v}_\delta,\theta_\delta)$
through $W$. By compactness of $\sph^2$, there exists a sequence
$\delta_i\downarrow 0$ such that $\vec{v}_{\delta_i}$ converges to some
direction $\vec{v}$ in $\sph^2$, and $\theta_{\delta_i}$ converges to some rotation $\theta$. By continuity, 
it follows that there exists a sliding collision-free motion of $K$ 
through $W$ in direction $\vec{v}$ from its view $(\vec{v},\theta)$. 

This completes the proof of Theorem \ref{thm:unbounded}.
$\Box$

We can therefore apply the machinery of Theorem~\ref{thm:slide:app}, and conclude that 
we can determine whether $K$ can be moved through $W$ by a collision-free motion in time 
$O(n^{8/3}{\rm polylog}(n))$.

We note that strong convexity of the approximating body is not needed 
for the analysis in this section, but only in the following Sections~\ref{sec:projection} 
and \ref{app:purely-translational}. To avoid duplication, we have used the same scheme for 
approximating a convex polytope, for the analysis both here and in the subsequent sections.

\section{From passing through an arbitrary convex window to \\
sliding through a gate} \label{sec:projection}

In this section we prove a similar yet different property of a convex polytope passing through 
an arbitrary compact planar convex window, not necessarily rectangular.

\begin{theorem} \label{thm:projection}
Let $W$ be an arbitrary compact convex region in the $xy$-plane.
Let $K$ be a convex polytope that can be moved by some collision-free motion 
(possibly full rigid motion, with six degrees of freedom) through $W$, and 
let $d$ be the diameter of $W$ (the maximum distance between any pair of points in $W$). 
Let $h$ be an arbitrary plane, and let $K_h$ be the orthogonal projection of $K$ on $h$. 
Then $K_h$ can be rigidly placed between two parallel lines at distance $d$. 
That is, for any fixed direction $\vec{v}$, $K$ can slide, from its (arbitrary)
initial placement, in direction $\vec{v}$ 
through a gate of width $d$, in a plane perpendicular to $\vec{v}$.
\end{theorem}

We provide two different topology-based proofs of Theorem~\ref{thm:projection}, both 
presented in full detail at the end of this section. We start by sketching one of these proofs. 
But first here is an interesting corollary of the theorem. 
\begin{corollary} \label{cor:slide_size_trade}
If $K$ can be moved through a rectangular window $W$ of dimensions $a\times b$ 
by some collision-free motion, then $K$ can slide through a rectangle of 
dimensions $\min{(a, b)}\times\sqrt{a^2 + b^2}$.
\end{corollary}

\begin{figure}[htbp]
\begin{center}
\includegraphics[scale = 0.2]{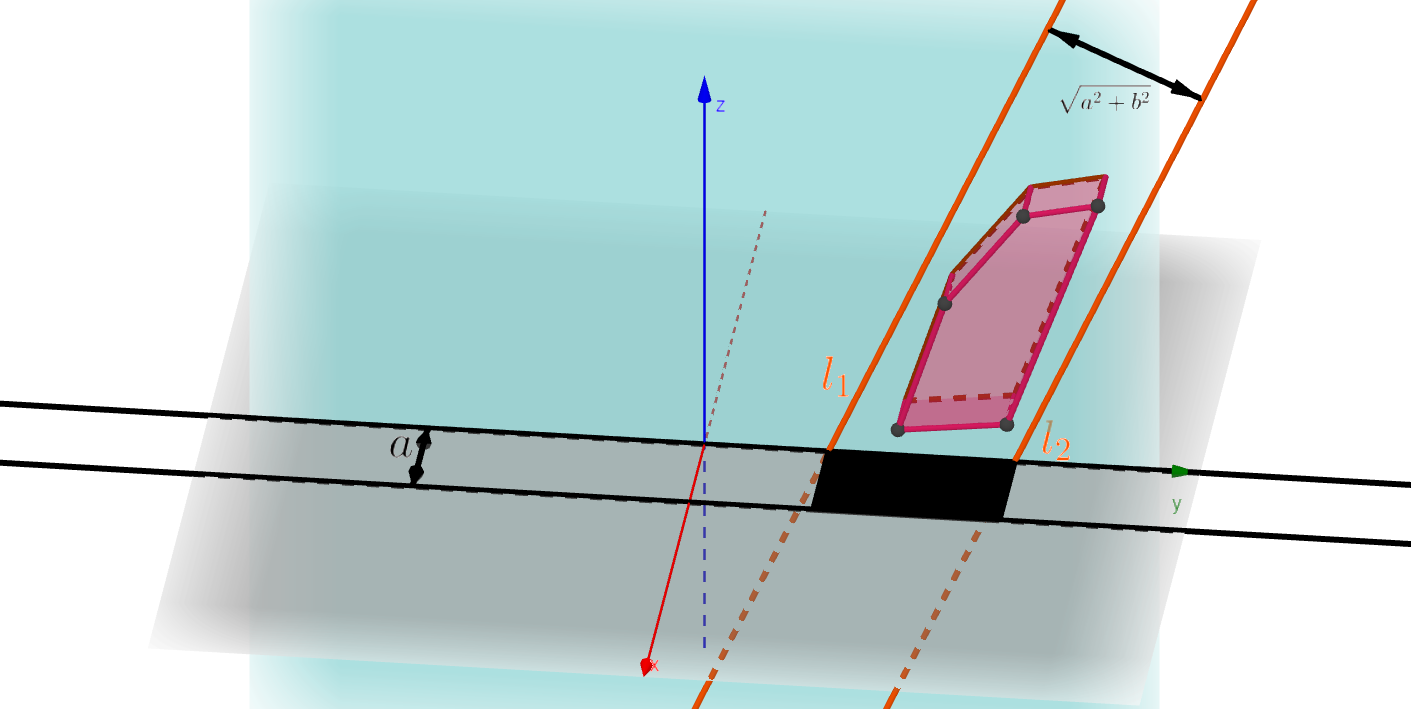}
\includegraphics[scale = 0.2]{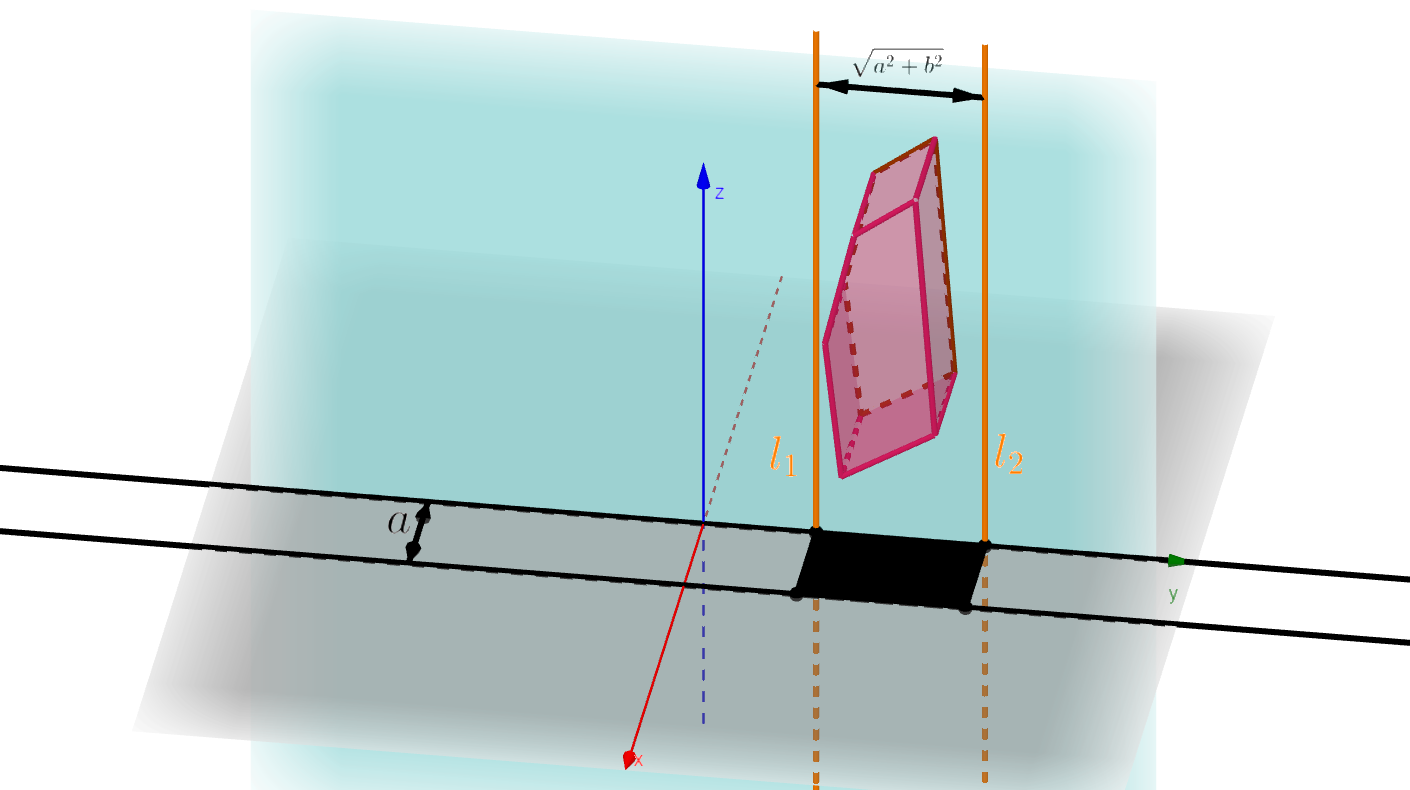}
\caption{\sf{$K$ can slide through a rectangle of dimensions $a\times \sqrt{a^2+b^2}$. 
Left: The projection of $K$ on the $xy$-plane is contained in a gate of width $a$ 
(the black lines), and the projection of $K$ on the $yz$-plane is contained between 
the lines $l_1,l_2$ (orange). 
Right: After rotating $K$ around the $x$-axis, $l_1,l_2$ are perpendicular to the 
$xy$-plane and the projection of $K$ on the $xy$-plane is contained in the desired rectangle.}}
\label{window-to-gate}
\end{center}
\end{figure}

\noindent{\bf Proof.}
Assume without loss of generality that $a < b$. Since $K$ can move
through a rectangle of dimensions $a\times b$, it can also move through
the gate, of width $a$, $[0,a]\times\reals$ (in the $xy$-plane). 
Now project $K$ on the $yz$-plane, and apply Theorem~\ref{thm:projection}, 
to conclude that the projection of $K$ can be placed in a slab in the
$yz$-plane, bounded by two parallel lines $l_1$, $l_2$ at distance 
$\sqrt{a^2+b^2}$ apart (which is the diameter of $W$). Rotate 3-space
around the $x$-axis so as to make $l_1$ and $l_2$ vertical (parallel
to the $z$-direction). Now the projected silhouette of $K$ on the
$xy$-plane is contained in a rectangle of dimensions 
$a\times\sqrt{a^2 + b^2}$, so $K$ can slide (vertically down) through
this rectangle. See Figure \ref{window-to-gate} for an illustration.
$\Box$

\noindent{\bf Two proofs of Theorem~\ref{thm:projection}.}
Similar to the previous section, we first prove the theorem for smooth strongly convex compact bodies, 
and then extend the result to polytopes the same way as before.
Consider the motion of $K$, now assumed to be a smooth, strongly convex, and compact body, 
during the time interval $[0,1]$. Assume that at $t=0$ 
(resp., at $t=1$), $K$ lies fully above (resp., below) the $xy$-plane. 

\paragraph{First proof.}
Fix some direction $\vec{v}$, and let $C=C(\vec{v})$ denote the silhouette of $K$ 
when viewed in direction $\vec{v}$. Let $h$ be some plane orthogonal to $\vec{v}$, 
and let $\pi_h$ denote the orthogonal projection onto $h$. Parameterize a point 
$u\in C$ by the orientation $\theta$ of the tangent at $\pi_h(u)$ to $K_h := \pi_h(K)$ 
which is well defined since $K$ is smooth, 
and let $\gamma_h$ be the inverse of $\pi_h$; that is, $\gamma_h(\theta)$ is the unique 
point $u \in C$ such that $\pi_h(u) = \theta$. Since $K$ is assumed to be strongly convex,
$K_h$ is also strongly convex, and $\gamma$ is a well-defined and continuous 
function on $\sph^1$. We extend $\gamma$ to a bivariate function 
$\gamma^* : \sph^1\times [0,1] \mapsto \reals^3$, so that $\gamma^*(\theta,t)$
is the position (in the ambient 3-space) of $\gamma(\theta)$ at time $t$ during the motion of $K$.

Let $\delta : \sph^1\times [0,1] \mapsto \reals$ be the function 
$\delta(\theta,t) = z(\gamma^*(\theta,t))$, namely, the $z$-coordinate of the
corresponding point $\gamma(\theta)$ of $C$ at time $t$. Note that at time $t=0$
(resp., at time $t=1$), $\delta$ is positive (resp., negative) at each $\theta$, 
since $K$ lies fully above (resp., below) the $xy$-plane at that time. Put
$M := \max_{\theta\in\sph^1} \delta(\theta,0)$ and
$m := \min_{\theta\in\sph^1} \delta(\theta,1)$. By our assumptions, $M > 0$ and
$m < 0$.

The functions $\delta_0(\theta) = \delta(\theta,0)$ and
$\delta_1(\theta) = \delta(\theta,1)$ are defined and continuous on $\sph^1$,
and we extend each of them to the closed unit disk $\B^1$ bounded by $\sph^1$, 
in polar coordinates, which, for technical reasons, we write in reverse order as $(\theta,r)$, by
\begin{align*}
\delta_0^*(\theta,r) & = r\delta_0(\theta) + (1-r)M \\
\delta_1^*(\theta,r) & = r\delta_1(\theta) + (1-r)m .
\end{align*}
It is easily checked that these extensions are well defined and continuous over $\B^1$.
Moreover, $\delta_0^*(\theta,r) > 0$ and $\delta_1^*(\theta,r) < 0$ for every $\theta$.

We now take our function $\delta$, which is so far defined on the side surface $S$
of the cylinder $\sph^1\times [0,1]$, and extend it to the entire boundary 
$S^* := S\cup B_0\cup B_1$ of the cylinder, so that $\delta$ coincides with $\delta_0^*$ 
on the base $B_0$ of the cylinder at $t=0$, and with $\delta_1^*$ on the base $B_1$ at $t=1$. 
Clearly, the extended $\delta$ is well defined and continuous over $S^*$.

\begin{figure}[htbp]
\begin{center}
\includegraphics[scale = 0.4]{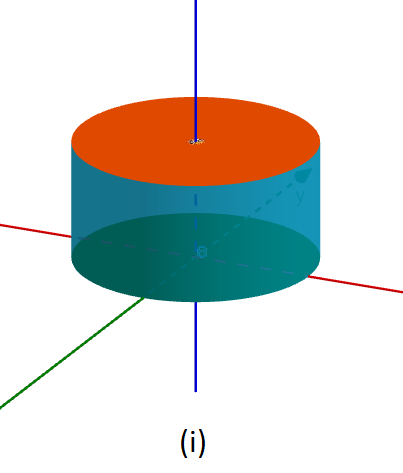}
\includegraphics[scale = 0.4]{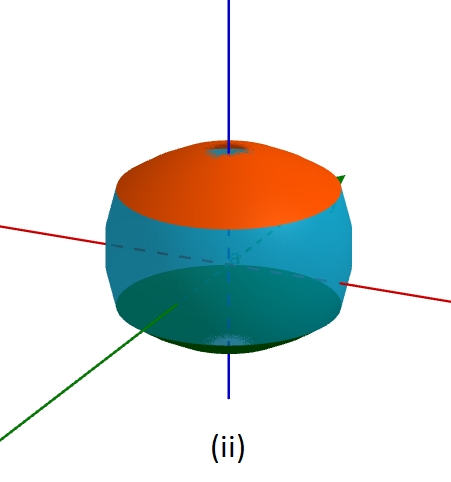}
\includegraphics[scale = 0.4]{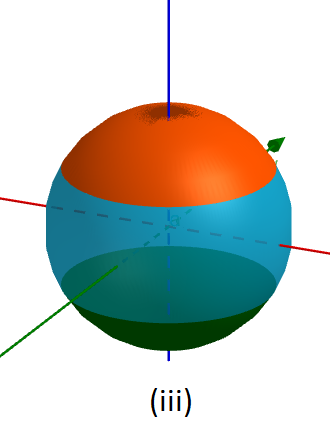}
\caption{\sf{Identifying $S^*$ with the unit sphere $\sph^2$. $B_0$ is shown in green, 
$B_1$ in orange, and $S$ in light blue. In (i) $S^*$ is depicted, in (ii) an intermediate 
snapshot of the deformation is shown, for visual convenience, and in (iii) the final 
unit ball is shown, divided into the three parts that correspond to $B_0$, $S$ and $B_1$.}}
\label{ss2}
\end{center}
\end{figure}

To simplify the forthcoming analysis, we identify $S^*$ with the unit sphere 
$\sph^2$, which we parameterize by $(\theta,z)$, where $\theta\in\sph^1$ is the 
horizontal orientation of the point on $\sph^2$ and $z$ is its $z$-coordinate
(so $\theta$ is not well defined at the north and south poles of $\sph^2$).
We use the simple homeomorphism $f$ that maps a point $(\theta,t)\in S$ 
to $(\theta,t-1/2)\in\sph^2$, maps a point $(\theta,r)\in B_0$ to
$(\theta,-1+r/2)\in\sph^2$, and maps a point $(\theta,r)\in B_1$ to
$(\theta,1-r/2)\in\sph^2$. See Figure~\ref{ss2} for an illustration.
In what follows, we will mostly use $\sph^2$
to represent $S^*$, except for a few technical observations.

Define a function $G$ from $\sph^2$ to $\reals^2$ by
$$
G(\theta,t) = \left(\delta(\theta,z),\;\delta(\theta+\pi,z) \right),\quad\text{for $(\delta,z)\in \sph^2$} .
$$
Our goal is to show that $G(\sph^2)$ contains the origin. Note that,
by construction, $G(f(B_0))$ is fully contained in the positive quadrant 
$Q_1 := \{(x,y) \mid x,y>0\}$, and $G(f(B_1))$ is fully contained in the 
negative quadrant $Q_3 := \{(x,y) \mid x,y<0\}$.
Thus, if $G(\sph^2)$ contains the origin then so does $G(f(S))$. Once this 
property is established, it provides us with a pair $(\theta,z)$ such that 
$\delta(\theta,z)=\delta(\theta+\pi,z)=0$, which means that there are two 
antipodal points $u,v\in C$ that pass through $W$ simultaneously. Therefore 
their distance must be at most the diameter of $W$, and hence also the distance 
between the parallel tangent planes through them, which is a slab parallel to 
$\vec{v}$ of width at most $d$ that contains $K$, as asserted.

Assume to the contrary that $G(\sph^2)$ does not contain the origin.
Then we can normalize $G$ to the function
$$
H(\theta,z) := \frac{G(\theta,z)}{\|G(\theta,z)\|} ,\qquad\text{for $(\theta,z)\in \sph^2$} ,
$$
which maps $\sph^2$ continuously to the unit circle $\sph^1$. The function $G$, and thus also 
the function $H$, are symmetric with respect to the line $y=x$ in $\reals^2$, meaning that
\begin{align*}
G(\theta+\pi,z) & = \Sigma(G(\theta,z)),\qquad\text{for $(\theta,z)\in \sph^2$} ,\quad\text{and thus also} \\
H(\theta+\pi,z) & = \Sigma(H(\theta,z)),\qquad\text{for $(\theta,z)\in \sph^2$} ,
\end{align*}
where $\Sigma$ is the reflection about $y=x$, that is, $\Sigma(x,y) = (y,x)$.

We now use the property that the real line is a \emph{covering space} of $\sph^1$,
in the specific (and easily verified) sense that the continuous map 
$p:\reals \mapsto \sph^1$, given by $p(x) = e^{2\pi i x}$, 
for $x\in\reals$, is surjective, and, for each $\zeta\in\sph^1$, there 
exists an open neighborhood $U$ of $\zeta$ such that $p^{-1}(U)$ is the 
disjoint union of open sets in $\reals$, each of which is mapped 
homeomorphically to $U$ by $p$. The map $p$ is called the \emph{covering map}.

A well known property of covering spaces is the \emph{lifting property}
(reviewed, e.g., in \cite{wiki}; see also \cite{Hatcher}), a special case 
of which asserts, in the specific context used here, that, if $\varphi$ is
any continuous map from $\sph^2$ to $\sph^1$ then $\varphi$ can be \emph{lifted}
to a map $\psi: \sph^2\mapsto \reals$, so that $p\circ \psi = \varphi$. 
(Technically, this property holds when the domain of $\varphi$ (and $\psi$),
which is $\sph^2$ in our case, is path connected, locally path connected, 
and simply connected, conditions that are trivially satsfied by $\sph^2$.
Hence the lifting $\psi$ does indeed exist.)

Applying the lifting property to the function $H$, we get a continuous mapping 
$T: \sph^2\mapsto\reals$, such that $p\circ T = H$, so we have the property that
$$
p(T(\theta+\pi,z)) = \Sigma(p(T(\theta,z))) ,\qquad\text{for $(\theta,z)\in \sph^2$} .
$$
As is easily checked, we have $\Sigma\left(e^{i y} \right) = e^{i(\pi/2-y)}$, 
and therefore, for a point $x\in\reals$, we have
$$
\Sigma(p(x)) = \Sigma\left(e^{2\pi ix}\right) =
e^{\pi i/2 - 2\pi ix} = p(1/4-x) , \qquad \text{so}
$$
$$
p(T(\theta+\pi,z)) = p(1/4-T(\theta,z)) ,\qquad\text{for $(\theta,z)\in \sph^2$} .
$$
This in turn implies, by the definition of $p$, that
$$
T(\theta+\pi,z) = 1/4 + k_{\theta,z} - T(\theta,z) ,
$$
for some integer $k_{\theta,z}$.
However, since $T$ is continuous, there must be a single integer $k$
such that
$k_{\theta,z} \equiv k$ for all $\theta$ and $z$. That is, we have
\begin{equation} \label{tsym}
T(\theta+\pi,z) + T(\theta,z) = 1/4 + k ,\qquad\text{for all $(\theta,z)\in \sph^2$} .
\end{equation}
By an easy application of the mean-value theorem (which is also a special case
of the Borsuk-Ulam theorem in dimension $1$), there exist $\theta_0$ and 
$\theta_1$ such that, recalling that the value $z=-1/2$ (resp., $z=1/2$) 
corresponds to points on the lower (resp., upper) circle bounding $S$,
\begin{align*}
T(\theta_0+\pi,-1/2) & = T(\theta_0,-1/2) \\
T(\theta_1+\pi,1/2) & = T(\theta_1,1/2) .
\end{align*}
Substituting in (\ref{tsym}), we get
$$
T(\theta_0,-1/2) = T(\theta_1,1/2) = 1/8 + k/2 .
$$
However, by construction,
$H(\theta_0,-1/2)$ lies in the first quadrant $Q_1$, and 
$H(\theta_1,1/2)$ lies in the third quadrant $Q_3$. Hence we have
$T(\theta_0,-1/2) \in (0,1/4) + \ZZ$ and $T(\theta_1,1/2) \in (1/2,3/4) + \ZZ$,
but $1/8+k/2$ can belong to only one of these sets (depending on whether $k$ is even or odd).
This contradiction shows that $G(\sph^2)$, and thus also $G(f(S))$, 
contains the origin, as asserted.

So far the proof was for smooth strongly convex compact bodies. The extension 
to the case of a convex polytope $K$ is done exactly as in 
the proof of Theorem~\ref{thm:unbounded}.
$\Box$ \medskip

\paragraph{Second proof.}
We provide an alternative proof of Theorem~\ref{thm:projection}, and we are grateful
to Boris Aronov for providing to us its main ingredients.

We use the same notations as in the first proof. Similar to that proof, the following, 
slightly more generally stated proposition is the main technical tool that we need.

\begin{proposition} \label{pro:include-origin}
Let $G\colon S \to \reals^2$ be a continuous map, interpreted as the homotopy 
of the closed curve $\delta_0 \colon \sph^1 \to \oone \subset \reals^2$, 
given by $\theta \mapsto G(\theta,0)$, to the closed curve 
$\delta_1\colon \sph^1 \to \othree \subset \reals^2$, given by 
$\theta \mapsto G(\theta,1)$. In addition, suppose that $G$~is symmetric,
in the sense that $G(\theta+\pi,t)=\Sigma(G(\theta,t))$, for all 
$\theta \in \sph^1$ and $t \in [0,1]$.  
Then there exist $\theta \in \sph^1$, $t\in\reals$ that satisfy 
$G(\theta,t)=O$, that is, $G$~cannot miss the origin.
\end{proposition}
\noindent{\bf Proof.}
Clearly, if $G(\theta,t) = O$ then we also have $G(\theta+\pi,t) = O$.
Hence it suffices to show that there exists $(\theta,t)$ in
$D:= [0,\pi] \times [0,1]$ (half the side surface of the cylinder)
such that $G(\theta,t) = O$. Let $\Pi$ be the image of $D$ under $G$.

Consider the curve $\gamma_0 \colon [0,1] \to S$ defined by $t \mapsto (0,t)$, 
and its image $\Gamma_0$ under $G$, i.e., $\Gamma_0(t)=G(0,t) \in \reals^2$.  
Let $\gamma_1$ and $\Gamma_1$ be defined similarly by $\gamma_1(t) := (\pi,t)$,
so that $\Gamma_1(t) =G(\pi,t)$. Let $\gamma_1'$ and $\Gamma_1'$ be the reverses 
of $\gamma_1$ and $\Gamma_1$, respectively~--- the same curves traversed in reverse direction.

\begin{figure}[htbp]
\begin{center}
\includegraphics[scale = 0.4]{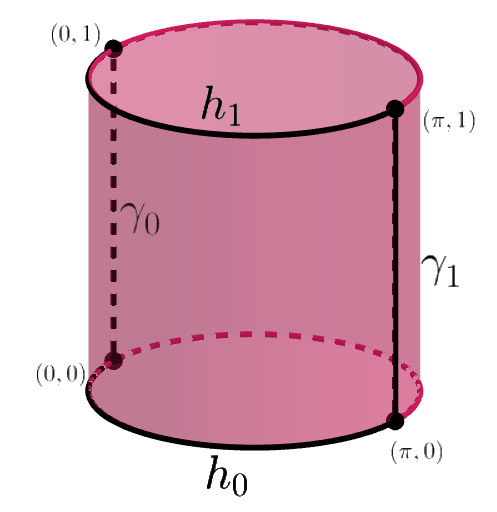}
\includegraphics[scale = 0.2]{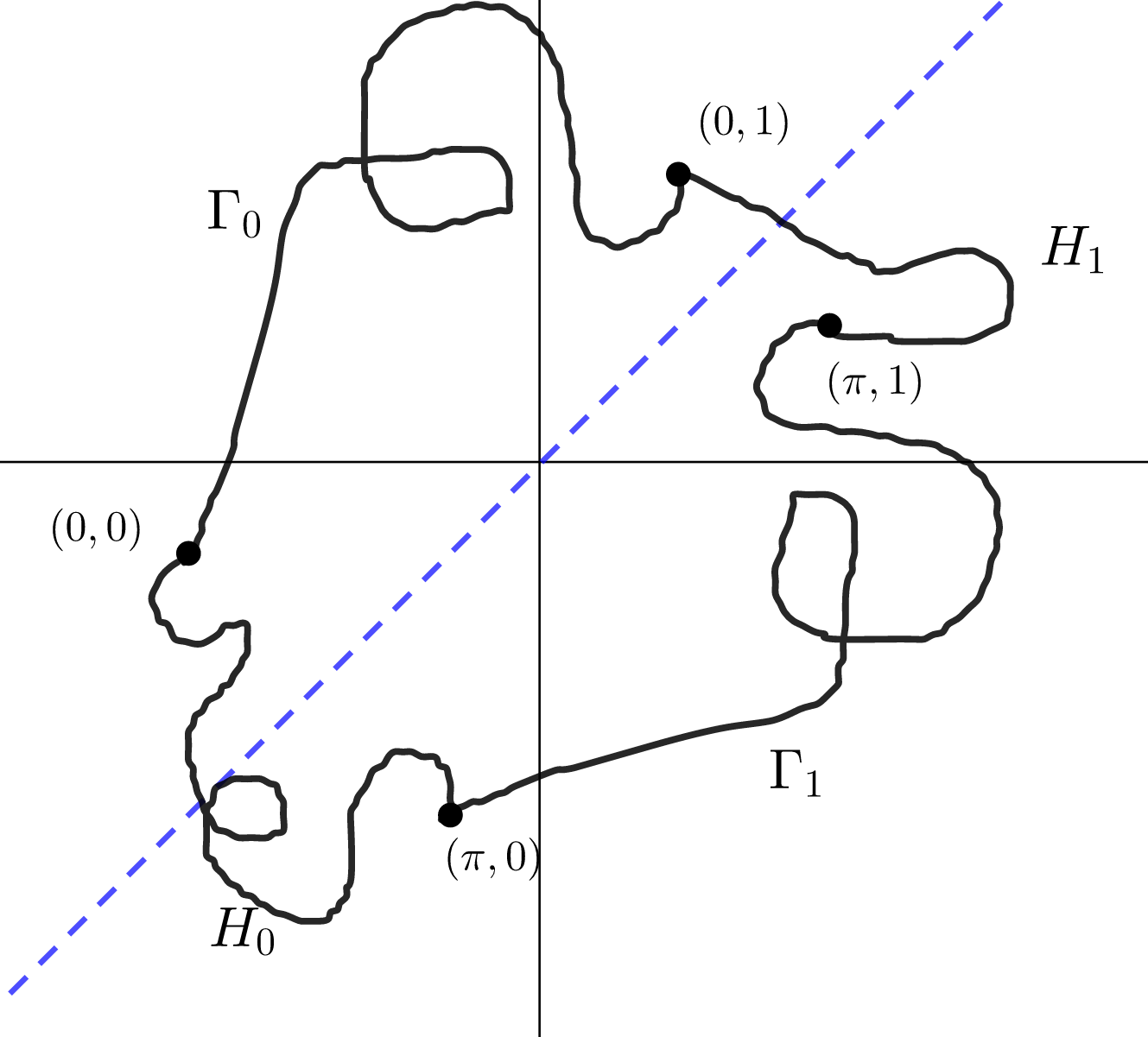}
\caption{\sf{The boundary of half the cylinder is mapped to a closed loop 
with a nonzero winding number around the origin. Note that $\gamma_0$ and $\gamma_1$ are 
symmetric about the axis of the cylinder, and therefore $\Gamma_0=\Sigma(\Gamma_1)$.}}
\label{winding}
\end{center}
\end{figure}

Additionally, let $\zeta_0, \zeta_1 \colon [0,\pi] \to S$ be the ``half-circles'' defined by 
$\theta \mapsto (\theta,0)$ and $\theta \mapsto (\theta,1)$, respectively, and 
$Z_i:=G \circ  \zeta_i$, for $i=0,1$. Let $\zeta_0'$ and $Z_0'$ be the reverses of $\zeta_0$ 
and $Z_0$, respectively. See Figure~\ref{winding} for an illustration.

Concatenating $\gamma_0$, $\zeta_1$, $\gamma_1'$, and $\zeta_0'$ in this order, we obtain 
a closed loop $\ell$ in $S$, which is the boundary of $D$, and its corresponding
image $L:= G(\ell)$ in the plane.  
By construction, $\ell$ bounds the topological disk $[0,\pi] \times [0,1]$ in $S$.  
and $L$ is a closed loop in $\reals^2$. We prove below that, if $O \notin L$, then 
$L$ has a non-zero winding number around $O$. Given this property, we claim
that $O$ must lie in $\Pi=G([0,\pi]\times[0,1])$. Indeed, if $O\notin\Pi$ then
$\Pi$ is contained in the punctured plane at the origin. Since $L$ has a 
nonzero winding number around $O$, it is homotopic, within $\Pi$, to a curve 
obtained by looping around the origin a nonzero number 
of times. This curve is not homotopy-trivial---it is not homotopic to a point
(within $\Pi$). On the other hand, $\ell$ is clearly homotopy-trivial within $D$, 
and therefore $L = G(\ell)$ is homotopic to a single point within $G(D) = \Pi$, 
a contradiction that establishes the proposition.
$\Box$ \medskip

To complete the proof, we thus show:
\begin{claim} \label{clm:winding-number}
In the notation of the above proof, if $L$ misses $O$, then the winding number of 
$L$ around $O$ is non-zero.
\end{claim}
\noindent{\bf Proof.}
Let $\arg (x,y)$ be the clockwise angle that the vector $(x,y)$ makes with the 
positive $x$-axis and let, for a section $\lambda$ of $L$, $\Delta \lambda$ be the 
integral of the change in $\arg \lambda (t)$ as $t$ traces out $\lambda$ from start to finish.

We will compute the winding number of $L$ around the origin by breaking $L$ into 
sections $\lambda$, computing the angle change $\Delta \lambda$ for each section, 
and adding up the numbers.
  
Let $\alpha:= \arg G(0,0) \in (0,\pi/2)$. Then by $\Sigma$-symmetry 
$\arg G(\pi,0)=\pi/2-\alpha$.  Similarly, put $\beta:= \arg G(0,1) \in (\pi, 3\pi/2)$, 
so that $\arg G(\pi,1) = 5\pi/2 - \beta \in (\pi,3\pi/2)$.  
Since $Z_0 \subset Q_1$ (so $Z_0$ cannot wind around $O$), 
$\Delta Z_0=\arg G(\pi,0) - \arg G(0,0) = (\pi/2 - \alpha ) - \alpha = \pi/2 - 2\alpha$ 
and $\Delta Z'_0= - \Delta Z_0 = 2\alpha - \pi/2$. Similarly, since $Z_1 \subset Q_3$, 
$\Delta Z_1=\arg G(\pi,1) - \arg G(0,1) = (5\pi/2 - \beta ) - \beta = 5\pi/2 - 2\beta$.

$\Gamma_0$ connects $G(0,0)$ to $G(0,1)$, so 
$\Delta \Gamma_0 = \arg G(0,1) - \arg G(0,0) + 2\pi k=\beta-\alpha+2\pi k$, 
for some integer $k$, over which we have no control as we do not know how 
many times $\Gamma_0$ winds around the origin 
(we use here the assumption that $\Gamma_0$ avoids the origin). Because of 
$\Sigma$-symmetry, we must have $\Delta \Gamma_1 = - \Delta \Gamma_0$ and therefore 
$\Delta \Gamma_1'=-\Delta \Gamma_1 = \Delta \Gamma_0$.

To summarize, the total change of the angle around $L$ is equal to
\begin{align*}
\Delta \Gamma_0 + \Delta Z_1 + \Delta \Gamma_1' + \Delta Z_0' 
& = 2 \Delta \Gamma_0 + \Delta Z_1 + \Delta Z_0'\\
& = 2 (\beta-\alpha+2\pi k) + (5\pi/2 - 2\beta) + (2\alpha - \pi/2)\\
& = 2\pi (2k+1).
\end{align*}
In particular, the total angle is not zero, no matter what the value of the 
integer~$k$ is, thereby completing the proof.

The remainder of the argument, namely that Proposition~\ref{pro:include-origin} 
implies the theorem, and the extension to the case of convex polytopes, is done 
exactly as in the first proof, thereby completing this second proof of the theorem.
$\Box$ \medskip

\section{Purely translational motions} \label{app:purely-translational}

In this section we study the case of translational motion. We show in Section~\ref{subsec:pure-trans} 
that purely translational motions of $K$ through a rectangular window $W$ are not more powerful than sliding,
in the sense that if a translational motion exists then a sliding motion exists as well, 
with the same orientation as that of the translational motion.
In Section~\ref{subsec:fixed-orientation} we consider the case where the orientation of $K$ is prescribed 
and we wish to find a sliding motion while maintaining the prescribed orientation. We also give a near-linear time 
algorithm for planning a purely translational motion of $K$ through an arbitrary flat (not necessarily convex)
polygonal window with a constant number of edges.

\subsection{Translational motion implies sliding}\label{subsec:pure-trans} 

\begin{figure}[htbp]
\begin{center}
\includegraphics[scale = 0.15]{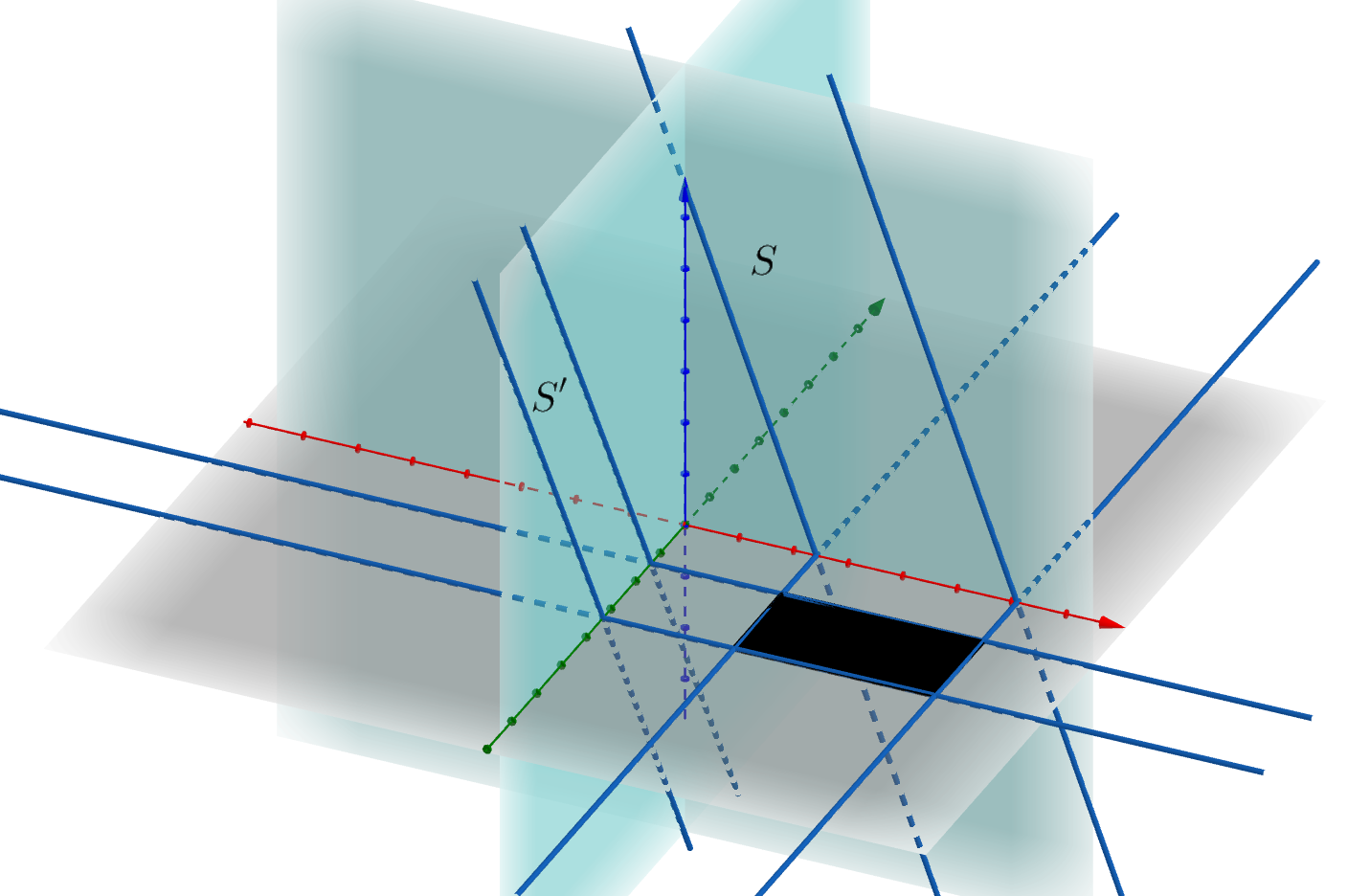}
\caption{\sf{Some translated copy of $K$ is contained within a slab $S$ perpendicular to the $xz$-plane, and some other translated copy of $K$
is contained in a slab $S'$ perpendicular to the $yz$-plane, so that the $x$-projection of the intersection of $S$ with the $xy$-plane is
contained in that of $W$, and the $y$-projection of the intersection of $S$ and the $xy$-plane is contained in that of $W$. We can thus
translate $K$ to a placement contained in the prism $S\cap S'$, from which $K$ can slide through the intersection of $S\cap S'$ and the
$xy$-plane (the black rectangle, which is contained in $W$).}}
\label{no-rotations-slide}
\end{center}
\end{figure}

We prove the following theorem, which is, 
in a sense, a strengthening of Lemma~\ref{lem:vertical}.
\begin{theorem} \label{thm:no_trans}
If $K$ can be moved through a rectangular window $W$ by a purely translational 
collision-free motion in some fixed orientation $\Theta$, then $K$ can be moved through $W$, 
possibly from some other starting position, 
by sliding while keeping the same orientation $\Theta$.
\end{theorem}
\noindent{\bf Proof.}
Again, we first carry out the proof for the case where $K$ is a smooth compact 
strongly convex body in three dimensions, and then extend the proof to
the case where $K$ is a convex polytope. When $K$ translates through $W$, 
its projection on the $xz$-plane is a fixed convex region that translates through
the interval $g$ on the $x$-axis, which is the $x$-projection of $W$.
Recall that we denote the projection on the $xz$-plane by $\pi$.
By the analysis in Section~\ref{sec:unb}, there is a time $t$ during the 
motion at which the tangents to $\pi(K(t))$ at the endpoints of $g\cap \pi(K(t))$ 
become parallel, and form, when extended in the $y$-direction, a (possibly slanted) 
slab $S$ that is orthogonal to the $xz$-plane, and that contains the placement 
of $K$ at time $t$, so that the intersection of $S$ with the $xy$-plane 
is a $y$-vertical strip of width at most $a$, whose $x$-projection is 
contained in that of $W$. Applying the same argument to the $yz$-plane 
(swapping the $x$- and $y$-directions), we get another time $t'$ at which 
$K$ is contained in another slab $S'$, orthogonal to the $yz$-plane, whose 
intersection with the $xy$-plane is an $x$-horizontal strip of width at most $b$, 
whose $y$-projection is contained in that of $W$ (see Figure~\ref{no-rotations-slide}).

Hence, the intersection $\tau = S\cap S'$ is a (slanted) prism, whose cross-section 
with the $xy$-plane is a rectangle contained in $W$. Moreover, as is easily verified, 
$\tau$ contains some translated copy $K_0$ of $K$. Hence, $K$ can slide through
$W$ from its placement $K_0$ in the unbounded direction of $\tau$. 

The case where $K$ is a convex polytope can be handled by the same limiting 
argument given in the proof of Theorem~\ref{thm:unbounded}.
$\Box$ \medskip

We remark that, by Lemma~\ref{lem:vertical}, the above lemma also implies
that $K$ can also slide through $W$ in the $z$-direction, 
from a different initial placement, possibly in a different orientation.

Theorem~\ref{thm:no_trans} together with Lemma~\ref{lem:vertical} imply the following
\begin{corollary}\label{cor:trans-means-z-sliding}
If $K$ can be moved through a rectangular window $W$ by a purely translational 
collision-free motion, then $K$ can be moved through $W$, possibly from some other 
(translated and rotated) starting position, by sliding in the $z$-direction.	
\end{corollary}
This leads to an efficient algorithm, with running time $O(n^{8/3}{\rm polylog}(n))$ (Section~\ref{sec:sliding-algo2}), for 
finding a translational motion for $K$ through $W$, if one exists, in the form of sliding in the $z$ direction, possibly at a different orientation.
If that algorithm notifies that no such sliding motion exists, then it follows from Corollary~\ref{cor:trans-means-z-sliding} that there is no translational motion for $K$ through $W$, at any orientation.
\subsection{Prescribed orientation} \label{subsec:fixed-orientation}

We now address a more restricted case where we are given a prescribed orientation $\Theta$ 
and we wish to find a purely translational motion for $K$ with this orientation. 
We denote the polytope $K$ at orientation $\Theta$ (ignoring translations) by $K^\Theta$. 

Notice that the algorithms of Section~\ref{sec:1d} are not immediately useful for answering the prescribed-orientation question. The algorithm
of Section~\ref{sec:sliding-algo1} gives us all the orientations of $K$ in which it can vertically slide through the window, while the
algorithm of Section~\ref{sec:sliding-algo2} gives us some orientations with a valid vertical sliding. But in either case these do not
necessarily include the desired orientation $\Theta$, which may require sliding in a different direction.

We designate an arbitrary vertex $v$ of $K^\Theta$ as a reference point.
Since the existence of a purely translational motion implies a sliding motion, we may require
the output of the prescribed-orientation motion-planning algorithm to be a sliding motion,
expressed as a line $L$ in space such that $K^\Theta$ slides through $W$ while $v$ moves along $L$, 
or an indication that no translational motion for $K^\Theta$ exists.



\begin{theorem}\label{them:fixed-orientation}
Given an orientation $\Theta$, we can determine whether a translational motion for $K^\Theta$ through the rectangular $W$ exists,
and if so find a sliding line for $K^\Theta$ through $W$ in $O(n)$ time.
\end{theorem}

\noindent{\bf Proof.} 
Let $\pi(K^\Theta)$ denote the orthogonal projection of $K^\Theta$ onto the $xz$-plane. 
We compute $\pi(K^\Theta)$ by traversing $K^\Theta$ from the topmost vertex to the bottommost vertx, in $O(n)$ time.

As before, let $g$ denote the projection of $W$ onto the $x$-axis. 
The proof of Theorem~\ref{thm:no_trans} (based on the analysis in Section~\ref{sec:unb})
shows that if there is a translational motion for $K^\Theta$  
through $W$ then there is a horizontal chord $\gamma$ of $\pi(K^\Theta)$ with endpoints $p_\ell$ and $p_r$ 
such that the length of the chord is not greater than $a$ (the length of $g$) and such that 
there exist tangents to $\pi(K^\Theta)$ at $p_\ell$ and $p_r$ that are parallel.
Such a chord, if exists, can be found in $O(n)$ time, and it will give us the slab $S$ 
of the proof of Theorem~\ref{thm:no_trans}. By an analogous procedure for the projection 
of $K^\Theta$ onto the $yz$-plane, we obtain the slab $S'$ of the theorem. If one of the 
two chords does not exist, then we conclude that there is no translational motion for $K^\Theta$ through $W$. 
Otherwise, we consider the intersection $\tau= S\cap S'$, which is a (slanted) prism. 
We then place a copy of $K^\Theta$ inside $\tau$, and let $L$ be the line through the 
reference vertex $v$ that is parallel to the unbounded direction of $\tau$. The theorem follows.
$\Box$\medskip

Finally, we observe that, for an arbitrary polygonal window $W$ with a constant number of edges, 
we can find a translational motion for $K^\Theta$ through $W$ (or determine that no such motion
exists) in $O(n\log n)$ time. More generally, we have:

\begin{theorem}\label{the:fixed-orientation-arbitrary-window}
Let $W$ be an arbitrary (not necessarily convex) polygonal window with $k$ edges, lying in the $xy$-plane.
Given a prescribed orientation $\Theta$ of $K$, we can determine whether a translational motion 
for $K^\Theta$ through $W$ exists, and, if so, find such a motion, in $O(nk\log k \log kn)$ 
randomized expected time.
\end{theorem}

\noindent{\bf Proof.} 
We triangulate the complement of $W$ within the
$xy$-plane into $s=O(k)$ triangles $A_1,\ldots,A_s$ with pairwise disjoint relative interiors.
We then construct the three-dimensional free configuration space $\F$ for the translational 
motion of $K^\Theta$ through $W$, using the technique of Aronov and Sharir~\cite{ArS97}. 
This technique asserts that the combinatorial complexity of $\F$ is $O(Nk\log k)$, where
$N$ is the overall complexity of the individual Minkowski sums $A_i \oplus (-K^\Theta)$,
and that $\F$ can be constructed by a randomized algorithm in $O(Nk\log k \log N)$ expected time.
As the complexity of each Minkowski sum $A_i \oplus (-K^\Theta)$ is $O(n)$, with an absolute
constant of proportionality, we have $N = O(kn)$, implying that the complexity of $\F$ is
$O(nk^2\log k)$, and that $\F$ can be computed in $O(nk^2\log k \log nk)$ expected time.
We now take an arbitrary point $p^+$ (resp., $p^-$) in $\F$ that represents a configuration 
where $K^\Theta$ is fully contained in the positive halfspace $z>0$ (resp., in the negative halfspace $z<0$),
and check whether $p^+$ and $p^-$ lie in the same connected component of $\F$. If indeed they lie in the same connected component of $\F$, we use the vertical decomposition of $\F$ constructed by the algorithm to extract a motion path for $K$ through $W$ within the same time bound. 
$\Box$\medskip

\section{Rotations are needed} \label{app:rotations-needed}

\begin{figure}[htbp]
\begin{center}
\includegraphics[scale = 0.24]{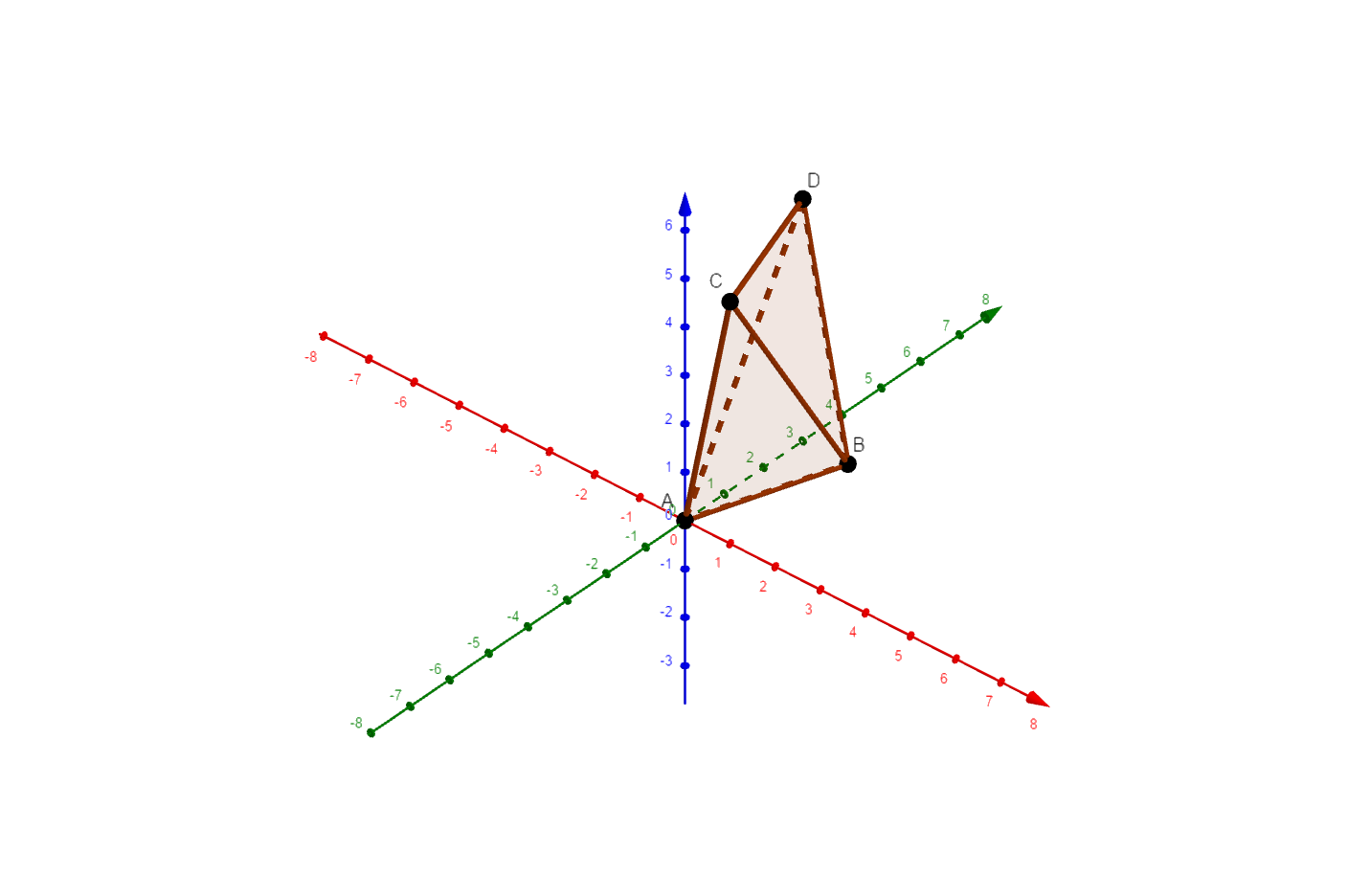}
\caption{\sf{The tetrahedron $K=ABCD$.}} 
\label{app:square-must-rotate}
\end{center}
\end{figure}

So far we have considered versions of the problem in which we were able to show that
the existence of an arbitrary collision-free motion of $K$ through $W$ implies that
$K$ can also slide through $W$ (or, in one instance, through another window related to $W$).
However, perhaps not very surprisingly, this is not the case in general. We show in 
this and the following section that in general rotations are needed to obtain a 
collision-free motion of the polytope through the window.
\begin{lemma} \label{lem:must_rotate}
Let $W$ be a square window with side length $\sqrt{5}$. Let $A=(0,0,0),B=(1,3,0),C=(1,0,h),D=(0,3,h)$ 
be four points, where $h\gg 1$ is a sufficiently large parameter, and let $K$ be the tetrahedron $ABCD$ 
(see Figure~\ref{app:square-must-rotate}). Then
\begin{enumerate}
\item 
$K$ cannot pass through $W$ by any purely translational collision-free motion (for sufficiently large $h\gg1$).
\item 
$K$ can pass through $W$ by a collision-free motion with only two degrees of freedom: 
translating in the $z$-direction combined with rotation around a $z$-vertical axis (for any value of $h>0$).
\end{enumerate}
\end{lemma}

\noindent{\bf Proof. (1)} Assume to the contrary that there exists a purely translational motion 
of $K$ through $W$. By Theorem~\ref{thm:no_trans}, there exists some placement $K_0$ of $K$ from 
which $K$ can slide through $W$ in the negative $z$-direction. Let $\pi(K_0)$ denote the vertical 
projection of $K_0$ onto the $xy$-plane. By the theorem, $\pi(K_0)$ can be rigidly placed inside 
$W$. Since we assume $h$ to be very large, it follows that, when transforming $K$ to $K_0$, the 
$z$-vertical direction turns by only a very small angle, for otherwise $\pi(K_0)$ would be very 
long and would not fit into such a square. More formally, for every $\eps>0$ there exists $h_0$ 
such that for every $h>h_0$ the angle by which the $z$-axis turns from $K$ to $K_0$ is at most 
$\eps$. As $\eps$ decreases to zero, the lengths of the projections of the segments $AB,CD$ 
grow to $\sqrt{10}$, which is their original length, and the angle between them converges to 
some $0<\phi<\frac{\pi}{2}$ (the exact angle is the angle obtained when the $z$-axis remains 
the same, which is then $\phi=2\sin^{-1}{\frac{1}{\sqrt{10}}}$). Therefore, the projection 
$\pi(K_0)$ is the convex hull of two segments of length sufficiently close to $\sqrt{10}$, 
which is the diagonal of $W$, where the angle between them is sufficiently far from $0$, $\pi/2$. 
Hence $\pi(K_0)$ cannot be placed inside a square with side length $\sqrt{5}$. 
This contradiction establishes the first part of the theorem.
\medskip

\begin{figure}[htbp]
\begin{center}
\includegraphics[scale = 0.12]{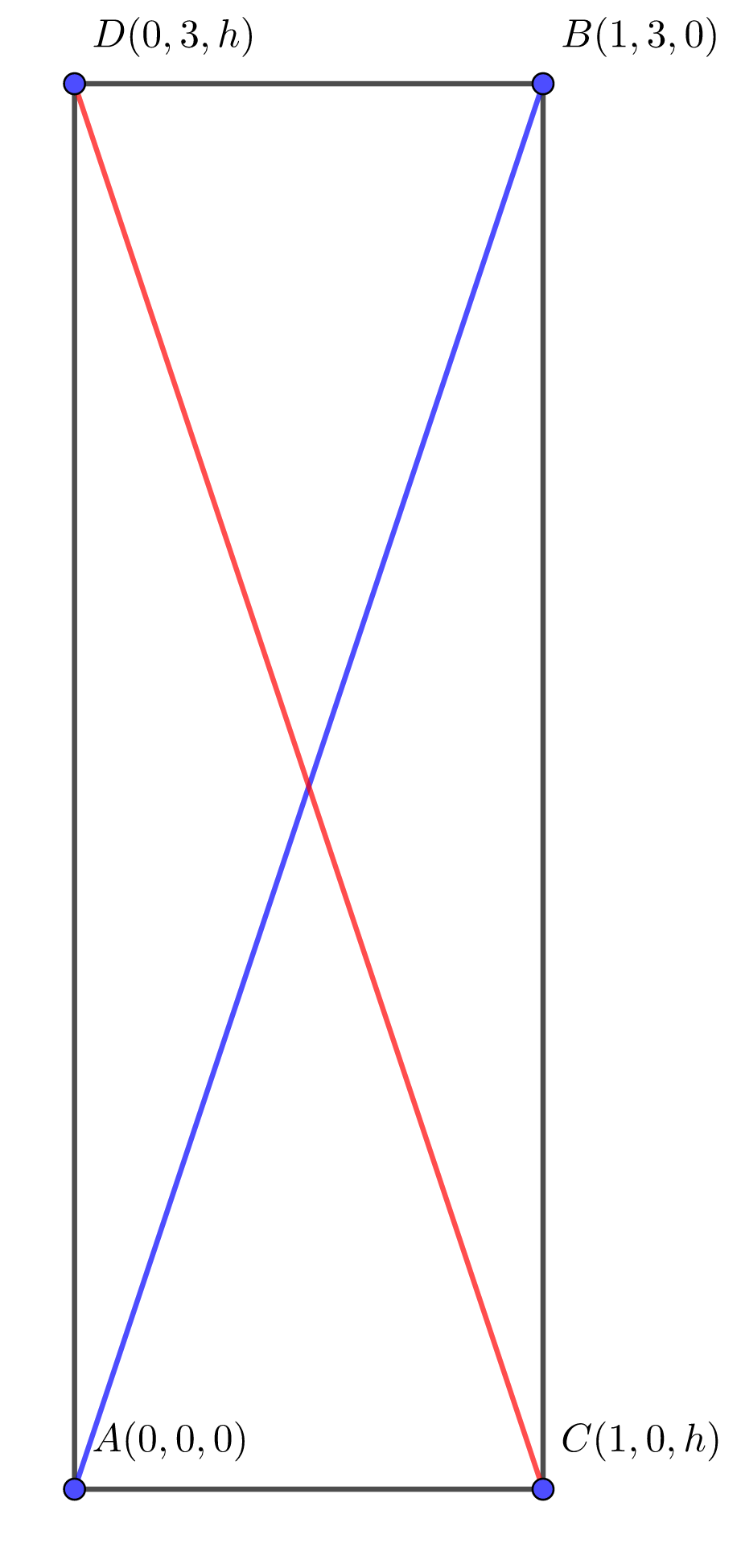}
\includegraphics[scale = 0.12]{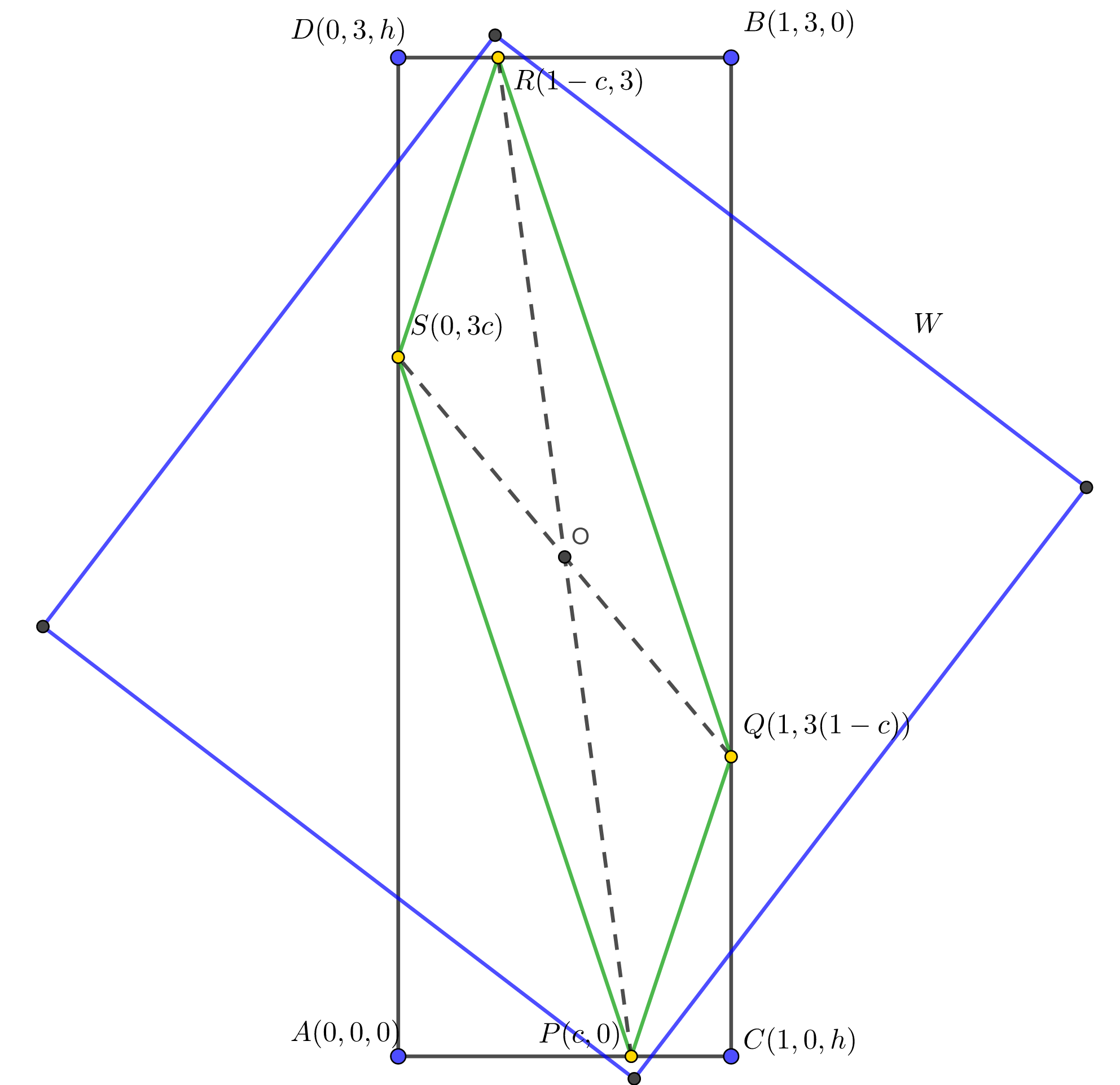}
\caption{\sf{Placing the cross section of $K$ inside $W$. 
Left: A schematic top view of $K$. Right: The cross section of $K$ at time $c$ (green), and a copy of $W$ that contains it.}}
\label{tetra}
\end{center}
\end{figure}

\noindent{\bf (2)} We move $W$ instead of $K$, allowing it only to translate in the $z$-direction
(so it always remains horizontal), and simultaneously rotate around its center
(so the motion of $W$ has only two degrees of freedom). More concretely, the center 
of $W$ moves up along the line $x=1/2$, $y=3/2$. We parameterize the motion by a 
parameter $c\in [0,1]$, so that at time $c$, $W$ lies on the plane $z=ch$ and its 
center is at $(1/2,3/2,ch)$. See Figure~\ref{tetra}(left) for a schematic top view of $K$.

The cross section $K_c$ of $K$ at time $c$ is shown (in green) in Figure~\ref{tetra}(right).
It is a quadrilateral $PQRS$, with $P = (c,0)$, $Q = (1,3(1-c))$, $R = (1-c,3)$ and
$S = (0,3c)$. We place $W$ around $K_c$ so that $PR$ lies at the middle of one diagonal of $W$
(so $W$ keeps rotating to align with this rotating segment). It is clear that the motion 
of $W$ is continuous, and it remains to show that $K_c$ always lies in (the placement 
at height $ch$, with the aligned diagonals, of) $W$.

It suffices to show that, at any time $c$ during the motion, $\Delta PRS$ is contained in the
isosceles right triangle with hypotenuse $PR$ (this triangle is half of $W$, and the
argument for the complementary half and for $\Delta PQR$ is fully symmetric). For this,
it suffices to show that each of the angles $\sphericalangle SPR$,
$\sphericalangle SRP$ is smaller than $\pi/4$. Note that the edges of $PQRS$ have fixed slopes, 
namely $3$ and $-3$, as they are parallel to the $xy$-projections of $AB$ and $CD$.
This implies that $\tan \sphericalangle SPQ = \tan \sphericalangle SRQ = \frac{3}{4} < 1$, so
$\sphericalangle SPQ = \sphericalangle SRQ < \pi/4$. We have thus shown that $K$ can move through $W$
by (the dual version of) this motion, of translation in the $z$-direction combined with horizontal rotation.
$\Box$ \medskip


\section{The case of a circular window} \label{sec:circular-window}

In this section we study the case where $W$ is a circular window. 
There are (at least) three possible types of motion of $K$ through $W$: sliding,
purely translational motion, and general motion with all six degrees of freedom.
In this section we show that these types are not equivalent, as spelled out in
the following theorem.
\begin{theorem} \label{thm:circle}
Let $K$ be the regular tetrahedron of side length $1$. Then there exist two 
threshold parameters $1 > \delta_1 \approx 0.901388 > \delta_2 \approx 0.895611$,
so that, denoting by $d$ the diameter of $W$, we have:
\begin{description}
\item[(i)]
$K$ can slide through $W$ if $d \ge 1$. 
\item[(ii)]
$K$ cannot slide through $W$, but can pass through $W$
by a purely translational motion, if $\delta_1 \le d <1$.
\item[(iii)]
$K$ cannot pass through $W$ by a purely translational motion, 
but can pass through $W$ by a general motion, if $\delta_2 \le d < \delta_1$.
\item[(iv)]
$K$ cannot pass through $W$ at all if $d < \delta_2$.
\end{description}
\end{theorem} 

\noindent{\bf Proof.}

\paragraph{$K$ can slide through $W$ if $d \ge 1$.}
\noindent{\bf (i)} 
In this case $K$ can slide through $W$, because $K$ can be enclosed in a 
cylinder of diameter $1$, whose axis is orthogonal to two opposite edges of $K$.

\paragraph{No sliding of $K$ is possible when $d < 1$.} 
This claim follows by showing that any circular cylinder that contains 
$K$ must have diameter at least $1$. The analysis below is taken from
\cite{exchange}, and is given here for the sake of completeness.

For any four vectors $\vec{v_1},\vec{v_2},\vec{v_3},\vec{v_4}$ the following identity holds:
$$
\frac{1}{2}\sum\limits_{i=1}^{4}\sum\limits_{j=1}^{4}|{\vec{v_i}-\vec{v_j}}|^2
= 3\sum\limits_{i=1}^{4}|{\vec{v_i}}|^2-2\sum\limits_{1\leq i<j\leq 4}\langle\vec{v_i},\vec{v_j}\rangle
= 4\sum\limits_{i=1}^{4}|{\vec{v_i}}|^2-\left|{\sum\limits_{i=1}^{4}\vec{v_i}}\right|^2.
$$
Let $K$ be the tetrahedron whose vertices are:
\begin{gather*}
\vec{v_1}=\frac{1}{\sqrt{24}}(0,0,3),\qquad \vec{v_2}=\frac{1}{\sqrt{24}}(\sqrt{8},0,-1),\\
\vec{v_3}=\frac{1}{\sqrt{24}}(-\sqrt{2},\sqrt{6},-1),\qquad \vec{v_4}=\frac{1}{\sqrt{24}}(-\sqrt{2},-\sqrt{6},-1).
\end{gather*}
It is indeed a regular tetrahedron of side length $1$:
\begin{gather*}
|{\vec{v_1}-\vec{v_2}}|^2=\frac{1}{24}({8+16})=1,\qquad |{\vec{v_1}-\vec{v_3}}|^2=\frac{1}{24}({2+6+16})=1,\\
|{\vec{v_1}-\vec{v_4}}|^2=\frac{1}{24}({2+6+16})=1,\qquad |{\vec{v_2}-\vec{v_3}}|^2=\frac{1}{24}({18+6})=1,\\
|{\vec{v_2}-\vec{v_4}}|^2=\frac{1}{24}({18+6})=1,\qquad |{\vec{v_3}-\vec{v_4}}|^2=\frac{1}{24}({24})=1.
\end{gather*}
Represent vectors in our 3-dimensional space as $3\times 1$ column vectors. By some more algebra, we obtain
\begin{gather*}
\sum\limits_{i=1}^{4}\vec{v_i}\vec{v_i}^T=\\ \frac{1}{24}\left[
\begin{pmatrix}
0 & 0 & 0\\
0 & 0 & 0\\
0 & 0 & 9
\end{pmatrix}+
\begin{pmatrix}
8 & 0 & -\sqrt{8}\\
0 & 0 & 0\\
-\sqrt{8} & 0 & 1
\end{pmatrix}\right]+\\+\frac{1}{24}\left[
\begin{pmatrix}
2 & -\sqrt{12} & \sqrt{2}\\
-\sqrt{12} & 6 & -\sqrt{6}\\
\sqrt{2} & -\sqrt{6} & 1
\end{pmatrix}+
\begin{pmatrix}
2 & \sqrt{12} & \sqrt{2}\\
\sqrt{12} & 6 & \sqrt{6}\\
\sqrt{2} & \sqrt{6} & 1
\end{pmatrix}\right]=
\end{gather*}
$$
\frac{1}{24}
\begin{pmatrix}
12 & 0 & 0\\
0 & 12 & 0\\
0 & 0 & 12
\end{pmatrix}=\frac{1}{2}I_3 .
$$
Therefore, for any unit vector $\vec{n}$ the following equation is satisfied:
$$
\sum_{i=1}^4 \langle \vec{v_i}, \vec{n} \rangle^2 =
\sum\limits_{i=1}^{4}|{\vec{v_i}^T\vec{n}}|^2 =
\sum\limits_{i=1}^{4}Tr\left((\vec{v_i}\vec{v_i}^T)\cdot(\vec{n}\vec{n}^T)\right) =
\frac{1}{2}Tr(\vec{n}\vec{n}^T)=\frac{1}{2} .
$$
Note that $\sum\limits_{i=1}^{4}v_i=0$, and hence:
$$
\sum\limits_{1\leq i<j\leq 4}{\langle \vec{n}, \vec{v_i} - \vec{v_j} \rangle}^2 =
\frac{1}{2}\sum\limits_{i=1}^{4}\sum\limits_{j=1}^{4}{\langle \vec{n}, \vec{v_i} - \vec{v_j} \rangle}^2 =
4\sum\limits_{i=1}^{4}{\langle \vec{n}, \vec{v_i} \rangle}^2 = 2 .
$$
Assume that the smallest cylinder that contains $K$ has diameter $d$. 
Let $h$ be a plane perpendicular to the axis of the cylinder, let $\vec{n_1},\vec{n_2}$ 
be two orthogonal unit vectors in $h$, let $\vec{u_i}$ be the projection of $\vec{v_i}$ 
on $h$, for $1\leq i\leq 4$, and put $\vec{l_{ij}}=\vec{u_i}-\vec{u_j}$. It is easy to see that
$$
|\vec{l_{ij}}|^2=\left|\vec{u_i}-\vec{u_j}\right|^2=
\langle \vec{n_1},\vec{u_i}-\vec{u_j}\rangle^2+\langle\vec{n_2},\vec{u_i}-\vec{u_j}\rangle^2=
\langle\vec{n_1},\vec{v_i}-\vec{v_j}\rangle^2+\langle\vec{n_2},\vec{v_i}-\vec{v_j}\rangle^2.
$$
We thus have $\sum\limits_{1\leq i<j\leq 4}|\vec{l_{ij}}|^2=4$. 
Consider the coordinate system in $h$ whose axes are parallel to $\vec{n}_1$ 
and $\vec{n}_2$, and whose origin is at the center of the intersection circle 
of $h$ and the cylinder. In this coordinate system we have $|\vec{u_i}|\leq \frac{d}{2}$ 
for each $i$. Note that $\vec{l_{ij}}$ remains the same and that 
$\sum\limits_{i=1}^{4}\vec{u}_i=0$, as the projection of $\sum_{i=1}^4 \vec{v_i} = 0$, and we thus obtain:
$$
\sum\limits_{1\leq i<j\leq 4}|\vec{l_{ij}}|^2 =
\sum\limits_{1\leq i<j\leq 4}\left|\vec{u_i}-\vec{u_j}\right|^2 =
4\sum\limits_{i=1}^{4}|{\vec{u_i}}|^2\leq 16\left(\frac{d}{2}\right)^2=4d^2.
$$
Finally we get that $4=\sum\limits_{1\leq i<j\leq 4}|\vec{l_{ij}}|^2\leq 4d^2$, 
so $d\geq 1$, but in our case the diameter of $W$ is strictly smaller than $1$.
We therefore conclude that $K$ cannot slide through $W$.

\paragraph{Purely translational motion through a circular window.} 

We next show that a purely translational motion of $K$ through a circular window
exists if and only if $d \ge \delta_1 \approx 0.901388$. 

Assume for now that the orientation of $K$ is fixed. 
We claim that $K$ can move through 
$W$ at this fixed orientation, by a purely translational motion, if and only if every 
horizontal cross section of $K$ can be enclosed in a disc
of diameter $d$; that is, the smallest enclosing disc of each
cross section has diameter at most $d$. We refer to this 
property as the \emph{small diameter property}. The `only if' 
part of this claim is obvious. We briefly explain the `if' part. 
Let $K(z)$ be the cross section of $K$ at height $h$. For every 
$x\in\partial K(z)$ let $c_x$ be a horizontal circle of diameter 
$d$ centered at $x$. That is, all the points within the plane of 
the cross section whose distance from $x$ is at most $\frac{d}{2}$. 
Clearly, the intersection $R(z)=\bigcap\limits_{x\in\partial K(z)}^{}{c_x}$ 
denotes the set of all available positions for the center of $W$ within 
that plane, such that it contains the cross section $K(z)$. $K(z)$ is a 
continuous function of $z$ in the Hausdorff metric of sets, and hence so 
is $R(z)$. This is easily seen to imply that we can choose the position 
of the center of $W$ for every cross section in a way that is continuous in $z$.

Assume without loss of generality that the initial placement of
$K$ is with its lowest vertex at $z=0$, and let $h$ denote the 
$z$-coordinate of the highest vertex. As above, denote by $K(z)$ the 
cross section of $K$ at height $z$, for $z\in [0,h]$. 
Assume without loss of generality that all four vertices have 
distinct $z$-coordinates, and that the order of increasing 
$z$-coordinates of the vertices is $A$, $B$, $C$, $D$; that is,
$z_A < z_B < z_C < z_D$.

We claim that the small diameter property holds if and only if 
it holds for $K(z_B)$ and $K(z_C)$. Indeed, observing that these two 
cross sections are triangles, assume without loss of generality 
that the radius $\rho$ of the smallest enclosing disc $D_B$ of 
$K(z_B)$ is larger than or equal to that of $K(z_C)$. 
Enclose $K(z_C)$ by some disc $D_C$ of radius $\rho$, 
and let $E$ be the convex hull of $D_B\cup D_C$, which
is a possibly slanted elliptic cylinder, each of whose horizontal
cross sections is a congruent copy of the disc $D_B$. Since $K$ has 
no vertices in the open slab $z_B < z < z_C$, it follows that
the portion of $K$ within the closed slab $z_B \le z \le z_C$
is the convex hull of $K(z_B)\cup K(z_C)$, and is
therefore fully contained in $E$. Hence, for every
$z_B < z < z_C$, $K(z)$ is contained in a disc of radius 
$\rho$. The cases of the slabs $z_A < z < z_B$ and $z_C < z < z_D$
are argued in the same manner. This establishes our claim.

In other words, we want to find orientations of $K$ for which
the (triangular) horizontal cross sections at the two middle vertices
of $K$ (in the $z$-direction) have smallest enclosing discs of 
diameters smaller than $1$.

\begin{figure}[htbp]
\begin{center}
\includegraphics[scale = 0.3]{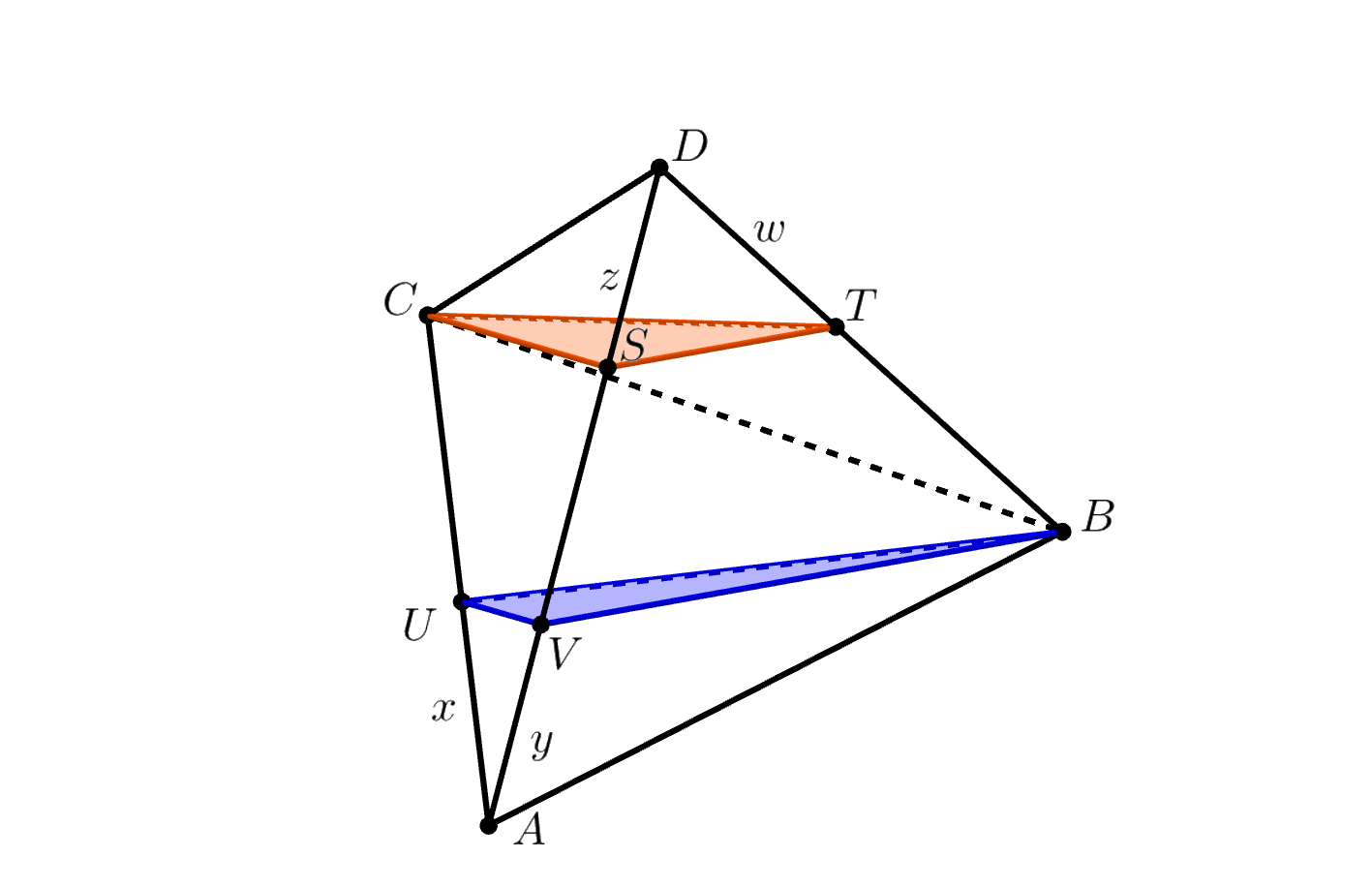}
\caption{\sf{The horizontal cross sections of a regular tetrahedron through its two middle vertices.}}
\label{tetrahedron-cross-section}
\end{center}
\end{figure}

Denote the cross section $K(z_B)$ through $B$ by $BUV$, where
$U$ is the point $AC\cap K(z_B)$ and $V$ is the point $AD\cap K(z_B)$.
Put $x = |AU|$ and $y = |AV|$, so $0\le x,y\le 1$.
Similarly, we write the triangular cross section $K(z_C)$ 
through $C$ as $CST$, where $S$ is the point $AD\cap K(z_C)$ 
and $T$ is the point $BD\cap K(z_C)$, and put
$z = |SD|$ and $w = |TD|$, so again $0\le z,w\le 1$.
See Figure~\ref{tetrahedron-cross-section} for an illustration.
Note that we must have $x > y$ and $w > z$, for otherwise
$A$ and $D$ would not have been the two $z$-extreme vertices of $K$.

The requirement that these two cross sections be parallel imposes
the following relations between $x$, $y$, $z$, and $w$. 

\begin{align} \label{eq:xyzw}
z & = \frac{x-y}{x} \\
w & = \frac{x-y}{x(1-y)} \nonumber .
\end{align}

Indeed, since the two cross sections are parallel, they intersect any plane
(not parallel to them) at parallel lines. In particular, we have $UV\parallel CS$
and $TS\parallel BV$, so the triangles $AUV$ and $ACS$ are similar, and so are
the triangles $DST$ and $DVB$. The first similarity implies that
$$
x = \frac{AU}{AC} = \frac{AV}{AS} = \frac{y}{AS} ,
$$
so $AS = y/x$, and then 
$$
z = AD - AS = 1-AS = \frac{x-y}{x} .
$$
The second similarity implies that
$$
w = \frac{DT}{DB} = \frac{DS}{DV} = \frac{z}{1-y} = \frac{x-y}{x(1-y)} ,
$$
thus establishing (\ref{eq:xyzw}).

Note that, once we enforce $x > y$, the second inequality $w > z$ trivially holds.

The goal is then to search for orientations of $K$ and for suitable choices of $x$ and $y$
(and thus of $z$ and $w$ too) for which the two cross sections have smallest enclosing
discs of diameters smaller than $1$. This is done as follows.

For a triangle $\Delta$ of side lengths $a,b,c$, the circumradius $r(\Delta)$
of $\Delta$ is given by the formula
$$
r(\Delta) = \frac{abc}{4\cdot{\rm Area}(\Delta)} .
$$
The area can be expressed by Heron's formula as
$$
{\rm Area}(\Delta)^2 = \pi(\pi-a)(\pi-b)(\pi-c) ,
$$
where $\pi = (a+b+c)/2$ is half the perimeter. That is, we have
\begin{align*}
{\rm Area}(\Delta)^2 & = \frac{1}{16} (a+b+c)(b+c-a)(a+c-b)(a+b-c) \\
& = \frac{1}{16} ((a+b)^2-c^2)(c^2-(a-b)^2) \\
& = \frac{1}{16} (2a^2b^2 + 2a^2c^2 + 2b^2c^2 - a^4 - b^4 - c^4) .
\end{align*}
Therefore,
\begin{equation} \label{circumrad}
r^2(\Delta) = \frac{a^2b^2c^2}{ 2a^2b^2 + 2a^2c^2 + 2b^2c^2 - a^4 - b^4 - c^4 } .
\end{equation} 
Assume that the triangles $BUV$ and $CST$ are both acute, so their smallest enclosing
discs coincide with their circumscribing discs.
Apply this formula to each of the triangles $BUV$ and $CST$.
An easy application of the Law of Cosines yields
\begin{align*}
|BU|^2 & = 1 - x + x^2 \\
|BV|^2 & = 1 - y + y^2 \\
|UV|^2 & = x^2 - xy + y^2 \\
|CS|^2 & = 1 - z + z^2 \\
|CT|^2 & = 1 - w + w^2 \\
|ST|^2 & = z^2 - zw + w^2 .
\end{align*}
Substituting these values in (\ref{circumrad}), once with
$a^2 = |BU|^2$, $b^2 = |BV|^2$, $c^2 = |UV|^2$, and once with
$a^2 = |CS|^2$, $b^2 = |CT|^2$, $c^2 = |ST|^2$, we get the values 
of the circumradii of the two triangles. If any of these triangles 
is obtuse, the radius of its smallest enclosing disc is half the longest edge.

The goal is, as said above, to find values of the parameters $x,y$
that minimize the larger of these two radii (note that the choice of 
$x$ and $y$ determines the orientation of $K$, up to rotation about the 
$z$-axis, because they determine a slice of $K$ (namely, $BUV$) that has 
to be horizontal). By numerically testing a dense grid of values for $x,y$ 
and running methods for finding the minimum of a function (computing the 
radius of the smallest enclosing disc using (\ref{circumrad}) for acute 
triangles, and half the longest edge for obtuse triangles), the optimizing 
parameters turned out to be $x\approx 0.43400$ and $y\approx 0.30265$, and 
the larger of the two diameters was $\approx 0.901388$. Setting $\delta_1$ 
to this value completes the argument. 

\paragraph{General motion of $K$.} 

We now complete the proof of Theorem~\ref{thm:circle} by showing that $K$ 
can move through $W$, by an arbitrary collision-free motion,
if and only if $d \ge \delta_2 \approx 0.895611$.
In other words, for diameters $\delta_2\le d < \delta_1$, the only way
to move $K$ through $W$ is via a motion that also involves rotations,
and for diameters $d < \delta_2$, no motion of $K$ through $W$ is possible.

We first construct the desired motion for $d\ge \delta_2$, which consists 
of five steps---sliding, rotation, sliding, rotation, and a final sliding.
We use the setup and notations introduced in the analysis of the preceding 
step in the proof, as depicted in Figure~\ref{tetrahedron-cross-section}.
As earlier, it is more convenient to consider $K$ as fixed, and $W$
as moving around $K$. 

\begin{figure}[htbp]
\begin{center}
\includegraphics[scale = 0.25]{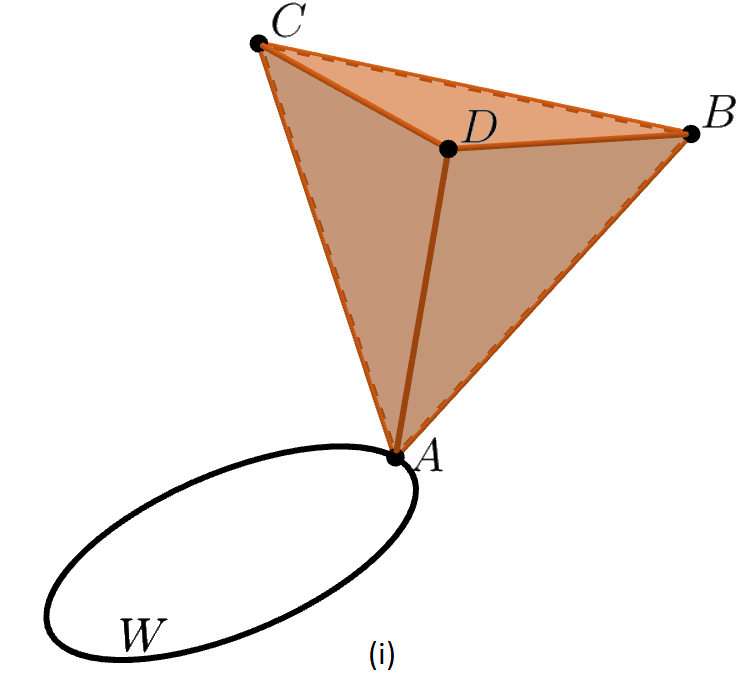}
\includegraphics[scale = 0.25]{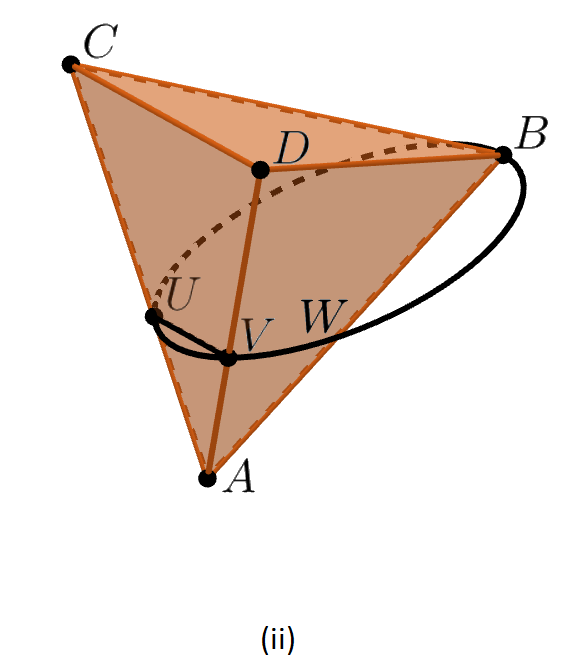}
\includegraphics[scale = 0.25]{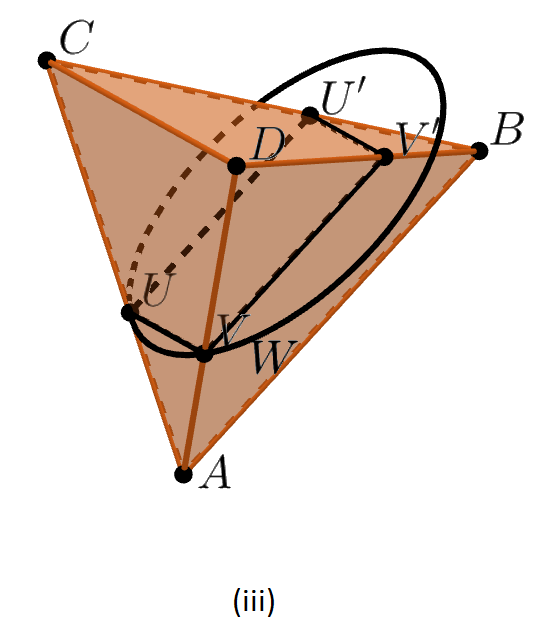}
\includegraphics[scale = 0.25]{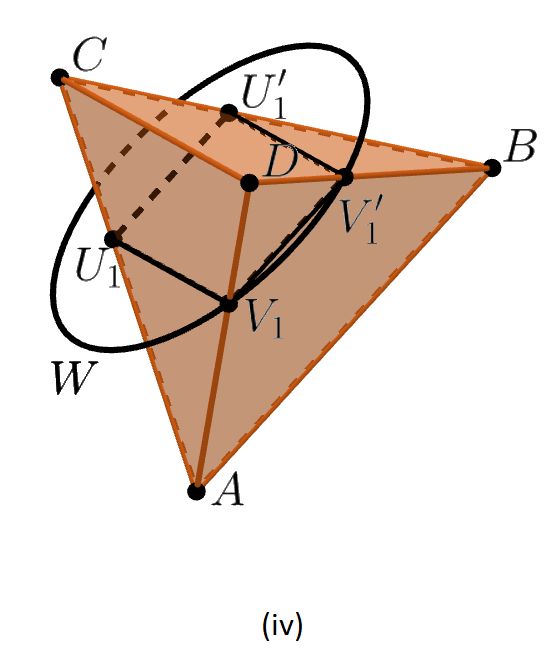}
\caption{\sf{Moving $W$ around $K$. (i) The initial configuration. 
(ii) $W$ contains the triangle $BUV$. (iii) $W$ contains the rectangle $UVV'U'$.
(iv) $W$ contains the symmetric rectangle, with edge lengths swapped. The remainder of the motion is 
a fully symmetric reversal of the first two steps.}}
\label{fig:circle_rotate}
\end{center}
\end{figure}

Assume that the lowest vertex $A$ lies on the $xy$-plane and inside $W$ 
(see Figure~\ref{fig:circle_rotate}(i)).
Start by sliding $W$ up, possibly in a slanted direction, ensuring that 
it keeps containing the cross section of $K$ with the plane supporting $W$, 
until $W$ comes to contain $B$;~see Figure~\ref{fig:circle_rotate}(ii).
We want to choose the initial orientation of $K$ so that the smallest 
enclosing disc of the horizontal (triangular) cross section of $K$ 
through $B$, namely the triangle $BUV$, is of diameter at most $d$.
As already noted, the orientation of $K$ is determined by $x$ and $y$, 
up to a possible rotation around the $z$-axis, as they determine
the vertical direction of $K$ (the one orthogonal to the triangle $BUV$).

We ran our numerical approximation scheme, and the smallest diameter
of the smallest enclosing disc of $BUV$ that we obtained was
$0.895611$, attained at $x = y = 0.391113$, and we take this value 
as our approximation of $\delta_2$. Note, incidentally, that this 
choice of parameters implies that the edge $CD$ of $K$ is horizontal.
It also implies that $|UV| = x = y = 0.391113$.

We now rotate $W$ about the line $UV$, in the direction that keeps
$A$ and $B$ on one side of it. The cross section of $K$ by the rotating 
plane is an isosceles trapezoid, and we keep rotating the plane until 
it becomes a rectangle $UVV'U'$. As is easily checked, we have 
$|UU'| = |VV'| = 1-x = 0.608887$, and the diameter of the smallest 
enclosing disc of $UVV'U'$, which is its diagonal, is $\approx 0.72368$, 
much smaller than $\delta_2$.
An easy adaptation of an argument used earlier shows that, during
this rotation of $W$ about $UV$, every cross section is contained
in the corresponding rotated copy of the disc of diameter $\delta_2$ 
whose bounding circle passes through $U$ and $V$.
See Figure~\ref{fig:circle_rotate}(iii).

We then slide $W$ in the direction perpendicular to $UVV'U'$.
During this sliding the cross section of $K$ remains rectangular,
so that $UV$ keeps increasing and $UU'$ keeps decreasing, while
the sum of their lengths remains $1$. We stop when we reach a `symmetric'
rectangle where the side parallel to $UV$ (resp., $UU'$) is of 
length $1-x$ (resp., $x$).
See Figure~\ref{fig:circle_rotate}(iv).

The situation that we have reached is fully symmetric to the one
after the first two steps, and we can now complete the motion by
a symmetric reversal of the first two steps.

\paragraph{No motion is possible when $d < \delta_2$.}

To complete the proof, for the case where $d < \delta_2$,
we observe that in this case $W$ cannot pass through any vertex of $K$,
because then, by definition of $\delta_2$, the smallest enclosing disc 
of any cross section through any vertex would have diameter larger than $d$.
$\Box$ \medskip

\section{Planning general rigid collision-free motion of a convex polytope through a rectangular window} \label{sec:6dofs}

Finally we deal with the general case, in which the motion of $K$ has all six degrees of freedom.
By standard (and general) arguments in algorithmic motion planning the free configuration space 
for this problem has complexity $O(n^6)$, and it can be computed in $O(n^{8+\eps})$ time~\cite{hs-a-18}, 
from which we can easily extract a solution path, when one exists, within the same time bound. 
We show here that we can exploit the special structure of the problem at hand to find a solution, 
or detect and notify that none exists, in time close to $O(n^4)$. We sketch below the main ideas; 
and then provide the full details.

If there is a solution path for $K$ to move through $W$ with all six degrees of freedom, then there is
also a canonical solution path where at all times at which $K$ intersects the plane of
$W$ (namely the $xy$-plane), $K$ touches the bottom and left edges of $W$ with two
edges $e_b$ and $e_\ell$ (possibly with the closure of these edges, namely with vertices of $K$, and possibly with more than one edge touching a side of $W$). 
During this motion, for every point on the path define $e_t$ to be the edge of $K$ whose intersection 
with the $xy$-plane has the largest $y$-coordinate, and $e_r$ to be the edge of $K$ whose intersection 
with the $xy$-plane has the largest $x$-coordinate. 
 
We now split the canonical solution path into maximal open segments,
along which the open edges $e_b,e_\ell,e_t$, and $e_r$ are fixed. We construct a collection 
of four-dimensional subspaces of the full-dimensional configuration space, one for each such quadruple
$e_b,e_\ell,e_t,e_r$ of four edges, consisting of those free placements that have those four edges 
as the extreme edges in the $x$- and $y$-directions within the $xy$-plane. 
This can be done in total $O(n^4)$ time since each of these subspaces has constant descriptive complexity. 

The major remaining problem is to efficiently detect the free connections among these $O(n^4)$ subspaces. 
The efficiency of our approach relies on the following lemma, which asserts that the total number of 
certain quintuplets of edges of $K$ that encode these connections is only $O(n^4)$, rather than $O(n^5)$, 
and that they can be computed efficiently:

\begin{lemma} \label{lem:no_of_quintuplets1}
The maximum number of quintuplets $(e_r,e_t,e_\ell,e_b,e_\xi)$, where $e_r,e_t,e_\ell,e_b$ are as defined above, 
and $e_\xi$ is another edge of $K$ whose intersection with the $xy$-plane $h_{xy}$ has the same $x$- (respectively, $y$-) 
coordinate as the intersection with $h_{xy}$ of $e_\ell$ or $e_r$ (respectively, $e_b$ or $e_t$), is $O(n^4)$. 
All these quintuplets can be computed in $O(n^3\lambda_q(n)\log n)$ time\footnote{$\lambda_q(n)$ is a near-linear function related to Davenport-Schinzel sequences~\cite{SA}.}, for some small constant $q$.
\end{lemma}
 
This in turn leads to the following summary result.
\begin{theorem}
Given a convex polytope $K$ with $n$ edges and a rectangular window $W$, we can 
construct a collision-free motion of $K$ through $W$, if one exists, or determine 
that no such motion exists, in time $O(n^3\lambda_q(n)\log n)$, for some small 
constant $q$. The algorithm requires $O(n^4)$ storage.
\end{theorem}

We now provide the full proofs and the algorithm.

\subsection{Planning the motion: Preliminaries}

\begin{figure}[htbp]
\begin{center}

\scalebox{1}{
  \begin{tikzpicture}[scale=0.5]
    \draw[draw=black,fill=cyan!20!white] (0,0) rectangle ++(8,6);
    \node[point=yellow,label={[label distance=0pt]east:$e_{\ell}$}] (l) at (0,2.5) {};
    \node[point=yellow,label={[label distance=0pt]north east:$e_t$}] (t) at (4,5) {};
    \node[point=yellow,label={[label distance=0pt]south:$e_b$}] (b) at (3,0) {};
    \node[point=yellow,label={[label distance=0pt]east:$e_r$}] (r) at (6,1) {};
    \coordinate (c1) at (1,1);
    \coordinate (c2) at (1.5,4);
    \coordinate (c3) at (5,4);
    \node[label=south east:$W$] at (0,6) {};
    \draw[red,thick] (l)--(c1)--(b)--(r)--(c3)--(t)--(c2)--(l);
  \end{tikzpicture}
}
\caption{\sf{A cross section of $K$ in $W$ during a canonical motion.}}
\label{fig:canon-in-W}
\end{center}
\end{figure}
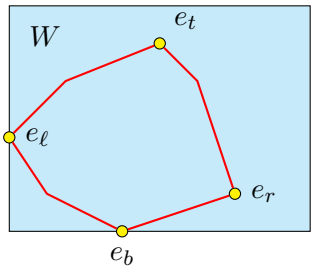

If there is a solution path for $K$ to move through $W$, then there is
also a canonical solution path $\pi$ where at all times at which $K$ intersects the plane of
W (namely the $xy$-plane), $K$ touches the bottom and left edges of $W$ with two
edges $e_b$ and $e_\ell$ (possibly with the closure of these edges, namely with vertices 
of $K$---we discuss this issue in detail below). During this motion, for every point on the path
define $e_t$ to be the edge of $K$ whose intersection with the $xy$-plane  has the largest 
$y$-coordinate and $e_r$ to be the edge of $K$ whose intersection with the $xy$-plane has 
the largest $x$-coordinate. In what follows we denote the $xy$-plane as $h_{xy}$. 
See Figure~\ref{fig:canon-in-W} for an illustration. 
Notice that the notation of the bottom (or top) edge of $W$ is with respect to the $y$-coordinate.  

We now split the canonical solution path into maximal open segments, whose union we denote by $\pi_Q$, 
along which the open edges $e_b,e_\ell,e_t$, and $e_r$ are fixed, and these edges are unique, 
namely we exclude path points where two edges of $K$ simultaneously touch one edge of $W$, 
or simultaneously attain the largest $x$- or $y$-coordinate of the intersection with $h_{xy}$. 
In-between those segments of $\pi$ there are points (or segments) along the path, where 
a vertex of $K$ touches the left edge or the bottom edge of $W$, or a vertex of $K$ 
that lies in $h_{xy}$ has the largest $y$-coordinate or $x$-coordinate within 
this cross-section, or a face of $K$ touches the left edge or the bottom edge of $W$, 
or the maximum in $y$ or $x$ of the intersection of $K$ with $h_{xy}$ is attained 
along a segment, which is the intersection of a face of $K$ with $h_{xy}$.

The maximal connected segments of $\pi_Q$, each falls into one of ${n \choose 4}$ 
categories according to the choice of the four edges $e_b,e_\ell,e_t$, and $e_r$.

The efficient planning scheme is as follows. Our problem induces a six-dimensional 
configuration space (C-space for short) for the motion of $K$. However, for efficiency, 
we will first construct a collection of four-dimensional sub-spaces of the full-dimensional 
configuration space---each of them will be represented explicitly, and together they will 
contain the path segments in $\pi_Q$ (so that each path segment is contained in one of these subspaces).
Then we will add connections between these four-dimensional C-spaces that will cover the motion along $\pi\setminus \pi_Q$. 
We will then construct a discretization of the free space, called the \emph{connectivity graph}~\cite{hss-amp-18},
which will capture the connectivity of the free space in which $\pi$ lies, and will 
enable planning canonical motions for $K$ through $W$.

\subsection{Four-dimensional C-spaces}

For each pair $e_\ell$, $e_b$ of edges of $K$, let $C(e_\ell,e_b)$ denote the four-dimensional configuration (sub-)space 
of our full six-dimensional C-space, consisting of those placements of $K$ at which (i) $e_\ell$ and $e_b$ intersect
the $xy$-plane $h_{xy}$, (ii) $e_\ell\cap h_{xy}$ lies on the left edge of $W$ (a portion of the $y$-axis) and is the 
leftmost point of $K\cap h_{xy}$, and (iii) $e_b\cap h_{xy}$ lies on the bottom edge of $W$ (a portion of the $x$-axis) 
and is the bottommost point of $K\cap h_{xy}$. (Top and bottom are with respect to the $y$-direction.) 
Note that $C(e_\ell,e_b)$ is indeed four-dimensional.

For each additional pair $e_r$, $e_t$ of edges of $K$, let $C_{e_r,e_t}(e_\ell,e_b)$ denote 
the portion of $C(e_\ell,e_b)$ in which (in addition to (i)--(iii)), we have
(iv) $e_r$ and $e_t$ intersect $h_{xy}$, (v) $e_r\cap h_{xy}$ is the 
rightmost point of $K\cap h_{xy}$, and (vi) $e_t\cap h_{xy}$ is the topmost point of $K\cap h_{xy}$. 
See Figure~\ref{fig:canon-in-W}.
%
Properties (i)--(vi) ensure that the C-space $C_{e_r,e_t}(e_\ell,e_b)$ has constant complexity; see Lemma~\ref{lem:constant} below.


Note that not all configurations in $C_{e_r,e_t}(e_\ell,e_b)$ are necessarily free, 
as this space may contain configurations in which the intersection of $K$ with $h_{xy}$ 
is not contained in $W$ (because $e_r\cap h_{xy}$ or $e_t\cap h_{xy}$ lies outside $W$). 
We denote by $F_{e_r,e_t}(e_\ell,e_b)$ the portion of $C_{e_r,e_t}(e_\ell,e_b)$ 
that represents \emph{free} (or \emph{valid}) configurations, namely configurations in which 
the intersection of $K$ with $h_{xy}$ is fully within the closure of the rectangle $W$. 
(As just discussed, this will be the case if and only if both
$e_r\cap h_{xy}$ and $e_t\cap h_{xy}$ lie inside $W$.) 
We call $F_{e_r,e_t}(e_\ell,e_b)$ the \emph{free space} of $C_{e_r,e_t}(e_\ell,e_b)$.

\begin{lemma}\label{lem:constant}
The combinatorial complexity of the free space $F_{e_r,e_t}(e_\ell,e_b)$ 
is bounded by a constant (independent of the complexity of $K$).
\end{lemma}
\noindent{\bf Proof.}
Consider a fixed quadruplet $q=\{e_r,e_t,e_\ell,e_b\}$ of edges of $K$. Let $\Phi_q$ be the set of at most eight facets of $K$ that are
incident to the edges in $q$. For each such facet, say a facet $g$ incident to $e_r$, we sweep a line parallel to $e_r$ on $g$ away from $e_r$
till it touches another vertex of $g$, where its intersection with $g$ is the line segment (or a point) $s$. Let $g'$ be the trapezoid (or
triangle) contained in $g$ and comprising the area swept by the line between $e_r$ and $s$. Let $\Phi'_q$ be the collection of these eight
trapezoids. Let $K_q$ be the convex hull of the trapezoids in $\Phi'_q$. We claim that $F_{e_r,e_t}(e_\ell,e_b)$ for $K$ and for $K_q$ are identical.
Indeed, we only need to show that if, at some placement, $K_q\cap h_{xy}\subset W$ then $K\cap h_{xy}$ is also contained in $W$. 
Since $K_q\cap h_{xy}\subset W$, the two edges of $K_q$ incident to $e_\ell\cap h_{xy}$ (resp., to $e_b\cap h_{xy}$, $e_r\cap h_{xy}$, 
$e_t\cap h_{xy}$) define a wedge that contains $K_q\cap h_{xy}$ and lies fully to the right of $e_\ell$ (resp., above $e_b$, 
to the left of $e_r$, below $e_t$). The intersection of these wedges is therefore contained in $W$. But this intersection
also contains $K\cap h_{xy}$, which implies our claim.
Since each of $K_q$ and $W$ has constant descriptive complexity, then by standard arguments in algorithmic motion planning,
the complexity of $F_{e_r,e_t}(e_\ell,e_b)$ for $K_q$ is bounded by a constant. 
Hence the complexity of $F_{e_r,e_t}(e_\ell,e_b)$ for $K$ is also bounded by a constant, as asserted.  
$\Box$

By the same arguments as in the proof of Lemma~\ref{lem:constant}, the decomposition of the C-space $C_{e_r,e_t}(e_\ell,e_b)$ into free and
forbidden cells is induced by a constant number of surface patches, each of  constant descriptive complexity~\cite{hs-a-18}.
Being of constant complexity, we can use standard tools, such as vertical decomposition~\cite{hs-a-18} or the Collins
decomposition~\cite{SS83b} to construct  the arrangement of surfaces defining $C_{e_r,e_t}(e_\ell,e_b)$ in time $O(1)$.  
It is easy to detect and maintain only the free portion $F_{e_r,e_t}(e_\ell,e_b)$ of this arrangement, in $O(1)$ time as well.

We thus obtain $O(n^4)$ constant-size arrangements. It remains to connect
between the boundaries of their free cells. 

Let $\F$ be the collection of these four-dimensional free spaces, for all quadruplets $(e_\ell,e_b,e_r,e_t)$ of edges 
of $K$ that can simultaneously be intersected by a plane. 

We wish to use $\F$ to plan a valid motion for $K$ from a configuration where $K$ is fully above the window 
(contained in the halfspace $z>0$) to a configuration in which it is fully below the window (contained in the 
halfspace $z<0$). We defer the handling of extreme configurations where $K$ intersects $h_{xy}$ for the 
first (resp.\ last) time to the sequel,
and focus first on transitions between free spaces where there is a 
three-dimensional volume of $K$ both below and above $h_{xy}$. 

Our description here assumes \emph{general position}. For example we preclude polytopes~$K$ that have two 
coplanar, or two parallel facets, or polytopes for which four vertices \emph{not all incident to the same face} 
lie on a single plane, and so on. We do allow the polytope~$K$ to be non-simple, namely have many edges incident 
to a vertex, and non-simplicial, namely have many edges and vertices on the boundary of a facet.
 
The free space $F_{e_r,e_t}(e_\ell,e_b)$ may comprise several connected four-dimensional cells. We apply the Collins
decomposition~\cite{SS83b} to these cells and construct a connectivity graph $G$, whose nodes represent cells of the 
decomposition over all free spaces in $\F$, and two nodes are connected by an arc in $G$ if the corresponding cells 
belong to the Collins decomposition of the same space $F_{e_r,e_t}(e_\ell,e_b)$ and share an artificial boundary added 
by the decomposition (so they are both part of the same connected component of $\F$). This process creates a (large) 
number of (at most $O(n^4)$) connected components of $G$, which we still need to connect.
The next subsection is devoted to making these connections.
 
\subsection{Connecting between cells of the free four-dimensional C-spaces}

Thus we turn to describing the provision in our data structure for the portions of the path $\pi$ 
that are not contained in the interior of the free spaces $\F$, 
namely the path portions $\pi\setminus\pi_Q$. 
Recall that these portions comprise configurations where one of the following situations occurs.

\begin{description}
\item[Case~(a)] 
Exactly one vertex of $K$ lies in $h_{xy}$, and it is extreme in either the $x$- or the $y$-direction.
\item[Case~(b)] 
Two or three vertices of $K$ lie in $h_{xy}$, and each one of them is extreme in either the $x$- or the $y$-direction.
\item[Case~(c)] 
Exactly one facet of $K$ that intersects $h_{xy}$ is parallel to the 
$x$- or to the $y$-axis, and no vertex of $K$ lies in $h_{xy}$. 
\item[Case~(d)] 
Either exactly one facet of $K$ that intersects $h_{xy}$ is parallel to the 
$x$- or the $y$-axis and one or more vertices of $K$ lie in $h_{xy}$, 
or two facets intersect $h_{xy}$, one parallel to the $x$-axis and 
one parallel to the $y$-axis; some vertices of $K$ may also appear
on $h_{xy}$ and replace some of the four extreme edges. 
\end{description}

\medskip
We next consider such configurations of $\pi\setminus\pi_Q$ in relation to the boundary of 
free cells in $\F$. Notice that Cases~(a) and~(c) correspond to 
three-dimensional surface patches bounding the four-dimensional cells in $\F$. 
The two other cases, (b) and~(d), correspond to lower-dimensional boundaries of these three-dimensional patches. 
Cases where only connections of types~(b) or~(d) are possible are degenerate and we ignore them.
We focus on Cases~(a) and~(c).
 
The free space $F_{e_r,e_t}(e_\ell,e_b)$ is bounded by three-dimensional surface patches of 
the following types, based on the type of the configurations that they comprise. 

\begin{description}
\item[Type~(i)] 
$K$ is placed such that an endpoint of one of the four edges $e_\ell,e_b,e_r,e_t$ is in $h_{xy}$. 
This type corresponds to configurations of Case~(a) above.
\item[Type~(ii)] 
A facet of $K$ incident to $e_\ell$ touches the left edge of $W$ (namely they overlap 
in a segment rather than a single point) or a facet of $K$ incident to $e_b$ touches the 
bottom edge of $W$. This type corresponds to configurations of Case~(c) above.
\item[Type~(iii)] 
A facet of $K$ incident to $e_r$ becomes parallel to the right edge of $W$ or a facet 
of $K$ incident to $e_t$ becomes parallel to the top edge of $W$ (while lying inside $W$). 
This type also corresponds to configurations of Case~(c) above.
\item[Type~(iv)] 
The intersection of $e_r$ with $h_{xy}$ lies on the right edge of $W$ or
the intersection of $e_t$ with $h_{xy}$ lies on the top edge of $W$.
This type is special in that it induces a \emph{constraint} in the standard sense of 
the study of motion planning: it is a three-dimensional surface patch such that on one side 
(locally, ignoring other $K$-$W$ contacts) it has valid configurations and on the other side 
it has forbidden configurations since for example $e_r$ penetrates (beyond) the frame of the window $W$.  
\end{description} 

Surface patches of Types~(i) through~(iii) induce connections between free four-dimensional cells 
(cells that are already recorded as nodes in our connectivity graph $G$), and will give rise to 
new nodes and arcs in $G$.
Type~(iv) surface patches bound free four-dimensional cells but they are impassable, 
and hence they will not introduce new features in $G$.

Given a pair of edge quadruplets $e_\ell^1,e_b^1,e_r^1,e_t^1$ and $e_\ell^2,e_b^2,e_r^2,e_t^2$, if their respective 4D free C-spaces,
$F_1=F_{e_r^1,e_t^1}(e_\ell^1,e_b^1)$ and $F_2=F_{e_r^2,e_t^2}(e_\ell^2,e_b^2)$ are candidates to be neighbors in the free space, then we need
to determine the overlap of their respective boundaries. Since each of $F_1$ and $F_2$ has constant complexity, this operation as well takes
constant time. The main goal of the remainder of this section is to show that the overall number of pairs $(F_1,F_2)$ that need to bo
considered, out of the potential $O(n^8)$ such pairs, is only $O(n^4)$. Furthermore, we will show that we can find all these neighboring pairs
in near-quartic time. We start with Type~(ii) surfaces since 
(a) they are simpler to handle, and (b) we will use the analysis of this case as part of the analysis of Type~(i) surfaces. 
The arguments for Type~(iii) surfaces are analogous to those for Type~(ii) surfaces. 
As mentioned above Type~(iv) surfaces induce no connections.

We also add new nodes and arcs to the connectivity graph for boundary surfaces that connect between cells in $\F$ and configurations where $K$ is fully contained in $z>0$ or in $z<0$. These will be needed to complete the motion path in its two endpoints. See Section~\ref{sec:details} for more details.  


\subsubsection{Connection through Type~(ii) surface patches} \label{ssec:typeii-surfaces}

We demonstrate these connections for the facet $f$ of $K$ incident to $e_b$ and to $e_{b'}$, where $e_{b'}$ 
is the other edge of $f$ that touches the bottom edge of $W$ at this critical event during the motion of $K$. 
The other Type~(ii) surfaces, where a facet of $K$ touches the left edge of $W$, can be handled similarly.

We connect the nodes in $G$ representing the boundary of the four-dimensional cell 
$\tau_1\in F_{e_r,e_t}(e_\ell,e_b)$ to the nodes
representing the boundary of the four-dimensional cell $\tau_2\in F_{e_r,e_t}(e_\ell,e_{b'})$. 
On the face of it, it may seem that over all of $\F$ this could lead to $O(n^5)$ connections. 
However, we show next that there are at most $O(n^4)$ such connections.

\setcounter{theorem}{0}
\begin{lemma}\label{lem:no_of_quintuplets}
The maximum number of quintuplets $(e_r,e_t,e_\ell,e_b,e_\xi)$, where $e_r,e_t,e_\ell,e_b$ are as defined above, 
and $e_\xi$ is another edge of $K$ whose intersection with the $xy$-plane $h_{xy}$ has the same $x$- (respectively, $y$-) 
coordinate as the intersection with $h_{xy}$ of $e_\ell$ or $e_r$ (respectively, $e_b$ or $e_t$), is $O(n^4)$. 
All these quintuplets can be computed in $O(n^3\lambda_q(n)\log n)$ time for some small constant $q$.
\end{lemma}

\noindent{\bf Proof.}
We prove the lemma for the quintuplets of the form $(e_r,e_t,e_\ell,e_b,e_{b'})$, where $e_{b'}$ touches, together
with $e_b$, the bottom edge of $W$ (in particular $e_b$ and $e_{b'}$ lie on the boundary of a common face of $K$). 
The proof for the other types is analogous. 
Let $f$ be a facet of $K$ and let $e_b$ be an edge of $f$. 
We place $K$ so that (i) both $f$ and $e_b$ intersect $h_{xy}$, (ii) $e_b\cap h_{xy}$ is the origin, and
(iii) the segment $f_{xy} := f\cap h_{xy}$ lies on the positive $x$-axis (and its left endpoint is at the origin).
Then $K$ has three degrees of freedom of motion: the dihedral angle between the plane $h_f$ that contains $f$ and 
$h_{xy}$, and two degrees of rotating $f$ about the origin and of sliding $f$ with $e_b$ touching the origin.
We denote this configuration space by $C_x(f;e_b)$. Notice that this is an artificial C-space, which is not 
a subspace of any of our C-spaces; we only use it here to bound the number of (and later on to find the) 
valid combinations of edges for which a configuration of Type~(ii) is possible. 
 
We decompose $C_x(f;e_b)$ by $O(n)$ surfaces, one surface for each vertex of $K$, 
and two surfaces for each face of $K$ other than $f$.
An edge of the quintuplet changes in one of the following cases: 

\medskip
\noindent
(i) A vertex of $K$ passes from one side of $h_{xy}$ to the other. To
account for this change, for each vertex $v$ of $K$ (including the vertices of $f$
that are not incident to $e_b$) we add a surface within $C_x(f)$ 
comprising all the configurations in which $v$ lies on $h_{xy}$. 
Notice that we account here for more changes than are necessary, as sometimes the vertex that crosses $h_{xy}$ is not extreme in the $x$- nor in the $y$-direction. However, this excess consideration has no effect on the asymptotic upper bound that we derive here.

\medskip
\noindent
(ii) A face $g$ of $K$ incident to the edge $e_t$ becomes parallel to the $x$-axis. 
Immediately after this occurrence, another edge on the boundary of $g$ may become the new $e_t$. 
To account for such changes, we consider the line $L_{f,g}$ of intersection between the planes
supporting $f$ and $g$. By the general position assumption this line is well defined. 
We now add a surface for all the configurations in which $L_{f,g}$ is parallel to $x$-axis. 
We add such a surface for each face of $K$ other than $f$. 

\medskip
\noindent
(iii) Similar to the change described in (ii), $e_\ell$ or $e_r$ may change when a face 
incident to them becomes parallel to the $y$-axis. To account for these changes we add a surface
comprising of configurations where $f$ and $g$ contain a pair of orthogonal lines that lie both on $h_{xy}$.
Again, we generate one surface for each face of $K$ other than $f$.

\medskip

An important property of these surfaces is that they are patches of algebraic surfaces of constant
description complexity. This follows from the fact that the left endpoint of $f_{xy}$ always lies
on the fixed edge $e_b$. Hence, these surfaces decompose $C_x(f;e_b)$ into $O(n^3)$ cells, where 
in each cell the three edges of $K$ whose intersection with $h_{xy}$ determine the leftmost, 
rightmost, and topmost vertices of the intersection of $K$ with $h_{xy}$, as well as the two 
edges of $K$ on the boundary of $f$ that touch the $x$-axis (the left of which is $e_b$),
are all fixed. Hence, the number of pairs of adjacent cells
is also $O(n^3)$. We repeat this argument for each pair $(f;e_b)$ of a facet $f$ of $K$ 
and the incident edge $e_b$, note that the number of these pairs is $O(n)$ (this number
is just twice the number of edges of $K$), and conclude that we have a total of $O(n^4)$ 
quintuplets, where each quintuplet, say $(e_r,e_t,e_\ell,e_b,e_{b'})$, translates into a 
pair of neighboring four-dimensional spaces $F_{e_r,e_t}(e_\ell,e_b)$ and
$F_{e_r,e_t}(e_\ell,e_b')$ (both abutting on the same Type~(ii) surface). 

The argument for quintuplets where the fifth edge $e_\xi=e_{t'}$ and $e_t, e_{t'}$ 
are such that their intersection with $h_{xy}$ simultaneously attain the largest 
$y$-coordinate among all edges of $K$ is similar.
Up till now, we took a solution path for $K$ through $W$ and turned it into a canonical 
path $\pi$, where at all times $K$ touches the bottom and left edges of $W$. 
Let us now consider another canonical path $\pi'$, where at all times $K$ touches 
the top and left edges of $W$. The quadruplets along the paths remain the same quadruplets. 
We only worry about which pairs of quadruplets are such that their subspaces need to be 
connected to one another. The arguments about the current type of quintuplets are 
analogous to those handled above, where $e_\xi=e_{b'}$, and the same bound can be derived.
Analogous arguments apply when the fifth edge $e_\xi=e_{r'}$ and $e_r, e_{r'}$ are such 
that their intersection with $h_{xy}$ simultaneously attain the largest $x$-coordinate among all edges of $K$.

As for the computation of these quintuplets, we construct the three-dimensional arrangement,
as described in the proof above, for every pair $(f;e_b)$ of a facet and an incident edge of
$K$, from which we obtain the quintuplets. It takes $O(n^2\lambda_q(n)\log n)$ per such
facet-edge pair~\cite{DBLP:journals/dcg/BergGH96}, where $q$ is a
constant parameter that is determined by these surfaces, for total time $O(n^3\lambda_q(n)\log n)$.
$\Box$

\subsubsection{Connection through Type~(i) surface patches} 

To facilitate these connections, we introduce additional four-dimensional subspaces,
which represent placements of $K$ at which a vertex of $K$ lies on $h_{xy}$.
Specifically, we demonstrate these connections when an endpoint $p$ of $e_\ell$ 
(namely $p$ is a vertex of $K$) lies in $h_{xy}$. All other Type~(i) surfaces can be handled similarly. 

Let $E_p$ denote the set of edges of $K$ incident to $p$. Our goal here is to make the right 
connections between the free spaces of the quadruplets \mbox{$\{(e,e_b,e_r,e_t) \mid e\in E_p\}$}.
We cannot afford to make connections between all possible pairs of edges of $E_p$, as this might
increase the overall number of connections to be $O(n^5)$ (which will be the case when the
vertices $p$ have large degree). Therefore we apply an infinitesimal transformation to $K$, 
which will allow us to use the analysis of Section~\ref{ssec:typeii-surfaces} 
and apply it to this case.  

\begin{figure}[htbp]
	\begin{center}
	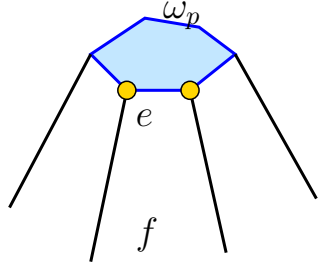
		\caption{\sf{The polygon $\omega_p$.}}
		\label{fig:omega_p}
	\end{center}
\end{figure}

We truncate $K$ into the still convex polytope $K_{\eps}$, for arbitrarily small $\eps>0$, as follows. 
We take a plane $\Omega_p$ supporting $K$ at $p$ (but not parallel to any facet of $K$), and move it 
parallel to itself, by distance $\eps$, into $K$, such that $p$ is on one side of the shifted $\Omega_p$
and all the other vertices of $K$ are on the other side. The intersection of $K$ and $\Omega_p$ 
is a tiny convex polygon, which we denote by $\omega_p$, with a vertex for each edge of $E_p$. 
We represent each edge $e$ of $\omega_p$ by the face of $K$ whose intersection with $\Omega_p$ is $e$. 
(The choice of the orientation of $\Omega_p$ is arbitrary, but it will have an artificial,
non-essential effect on the concrete connections around $p$ that we will construct.)
See Figure~\ref{fig:omega_p} for an illustration.

We repeat this process for every vertex of $K$, thereby obtaining $K_\eps$, which formally is
the intersection of $K$ with all the halfspaces bounded by the shifted planes $\Omega_p$ and not containing
the respective vertices $p$ 
Notice that the polytope $K_\eps$ has $O(n)$ edges, which we classify
either as \emph{real edges}, if they are truncated versions of the edges of $K$, or as
\emph{virtual edges}, if they are edges of the tiny  new facets $\omega_p$.
Let $\F_\eps$ denote the collection of the four-dimensional free C-spaces
$F_{e_r,e_t}(e_\ell,e_b)$ for $K_\eps$. Each quadruplet
$(e_r,e_t,e_\ell,e_b)$ of edges of $K_\eps$ may involve any 
combination of real and virtual edges, as long as there exists a plane that meets all four edges.

We apply Lemma~\ref{lem:no_of_quintuplets} to $K_\eps$, to ascertain that there are only $O(n^4)$ 
possible pairs of edge quadruplets for $K_\eps$ that encode connections between pairs of subspaces
in $\F_\eps$. As before, we focus for concreteness on quintuplets
$(e_r,e_t,e_\ell,e_b,e_{b'})$ that encode connections in which $e_b$ is replaced by $e_{b'}$.
The new situations that we need to consider are when one or both edges
$e_b,e_{b'}$ belong to a virtual facet $\omega_p$, for some vertex $p$ of $K$.

Note that while $F_{e_r,e_t}(e_\ell,e_b)$ is four-dimensional for $K_\eps$, it may shrink 
to a lower-dimensional space for $K$ itself, as we let $\eps$ tend to zero,
when some of these four edges are virtual.

We claim that
if $K_\eps$ has a valid path $\pi_\eps$ 
through $W$ for every $\eps>0$, then $K$ has a similar valid path $\pi$, which is 
a limit path of the set $\{\pi_\eps\}_{\eps>0}$, in (a suitably defined compact subspace of) 
the space of paths in the configuration space of $K$, equipped with the Hausdorff metric. 
By passing to a subsequence if needed, we may assume that the sequence of quadruplets of 
edges of $K_\eps$ that is assigned to $\pi_\eps$ is fixed as $\eps$ converges to $0$.
Denote this sequence as $\sigma$.

For each quadruplet $(e_r,e_t,e_\ell,e_b)$ in $\sigma$, some of its edges are real edges 
of $K$, and some 
are virtual edges of the polygons $\omega_p$. 
Assume first that just one of these edges, say $e_b$, is virtual, bounding some
virtual facet $\omega_p$, for some vertex $p$ of $K$. In the limit, as $\eps\to 0$,
the free four-dimensional space $F_{e_r,e_t}(e_\ell,e_b)$ becomes three-dimensional,
and consists of placements of $K$ at which $p$ lies on the bottom edge of $W$. 
Similarly, the space becomes two-dimensional (resp., one-dimensional, zero-dimensional) when the number
of virtual edges is two (resp., three, four).


Our strategy is as follows. We use symbolic computation, taking $\eps$ to be infinitesimal (see, 
e.g.,~\cite{DBLP:journals/jcss/Yap90}).
We construct the closures of the virtual subspaces $F_{e_r,e_t}(e_\ell,e_b)$ (namely those
for which at least one edge is virtual) and their Collins decompositions in constant time
per subspace, for a total cost of $O(n^4)$. The resulting subcells of the decompositions
are added as new vertices of our connectivity graph.
We then compute all the $O(n^4)$ connecting quintuplets, such as
$(e_r,e_t,e_\ell,e_b,e_{b'})$, as provided by Lemma~\ref{lem:no_of_quintuplets},
in $O(n^3\lambda_q(n)\log n)$ time, focusing only on quintuplets where at least one
of the edges is virtual. For each such quintuplet we determine, in constant time, whether
there is a free connection between $F_{e_r,e_t}(e_\ell,e_b)$ and $F_{e_r,e_t}(e_\ell,e_{b'})$,
and if so add the corresponding edge to the connectivity graph.

All these computations are performed symbolically (on $\eps$). If we think of $\eps$
as infinitesimal but still nonzero then the machinery developed here is not significantly
different from that developed in Section~\ref{ssec:typeii-surfaces}.
Nevertheless, the connections formed by this procedure can also be given
concise geometric interpretation in terms of the limit (original) polytope $K$. For example,
assume that $e_b$ is a real edge with endpoint $p$ and $e_{b'}$ is a virtual edge of 
$\omega_p$. Connecting $F_{e_r,e_t}(e_\ell,e_b)$ and $F_{e_r,e_t}(e_\ell,e_{b'})$
means that $K$ moves with $e_b$ touching the bottom edge of $W$ until the contact point
reaches $p$. We then switch to placements at which the virtual edge $e_{b'}$ contacts
the bottom edge of $W$. The meaning of this latter scenario is that $p$ touches the 
bottom edge of $W$, and its two incident edges that bound the incident face 
corresponding to $e_{b'}$ (one of which is $e_b$) are such that one of them
lies in the positive halfspace $z>0$ and the other lies in the negative halfspace $z<0$.

Similarly, a connection between $F_{e_r,e_t}(e_\ell,e_b)$ and $F_{e_r,e_t}(e_\ell,e_{b'})$,
where both $e_b$ and $e_{b'}$ are virtual edges of the same virtual facet $\omega_p$
means that $K$ is positioned so that $\omega_p$ overlaps the $x$-axis. (Such placements
depend on the artificial choice of the orientation of the supporting plane $\Omega_p$ at $p$.)
The fact that $e_b$ and $e_{b'}$ are the edges of $\omega_p$ that touch the $x$-axis
simply means that the pencil of edges incident to $p$ is split into two contiguous subsequences,
where the edges in one subsequence lie in the positive halfspace $z>0$ and those in the other
subsequence lie in the negative halfspace $z<0$, and the split occurs at the real facets of $K$
adjacent to (and representing) $e_b$ and $e_{b'}$. 

Omitting further details about these virtual free subspaces, we obtain the desired
connectivity graph $G$, and can then use it to plan a collision-free motion of $K$ through $W$ in
a standard fashion. Typically, the path in $G$ will have subpaths that start at a real
subspace $F_{in}$, continue along a sequence of virtual subspaces, and end at another
real subspace $F_{out}$. This means that the motion reaches a placement at which $p$
lies on the, say, bottom edge of $W$, and then $K$ rotates around $p$ (with $p$ possibly
sliding along the $x$-axis), switching between virtual edges of $\omega_p$ in the manner
explained above, until $p$ leaves the $x$-axis and another edge incident to $p$ starts
touching the axis. In the notations introduced earlier, the portions of the path $\pi$
that traverse real free subspaces comprise $\pi_Q$ and those
that traverse virtual free subspaces comprise $\pi\setminus\pi_Q$.

\subsubsection{Connection through Type~(iii) surface patches} 

The connections are similar to those for Type~(ii). The number of such connections among 
all decomposition cells in $\F$ can be bounded by $O(n^4)$ (and can be computed in 
$O(n^3\lambda_q(n)\log n)$ time), arguing as in the analysis of Type~(ii) connections.


\subsection{More algorithmic details}\label{sec:details}

We construct the closure of $F_{e_r,e_t}(e_\ell,e_b)$ and its Collins decomposition in constant time. This decomposition induces a subgraph of
$G$, with a node for each cell in the decomposition and arcs connecting nodes that represent cells of the decomposition that are adjacent
through an artificial wall added by the decomposition. This subgraph has constant complexity and each cell contains the full list of
constraints (polynomials) defining it. Notice that this decomposition contains lower-dimensional cells as well.

We store these subgraphs at an $n\times n \times n \times n$ matrix $M$, indexed by quadruplets of edges of $K$. 
(It is $n$ for the number of original edges of $K$, by definition, but will increase to $cn$, for some constant $c$, for $K_\eps$.)



When we check for a possible connection (out of the $O(n^4)$ possible connections computed by the preceding algorithms)
between, say, the free space $F_{e_r,e_t}(e_\ell,e_b)$ and the free space $F_{e_r,e_t}(e_\ell,e_b')$, 
we consider the two relevant entries in $M$, and apply a constant-time procedure to find all the 
overlapping elements (namely, cells in the Collins decomposition of the free space) of the boundaries 
of both free spaces. We connect the corresponding nodes in the two graphs, one for each of the free spaces, 
by adding an edge to $G$.

We also add two special nodes to the connectivity graph $G$: $v_{\rm above}$ and $v_{\rm below}$, 
which represent the free space where $K$ is strictly above (respectively below) the $xy$-plane. 
We connect $v_{\rm above}$ to every node in $G$ representing (free) configurations in which $K$ 
is fully contained in the closed half-space $z\geq 0$: We traverse the Collins decomposition cells 
of the boundaries of the closure of all the free spaces $\F$ and exhaustively look for cells that 
contain configurations of this type. If we find such a configuration $c$ inside a cell $\eta$, 
we mark the edge $e$ in the connectivity graph between the node of $\eta$ and $v_{\rm above}$ 
with the configuration $c$---it means that motion along $e$ is a simple rigid collision-free 
motion that will bring $K$ from an arbitrary placement in $z>0$ to the configuration $c$.
Similarly, we connect $v_{\rm below}$ to every node in $G$ representing configurations in 
which $K$ is fully contained in the closed half-space $z\leq 0$.

Finally we look for a path in $G$ between $v_{\rm above}$ and $v_{\rm below}$. 
If such a path in $G$ exists it can be easily transformed into a path $\pi$ that 
will take $K$ through the window. If such a path in $G$ does not exist, we can 
safely notify that $K$ cannot pass through $W$ by any rigid motion. 


Adding all the connections between the four-dimensional free spaces $F_{e_r,e_t}(e_\ell,e_b)$
takes $O(n^4)$ time. This is subsumed by the time required to compute all the quintuplets as 
in Lemma~\ref{lem:no_of_quintuplets}. In summary, we have:

\begin{theorem}
Given a convex polytope $K$ with $n$ edges and a rectangular window $W$, we can find a 
collision-free righd motion of $K$ through $W$, with all six degrees of freedom, or determine 
that no such path exists, in time $O(n^3\lambda_q(n)\log n)$, for some small constant $q$.
The algorithm requires $O(n^4)$ storage.
\end{theorem}

%
\setcounter{theorem}{3}

\section{Conclusion and further research} \label{sec:conclusion}

In this paper we have studied a variety of problems concerning collision-free motion 
of a convex polytope through a planar window, under several kinds of allowed motion --- 
sliding (translating in a fixed direction), purely translational motion, and general motion. 
We have presented several structural properties and characterizations of such motions, 
and obtained efficient algorithms for several special cases, as well as for the general case.

There are several open problems and directions for further research. 
One such direction is to show that the near-quartic upper bound, established in 
Section~\ref{sec:6dofs}, on the cost of the general motion planning problem for 
$K$ and a rectangular window $W$ is almost tight in the worst case, in the specific sense 
of establishing a lower bound $\Omega(n^4)$ on the worst-case complexity of the 
resulting free configuration space, a property that we conjecture to hold, and
have in fact an initial plan for establishing this bound.

In addition, in Section~\ref{app:rotations-needed}, we presented an example in 
which a rotation is needed to pass a polytope through a rectangular window. 
However, in this construction we only used rotation about the line perpendicular 
to the plane that contains the window. This suggests the conjecture that every 
convex polytope that can pass through a rectangular window $W$ can also pass 
through $W$ by a motion consisting of arbitrary translations and rotations only 
about the line perpendicular to the plane of $W$. The results of 
Section~\ref{sec:circular-window} show that for circular windows this claim is 
false in general, but the status of the conjecture is still open for a rectangular window.

It is also not clear what can be said about the motion of a general non-convex 
polytope through a rectangular, general convex, or even non-convex window. 
There are several variants of this question, depending 
on the type of motion that we allow, both in terms of structural properties of the 
motion, and of the efficiency of algorithms for performing it.

\paragraph{Acknowledgements.}
We deeply thank Pankaj Agarwal and Boris Aronov for useful interactions concerning this work. 
In particular, Boris has suggested an alternative proof of the key topological property in the analysis in 
Section~\ref{sec:projection}, and Pankaj has been instrumental in discussions concerning the range searching 
problem in Section~\ref{sec:1d}. We are also grateful to Lior Hadassi for interaction involving the topological 
property presented in Section~\ref{sec:projection}, and to Eytan Tirosh for help with the alternative proof 
of Lemma~\ref{lem:vertical}.


\end{document}